\documentclass[aps,pra,amssymb,twocolumn,showpacs,supersriptaddress,groupeaddress]{revtex4-1}
\usepackage{amssymb}
\usepackage{amsmath}
\usepackage{graphicx}
\usepackage{float}
\usepackage{dcolumn}
\usepackage{bm}
\usepackage{natbib,hyperref}
\usepackage{hyperref}

\usepackage{color}
\hypersetup{colorlinks=true, 
    linkcolor=red,          
    citecolor=magenta,        
    filecolor=gree,      
    urlcolor=blue           
}

\usepackage[normalem]{ulem}

\allowdisplaybreaks[4]

\begin{document}


\title{Fluctuations and stability of a fast driven Otto cycle}

\author{Ana Laura Gramajo$^{1}$, Elisabetta Paladino$^{2}$, Jukka Pekola$^{3}$, Rosario Fazio$^{1,4}$}

\affiliation{$^{1}$ The Abdus Salam International Center for Theoretical Physics, Strada Costiera 11, 34151 Trieste, Italy\\
$^{2}$ Dipartimento di Fisica e Astronomia Ettore Majorana, Universit\`a di Catania, Via S. Sofia 64, 95123 Catania, Italy\\
$^{3}$ QTF Centre of Excellence, Department of Applied Physics, Aalto University, P.O. Box 15100, FI-00076 Aalto, Finland\\
$^{4}$ Dipartimento di Fisica, Universit\`a di Napoli ``Federico I'', Monte S. Angelo, I-80126 Napoli, Italy
}

\date{\today}%
\begin{abstract}

We investigate the stochastic dynamics of a thermal machine realized by a fast-driven Otto cycle. By employing a stochastic approach, we find that system coherences strongly affect fluctuations depending on the thermodynamic current. Specifically, we observe an increment in the system instabilities when considering the heat exchanged with the cold bath. On the contrary, the cycle precision improves when the system couples with the hot bath, where thermodynamic fluctuations reduce below the classical Thermodynamic Uncertainty Relation bound. Violation of the classical bound holds even when a dephasing source couples with the system. We also find that coherence suppression not only restores the cycle cooling but also enhances the convergence of fluctuation relations by increasing the entropy production of the reversed process. An additional analysis unveiled that the stochastic sampling required to ensure good statistics increases for the cooling cycle while downsizes for the other protocols. Despite the simplicity of our model, our results provide further insight into thermodynamic relations at the stochastic level.

\end{abstract}
\maketitle

\section{Introduction}

The constantly growing field of emerging new technologies has prompted the interest in understanding of quantum effects of energy manipulation. 
The miniaturization of thermal machines due to the fast development of new technologies has led to an intense investigation on how thermodynamics manifests at nanoscales \cite{Vinjanampathy_2016,Ali_2020,Kosloff_2013,Pekola_2015,Uzdin_2015,Hanggi_2009,Goold_2016}. In this context, quantum thermal machine models offer a simple route to understand 
(and consequently exploit) how small quantum systems exchange energy with thermal reservoirs. Quantum engines have been already explored  in a large variety of physical systems 
like spins \cite{Bouton_2021,Peterson_2019,Ono_2020,Ji_2022,Lindenfels_2019}, cold atoms \cite{Zou_2017,Brantut_2013}, diamonds \cite{Klatzow_2019}, superconducting devices
 \cite{Guthrie_2022,Ronzani_2018,Pekola_2019,Karimi_2016}, trapped ions \cite{Maslennikov_2019,Niskanen_2007,Robnagel_2016,Horne_2020}, and optomechanical devices \cite{Zhang_2014}.
 Furthermore, recent works have shown that quantum thermal machines may also offer a promising route for the optimization of quantum hardware, examples are related to methods to 
 purify a qubit on a quantum processing unit \cite{Solfanelli_2022} or the verification of certain thermodynamics in a related setting \cite{Solfanelli_2021}.  

As physical systems are scaled down, fluctuations in thermodynamic quantities, such as heat or work, may become significant and cannot any longer be 
disregarded \cite{Esposito_2009,Campisi_2011,Verley_2014}. Stochastic schemes offer a well-suited description of these systems allowing the treatment of 
thermodynamics quantities as random variables \cite{Horowitz_2012,Seifert_2005,Seifert_2008}, which may be described, in the quantum realm, in terms of quantum trajectories \cite{Dalibard_1992,Molmer_1996,Brun_2002,Gardiner_1992}. Within this approach, for a specific trajectory, the system dynamics breaks down in a subtle evolution governed by a 
non-hermitian Hamiltonian disrupted by quantum jumps between the system states. Experimentally, quantum trajectories can be accessed by continuously monitoring the 
system \cite{Manzano_2022}, and their observation has already been demonstrated in experiments based on superconducting devices \cite{Minev_2019,Murch_2013,Vijay_2011}. 
Among various proposals, Ref.\cite{Karimi_2020} recently reported a novel measurement protocol to access heat exchanges. The authors studied the main characteristics of the 
jump trajectories in a superconducting setup consisting of a qubit coupled to a heat bath realized as a resistor. Here, the resistor behaves as a nanocalorimeter continuously monitored 
using fluorescent measurements, where changes in the resistor temperature unveil whether a photon is absorbed or emitted.

In the past few years, stochastic schemes have become a powerful tool for studying non-equilibrium quantum systems \cite{Friedman_2018,Campisi_2015,Martinez_2016}. Several 
works studied a set of universal fluctuation relations (FRs) that imposes strict restrictions on the stochastic distribution of thermodynamic quantities \cite{Campisi_2011,Campisi_2011,Leggio_2013,Eposito_2009,Gupta_2020,Manzano_2018}. Although FRs are universal, they require sufficient sampling from the initial ensemble, 
causing poor convergence in many situations \cite{Jarzynski_1997,Yukawa_2000,Buffoni_2022}.  
Recent findings have shown that the thermodynamic cost of generating a specific dissipative process restricts the dispersion of observables \cite{Barato_2015,Horowitz_2020,Pietzonka_2016,Pietzonka_2017,Pietzonka_2018,Gingrich_2016,Liu_2019}. The irreversible entropy production thus sets a lower bound to the 
signal-to-noise ratio, better known as the Thermodynamic Uncertainty Relation (TUR). Indeed, the stability and dispersion of thermodynamic currents play a relevant role in determining 
the thermal machine performance \cite{Bret_2021,Holubec_2017,Holubec_2014,Souza_2022}.

Quantum thermal machines operate at scales where the quantum mechanics dominates the system dynamics. One may expect that quantum coherence may play an important role 
in energy exchanges. Nevertheless, it has not yet been well-established whether coherence offers an advantage in thermal machine performance 
\cite{Latune_2021,Streltsov_2017,Park_2013,Brandner_2017}. An increasing number of studies have found that quantum coherence may enhance the collective capabilities of heat engines \cite{Hammam_2021,Manzano_2019,Camati_2019,Uzdin_2015,Killoran_2015,Scully_2011} and refrigerators \cite{Hammam_2021,Holubec_2018,Kilgour_2018,Du_2018,Correa_2014}. It 
has also shown that quantum coherences may reduce thermodynamic fluctuations below the classical bound \cite{Agarwalla_2018,Kalaee_2021,Menczel_2021,Liu_2021,Timpanaro_2019, 
Arash_2021,Ptaszynnski_2018}. In particular, recent studies discussed the optimization of thermal machines in fast driving regimes \cite{Erdman_2019,Cavina_2021}. Via optimal control 
of sudden quenches, the authors demonstrated that an Otto cycle operating as a heat engine or as a refrigerator universally achieves the maximum power and the maximum cooling rate, 
respectively. In this direction, Ref.\cite{Pekola_2019_2} proposed a driving protocol based on a sudden cycle scheme that avoids the creation of coherence and restores the cooling in a fast Otto cycle.  

Motivated by the works of Refs. \cite{Erdman_2019,Pekola_2019_2} , we study the stochastic characteristics of an Otto cycle realized by a two-level system (TLS) driven by sudden 
quenches. We focus on studying the effects of coherence in the stochastic properties of our working medium. We describe the stochastic dynamic using the well-known Monte Carlo Wave 
Function (MCWF) method \cite{Molmer_1996}, which allowed us to successfully compute the probability distribution of heat exchanges and entropy production.  

This work is organized as follows. In Sec.\ref{sec:2}, we present the basic principle design of our fast-driven Otto cycle. The open-system dynamics, using a Lindblad equation, is described in 
Sec.\ref{sec:3}. Here, we compute numerically and analytically the averaged energetic exchanges and study the role of system coherences in cooling in our fast Otto cycle. We identify the 
different operating regimes and briefly discuss the effects of adding a dephasing noise source into the cycle dynamics. The stochastic approach is described in Sec.\ref{sec:4}. We start by 
presenting the main features of the MCWF method. We then move forward and study the coherence effects on the stochastic characteristics of  the fluctuation realtions. The stability of the machines is 
analyzed by looking at the TURs. The final remarks and conclusions are given in Sec.\ref{sec:5}.

\section{The Otto engine}
\label{sec:2}

\begin{figure}[!htb]
\centering
   \includegraphics[width=8.5cm]{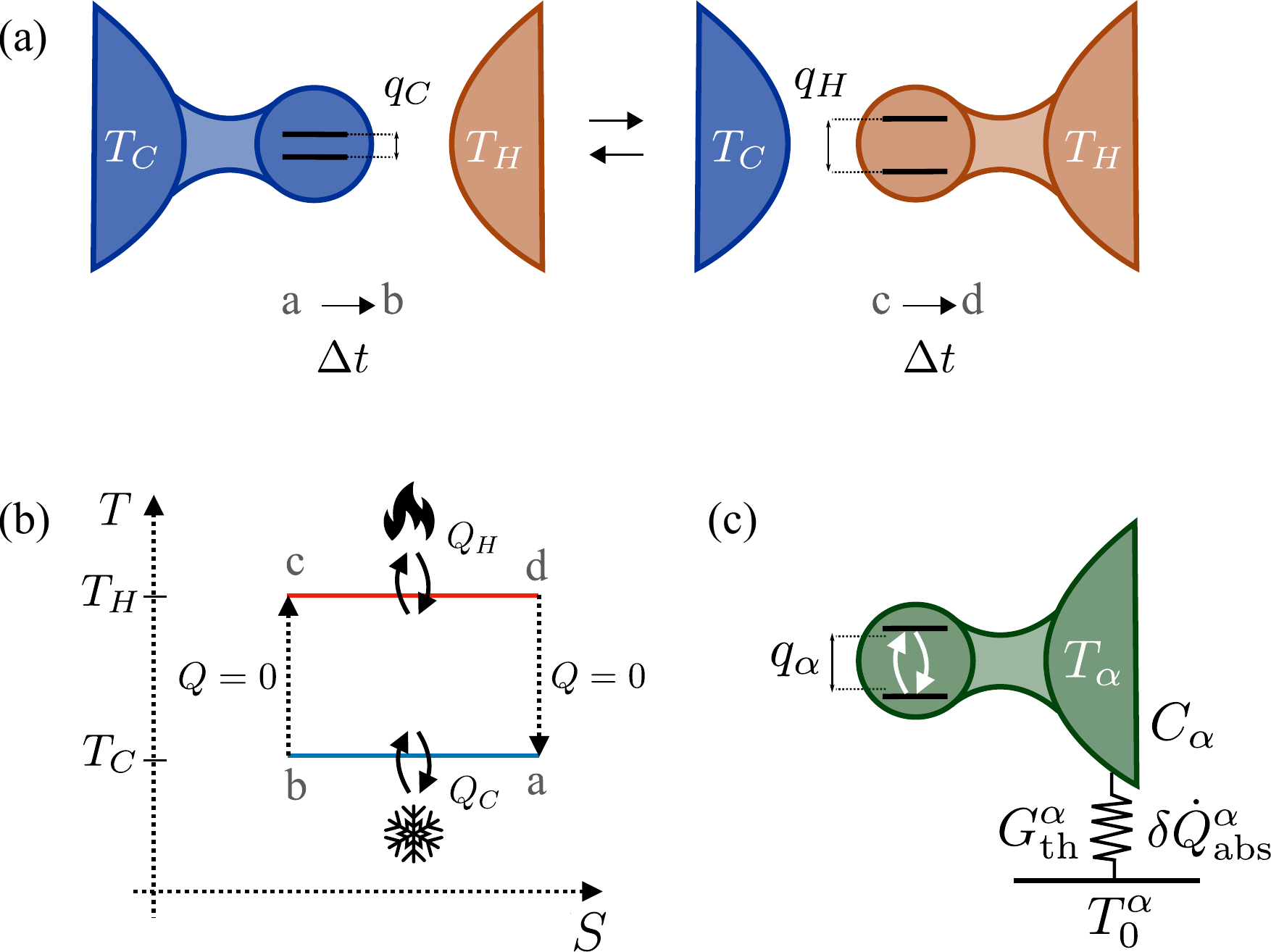} 
   \caption{(a) Schematic plot of the cooling cycle. The TLS couples alternately to one of the baths at a time. The interaction of the TLS with each bath is controlled by the level separation. (b) Cycle diagram. Temperature versus entropy. The system contacts the thermal reservoirs only in the strokes a$\to$b and c$\to$d. There is no dissipation of heat during the strokes b$\to$c and d$\to$a. (c) Model of the measurement setup for each $\alpha$th-bath. $C_{\alpha}$ and $G^{\alpha}_{\text{th}}$ are the heat capacity of the absorber and the thermal conductance to the super bath, respectively. $\delta \dot{Q}^{\alpha}_{\text{abs}}$ is the instantaneous heat current on the absorber. See text for further details.} 
   \label{fig:1}
\end{figure}

The  Otto cycle we are going to analyze in the paper shown in Fig.\ref{fig:1}. It consists of a Two-Level System (TLS) alternatively coupled to cold and hot thermal baths at temperatures 
$T_{C}$ and $T_{H}$, respectively. The Hamiltonian of the working substance (the two level system) is given by
\begin{equation}
	\begin{aligned}
	H_{\alpha} &= -E_{0} ( q_{\alpha} \sigma_{z} + \Delta \sigma_{x} ),
	\label{eq:H_q}
	\end{aligned}
\end{equation} with $\alpha=H,C$, $E_{0}$ the overall energy scale, $\Delta$ the splitting energy, and $q_{\alpha}$ the control parameter. The eigenstates of $H_{\alpha}$ are 
 $|g\rangle_{\alpha} = \Big( \sqrt{1 - \eta_{\alpha}} |+\rangle  + \sqrt{1 + \eta_{\alpha}} |-\rangle  \Big) /\sqrt{2}$ and $|e\rangle_{\alpha} = \Big( \sqrt{1 + \eta_{\alpha}}|+\rangle  - 
 \sqrt{1 - \eta_{\alpha}} |-\rangle \Big) /\sqrt{2}$ respectively , where $|\pm\rangle$ are eigenstates of $\sigma_{z}$ with eigenvalues $\pm 1$. Here, $\eta_{\alpha}= 
 (q_{\alpha}/\Delta) /\sqrt{ 1 + (q_{\alpha}/\Delta)^2} $. The energy level spacing is given by $\Delta E_{\alpha} = 2 E_{0} \sqrt{ q_{\alpha}^2 + \Delta^2 }$.

As illustrated in Fig.\ref{fig:1}(a), the Otto cycle is composed by four strokes: (i) the control parameter is initially set to $q_{\alpha}=q_{C}$ and the TLS couples to the cold bath for the time 
interval $\Delta t$ ($\text{a} \to \text{b}$), (ii) after that time, the energy level spacing is expanded by a sudden change on the control parameter $q_{\alpha}: q_{C} \rightarrow q_{H}$ ($\text{b} \to \text{c}$), 
(iii) the TLS now couples to the hot bath for the same time interval $\Delta t$  ($\text{c} \to \text{d}$), after which, (iv) the energy level spacing is abruptly compressed, 
i.e., $q_{\alpha}: q_{H} \rightarrow q_{C}$ ($\text{d} \to \text{a}$). An equivalent description of our Otto cycle can be done using a temperature-entropy (T-S) diagram. 
Furthermore, as sketched in Fig.\ref{fig:1}(b), the system exchanges energy with thermal baths only in isothermal strokes, i.e.,  $ \text{a} \to \text{b}$ and $\text{c} \to \text{d}$. Alternatively, the 
sudden changes on the control parameter $q_{\alpha}:q_{C} \leftrightarrow q_{H}$ ($\text{b} \to \text{c}$ and $\text{d} \to \text{a}$) are described as isentropic processes. 
Finally, Fig.\ref{fig:1}(c), illustrating some details of the measurement protocols \cite{Karimi_2020}, will be discussed later in the paper. 

\section{Average heat and work - Lindblad dynamics}
\label{sec:3}

If one is interested in the average energy exchanges taking place during the cycle, it is sufficient to consider the dynamics of the density matrix $\rho$ of the TLS.
The thermalization processes taking place in the isothermal strokes $\text{a} \to \text{b}$ and $\text{c} \to \text{d}$ can be described using a simple Lindblad evolution,
\begin{equation}
	\begin{aligned}
	\dot{\rho}=& -\frac{i}{\hbar}[H_{\alpha},\rho] + \sum_{k} \Big[ L_{k,\alpha} \rho L^{\dag}_{k,\alpha} - \frac{1}{2} \{ L^{\dag}_{k,\alpha}L_{k,\alpha},\rho \} \Big]\\
	& + \Gamma_{\phi,\alpha} \Big[ \rho - \sigma_{z} \rho \sigma_{z} \Big],
	\label{eq:lindblad}
	\end{aligned}
\end{equation} with $\rho$ the density matrix of the TLS and $L_{k,\alpha}$ the jump operators. In this picture, among all possible transitions that the system may
undergo due to interactions with the reservoir, we consider the following jump operators
\begin{equation}
	\begin{aligned}
	L_{\uparrow,\alpha}&= \sqrt{\Gamma_{\uparrow,\alpha}}  |e\rangle_{\alpha} \langle g|_{\alpha},\\
	L_{\downarrow,\alpha}&= \sqrt{\Gamma_{\downarrow,\alpha}} |g\rangle_{\alpha} \langle e|_{\alpha},
	\label{eq:A_up_down_eg_new}
	\end{aligned}
\end{equation}  with $\{|g\rangle, |e\rangle \}$ the eigenstates basis previously defined. $\Gamma_{\downarrow,\alpha}$ and $\Gamma_{\uparrow,\alpha}$ are the corresponding transition 
rates, with $\Gamma_{\uparrow,\alpha}=\Gamma_{\downarrow,\alpha} \text{e}^{\beta_{\alpha}\Delta E_{\alpha}}$. Note that $L_{\downarrow,\alpha}$ and $L_{\uparrow,\alpha}$ will, in fact, 
induce relaxation through population transfer between the ground and excited states.  In Eq.\eqref{eq:lindblad}, we also consider a dephasing mechanism described by the last term. 
This mechanism simply leads to a decay of the coherence terms given by the off-diagonal terms of the density matrix $\rho$. $\Gamma_{\phi,\alpha}$ describes the corresponding decay rate.

After considering the previous description, and using the parametrization $\mathcal{D}_{\text{i},\text{f}} =  \rho^{\text{i},\text{f}}_{gg}-1/2$, $\mathcal{R}_{\text{i},\text{f}} = \text{Re} (\rho^{\text{i},\text{f}}_{ge})$, 
and $\mathcal{I}_{\text{i},\text{f}} = \text{Im} (\rho^{\text{i},\text{f}}_{ge})$ referred to the initial (i) and final (f) states of the density matrix, the Lindblad equation 
\eqref{eq:lindblad} changes to
\begin{equation}
	\begin{aligned}
	\mathcal{D}_{\text{f}} &= \frac{\Gamma_{\downarrow,\alpha}}{\Gamma_{\Sigma,\alpha}} + \frac{ (-\Gamma_{\downarrow,\alpha}  + \Gamma_{\Sigma,\alpha} (\mathcal{D}_{\text{i}} + \frac{1}{2})) }{\Gamma_{\Sigma,\alpha}} \text{e}^{-\Gamma_{\Sigma,\alpha} \Delta t} - \frac{1}{2},\\
	\mathcal{R}_{\text{f}} &=\text{e}^{-(\frac{\Gamma_{\Sigma,\alpha}}{2} + 2\Gamma_{\phi,\alpha}) \Delta t} \Big( \mathcal{R}_{\text{i}} \cos(\varphi_{\alpha}) -  \mathcal{I}_{\text{i}} \sin(\varphi_{\alpha}) \Big),\\
	\mathcal{I}_{\text{f}} &=\text{e}^{-(\frac{\Gamma_{\Sigma,\alpha}}{2} + 2\Gamma_{\phi,\alpha}) \Delta t} \Big( \mathcal{R}_\text{{i}} \sin(\varphi_{\alpha}) +  \mathcal{I}_{\text{i}} \cos(\varphi_{\alpha}) \Big),
	\label{eq:SOE_5}
	\end{aligned}
\end{equation} with $\Gamma_{\Sigma,\alpha} = \Gamma_{\downarrow,\alpha} + \Gamma_{\uparrow,\alpha}$ and  $\varphi_{\alpha}= \int_{0}^{\Delta t} \Delta E_{\alpha} dt$, which represents 
the dynamic phase acquired in each isothermal stroke. 

We assume that isentropic processes can be modeled by imposing the continuity condition of the density matrix, i.e., $\rho^\text{f} = \rho^\text{i}$, and express it in the respective eigenstates basis 
before and after the isentropic stokes $\text{b} \to \text{c}$ and $\text{d} \to \text{a}$, see Appendix \ref{ap:A}.

Due to the absence of an input work, the heat exchanged from the system to each reservoir $\langle Q_{\alpha} \rangle$ equals the change of internal energy $\Delta \mathcal{E}_{\alpha}$ and 
can be simply computed as
\begin{equation}
	\begin{aligned}
	\langle {Q}_{\alpha} \rangle = \Delta \mathcal{E}_{\alpha} = \Delta E_{\alpha} \Big( \mathcal{D}_{\text{f}} - \mathcal{D}_{\text{i}} \Big).
	\label{eq:E_alpha_1}
	\end{aligned}
\end{equation} Note the change of internal energy is given by $\Delta \mathcal{E}_{\alpha} = \langle \mathcal{E}_{\alpha}(t_{\text{f}}) \rangle - \langle \mathcal{E}_{\alpha}(t_{\text{i}}) \rangle$, 
where $\langle \mathcal{E}_{\alpha}(t) \rangle =  \text{Tr}( \rho  H_{\alpha} )$ describes the system energy at time $t$, with $\rho$ the density matrix of the system. In particular, $t_{\text{i}}$ and 
$t_{\text{f}}$ are the initial and final times for each isothermal stroke of the cycle. We then get $\langle {Q}_{C} \rangle = \Delta E_{C} (\mathcal{D}_{\text{b}} - \mathcal{D}_{\text{a}}) $ and 
$\langle {Q}_{H} \rangle = \Delta E_{H} (\mathcal{D}_{\text{d}} - \mathcal{D}_{\text{c}}) $.

From now on, all numerical and analytical plots are obtained for $T_H = 0.42 [E_{0}/k_{\text{b}}]$, $T_C=0.32  [E_{0}/k_{\text{b}}]$, and $\Gamma =0.5 [E_{0}] $. These values correspond to typical numbers for superconducting qubits \cite{oliver_2005,oliver_2009,Koch_2007,berns_2008,Clarke_2008}. For simplicity, we have chosen 
$\Gamma_{\downarrow,H}=\Gamma_{\downarrow,C}=\Gamma$. Depending on the choice of the other parameters the engines can operate indifferent regimes. 

\textit{Refrigerator (incoherent) regime}. We first consider protocol $q_{\alpha}(t):q_{C} \to q_{H}$ and $\Delta=0$. Besides the cold and hot reservoirs, we assume the system also couples to a 
dephasing noise source. For the sake of simplicity, we fixed $\Gamma_{\phi,C}=\Gamma_{\phi,H}=\Gamma_{\phi}\neq 0$. Within this setup, the solution of Eq.\eqref{eq:E_alpha_1} can be easily computed as
\begin{equation}
	\begin{aligned}
	\langle Q_{C,H} \rangle =& -\Delta E_{C,H} \mathcal{F}[\Gamma_{C},\Gamma_{H},\Delta t],\\
	\label{eq:Q_CIncoh}
	\end{aligned}
\end{equation} with
\begin{equation}
	\begin{aligned}
	\mathcal{F}[\Gamma_{C},\Gamma_{H},\Delta t] &= \frac{ (\Gamma_{\downarrow}^{C}\Gamma_{\Sigma,H} - \Gamma_{\downarrow}^{H}\Gamma_{\Sigma,C})}{ \Gamma_{\Sigma,C}  \Gamma_{\Sigma,H}  (\text{e}^{ ( \Gamma_{\Sigma,C} + \Gamma_{\Sigma,H}) \Delta t}-1 ) }\\
    &\times (\text{e}^{\Gamma_{\Sigma,C}\Delta t}-1 ) (\text{e}^{\Gamma_{\Sigma,H}\Delta t}-1 ).
	\label{eq:f_incoh}
	\end{aligned}
\end{equation}
\begin{figure}[!htb]
  \centering
   \includegraphics[width=8.5cm]{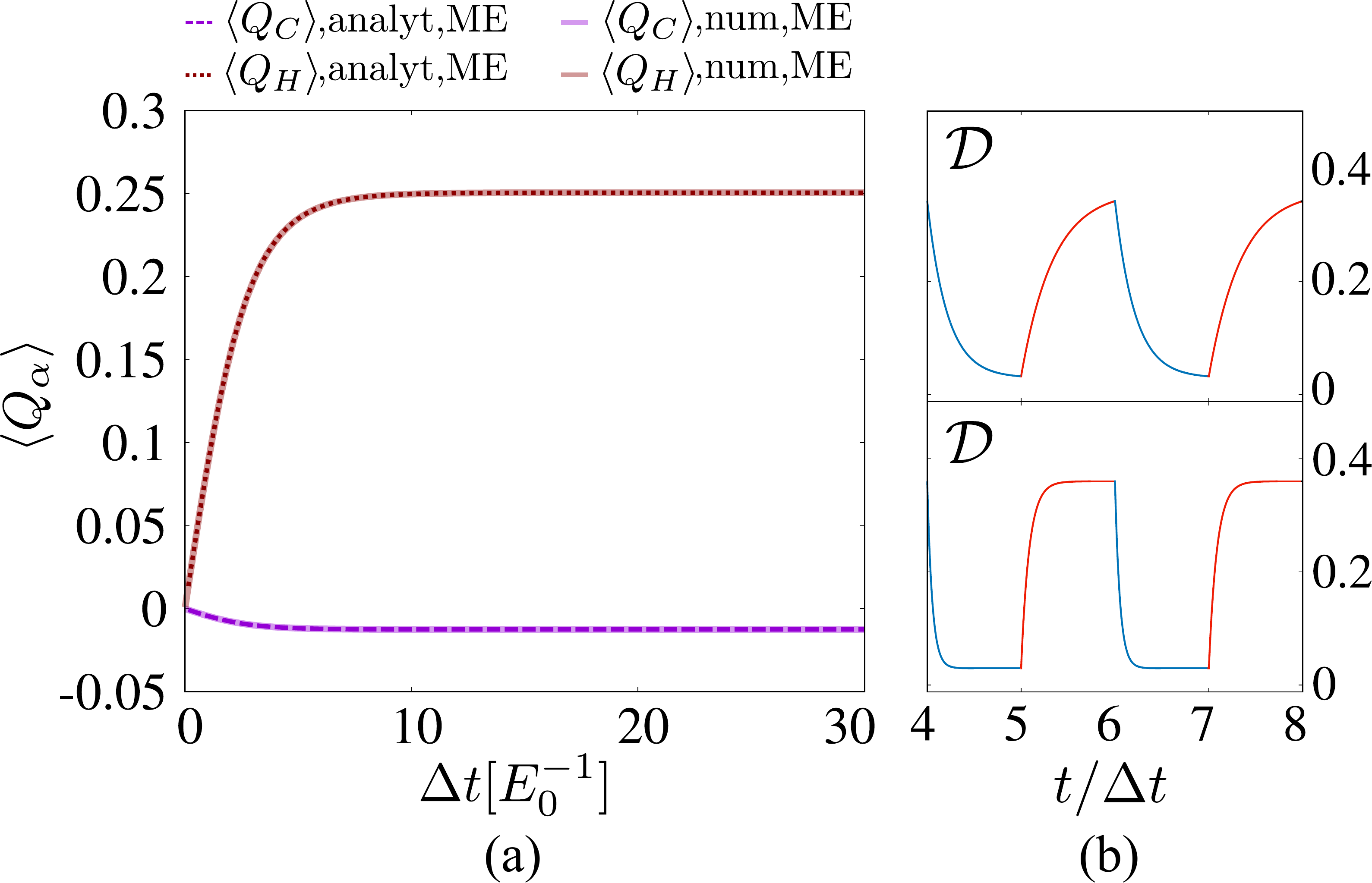}
   \caption{\textbf{Incoherent cycle}. (a) Numerical and analytical plots of heat exchanges $\langle Q_{\alpha} \rangle $ as a function $\Delta t$ without dephasing $\Gamma_{\phi}=0$. Analytical results are calculated using Eq.\eqref{eq:Q_CIncoh} and are plotted using dashed lines. (b) Numerical plots of $\mathcal{D}$ as a function of $t/\Delta t$ for $\Delta t = 5 [E_{0}^{-1}]$ (top plot) and $\Delta t = 20 [E_{0}^{-1}]$ (bottom). The other numerical parameters are $T_H = 0.42 [E_{0}/k_{\text{b}}]$, $T_C=0.32 [E_{0}/k_{\text{b}}]$, and $\Gamma =0.5 [E_{0}]$, with  $\Gamma_{\downarrow,H}=\Gamma_{\downarrow,C}=\Gamma$.}
   \label{fig:2}
\end{figure} 
If $\beta_{C}\Delta E_{C} < \beta_{H}\Delta E_{H}$, the Otto cycle always operates as a refrigerator, i.e., $\langle Q_{C} \rangle < 0$ and $\langle Q_{H} \rangle >0$ \cite{Solfanelli_2020}. The specific behavior 
of $\langle Q_{\alpha} \rangle$ as a function of $\Delta t$ can be seen in Fig.\ref{fig:2}(a) for $q_{C}=0.019[E_{0}]$ and $q_{H}=0.38[E_{0}]$. As the time-interval $\Delta t$ increases, the 
absolute value of each $\langle Q_{\alpha}\rangle$ exhibits an exponential increment. Essentially, the system-bath interaction progressively lasts longer, which continuously induces an 
increase of energetic exchanges. The subsequent saturation of the energy flow for large values of $\Delta t$ reflects the system thermalization. 
This is also  illustrated in Fig.\ref{fig:2}(b), where we plot the steady solution $\mathcal{D}$ as function of $t/\Delta t$  for $\Delta t = 5 [E_{0}^{-1}]$ (top) and 
$\Delta t = 20 [E_{0}^{-1}]$ (bottom).  The results are plotted using blue and red colors to indicate which isothermal strokes they belong. The blue color corresponds to the stroke 
$\text{a}\to\text{b}$, while red corresponds to $\text{c}\to\text{d}$.  

The heat exchanges, see Eq.\eqref{eq:Q_CIncoh}, are independent on the dephasing noise power $\Gamma_{\phi,\alpha}$ and the dynamic phase $\varphi_{\alpha}$. Indeed, the coherences vanish 
in the steady state regime. Specifically, we obtain  $\mathcal{R}_{\text{a,b,c,d}}(t) = \mathcal{I}_{\text{a,b,c,d}}(t) = 0$ for any value of $ \Gamma_{\phi,\alpha}$ and $\varphi_{\alpha}$ thus implying that 
the cycle operates as an \textit{incoherent} refrigerator.

\begin{figure}[!htb]
  \centering
   \includegraphics[width=8.5cm]{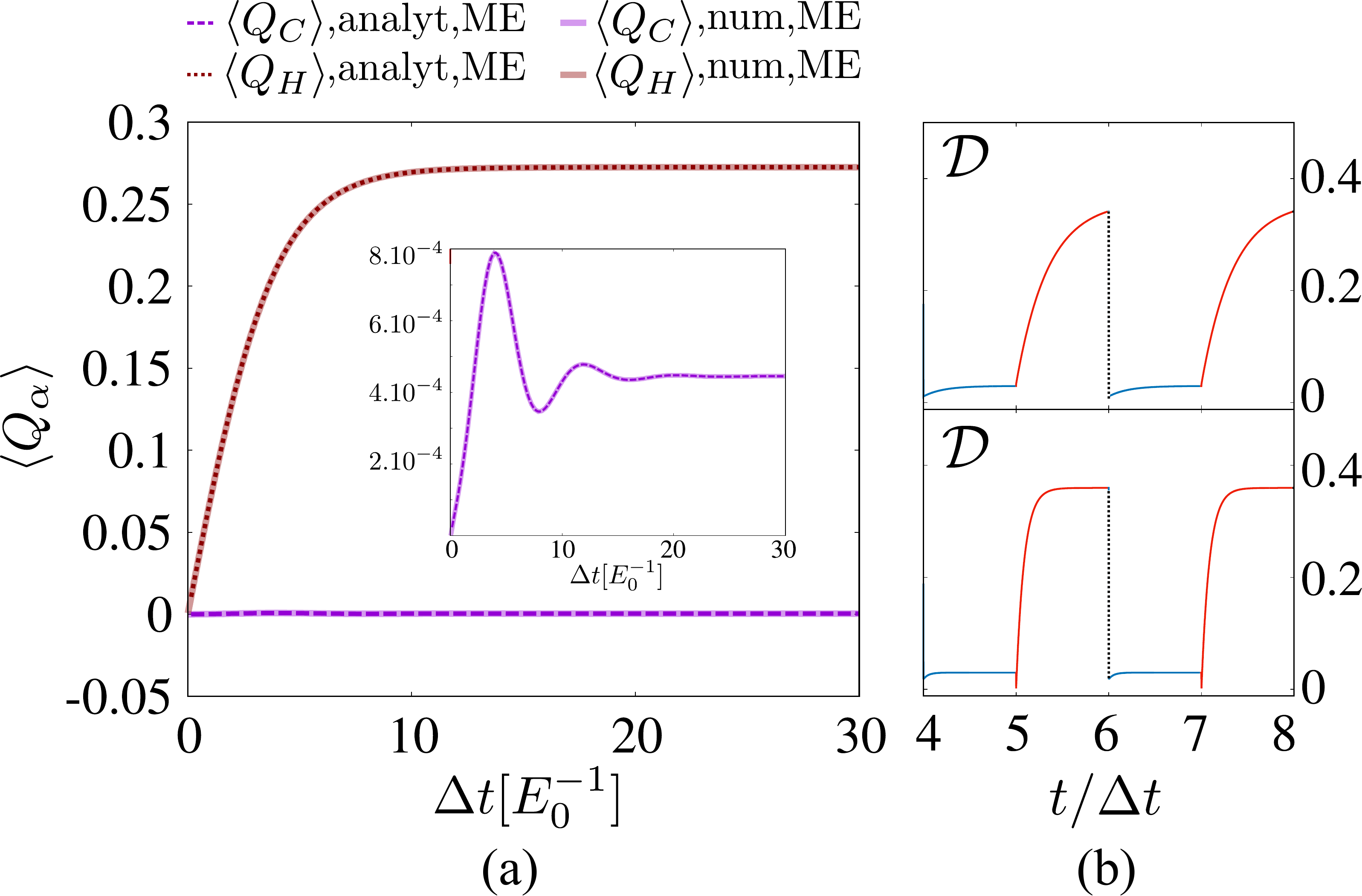}
   \caption{\textbf{Coherent cycle}. (a) Numerical and analytical plots of heat exchanges $\langle Q_{\alpha} \rangle$ as a function $\Delta t$ without dephasing $\Gamma_{\phi}=0$. In the inset, we report $\langle Q_{\alpha} \rangle $ as a function of $\Delta t$. Analytical results are calculated using Eq.\eqref{eq:Q_CIncoh} and are plotted using dashed lines. (b) Numerical plots of $\mathcal{D}$ as a function of $t/\Delta t$ for $\Delta t = 5 [E_{0}^{-1}]$ (top plot) and $\Delta t = 20 [E_{0}^{-1}]$ (bottom). The other numerical parameters are the same in Fig.\ref{fig:2}.}
   \label{fig:3}
\end{figure}

\textit{Coherent regime}. Another interesting case to consider is $q_{\alpha}(t):0 \to q_{H}$ and $\Delta\neq0$. Unlike the previous cycle, we now assume that the system is no longer coupled with a dephasing noise source ($\Gamma_{\phi,\alpha} = 0$). In Fig.\ref{fig:3}, we present the analytical solution of Eq.\eqref{eq:SOE_5} for $\Delta=0.019[E_{0}]$ and $q_{H}=0.38[E_{0}]$.

\begin{figure}[!htb]
  \centering
   \includegraphics[width=8.5cm]{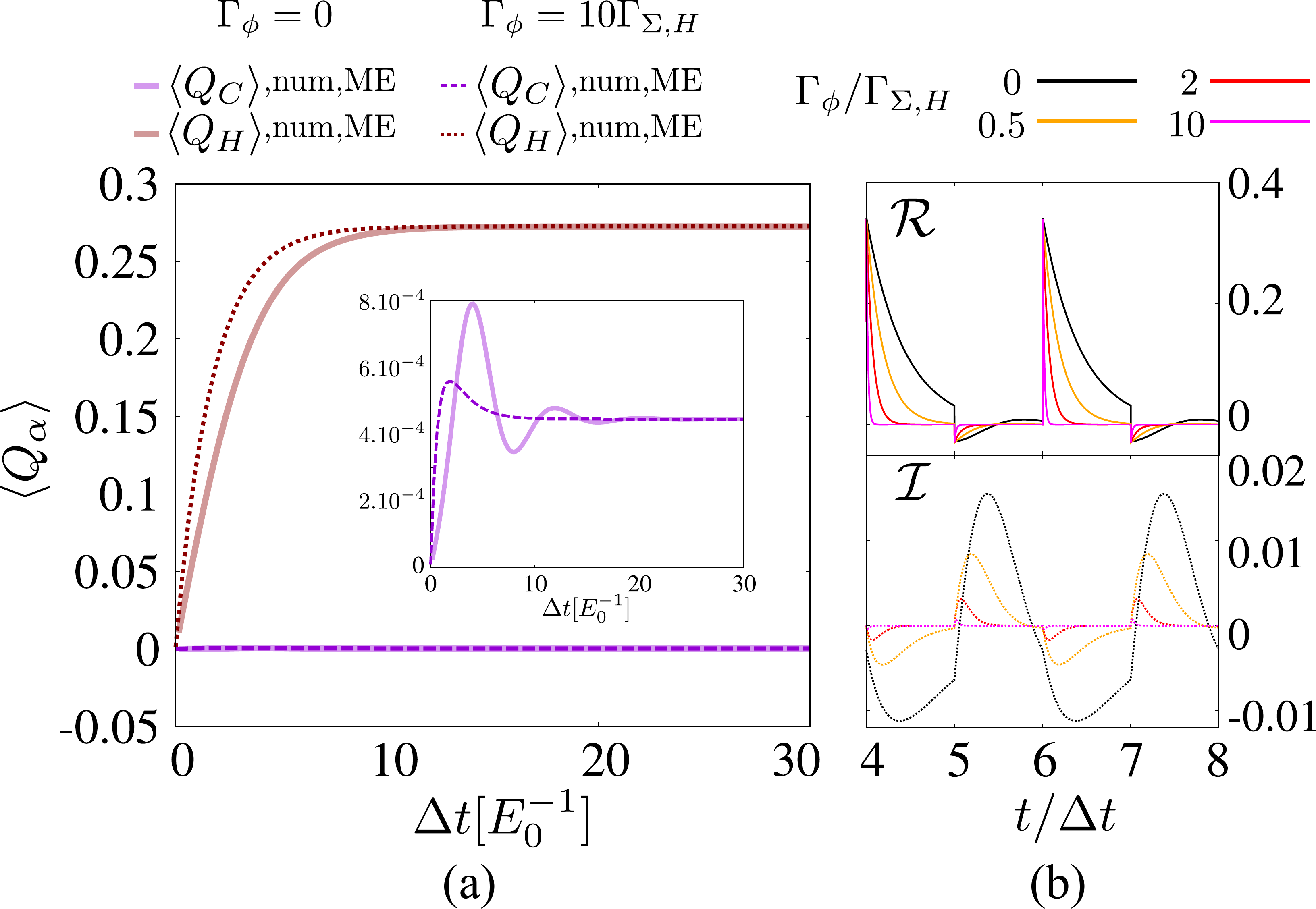}
   \caption{\textbf{Coherent cycle}. (a) Numerical plots of heat exchanges $\langle Q_{\alpha} \rangle$ as a function $\Delta t$ including dephasing $\Gamma_{\phi}=10\Gamma_{\Sigma,H}$. Comparatively, numerical results from Fig.\ref{fig:3}(a) are also plotted using bold lines in light colors. In the inset, we report $\langle Q_{\alpha} \rangle $ as a function of $\Delta t$. (b) Numerical plots of $\mathcal{R}$ (top) and $\mathcal{I}$ (bottom) as a funtion of $t/\Delta t$ for $\Delta t = 5 [E_{0}^{-1}]$ and different values of $\Gamma_{\phi}$. All numerical results are obtained using the ME approach. The other numerical parameters are the same in Fig.\ref{fig:2}.}
   \label{fig:4}
\end{figure}

Opposite to the previous case, both baths are now heated, that is, $\langle Q_{C} \rangle > 0$ and $\langle Q_{H} \rangle >0$. A striking feature of this cycle is the oscillating behavior that the heat exchange $\langle Q_{C} \rangle$ exhibits as a function of $\Delta t$. As shown in Fig.\ref{fig:3}, for small values of $\Delta t$, the system reaches a partially thermalized state, for which the dynamic phase $\varphi_{C} = \Delta E_{C} \Delta t$ survives dissipation effects. This last result suggests that the system may exhibit coherences, which ultimately facilitates the oscillating behavior of $\langle Q_{C} \rangle$ as a function of $\Delta t$. This is well-illustrated in Fig.\ref{fig:4}(b), where we plot the time evolution of the off-diagonal elements $\mathcal{R}$ and $\mathcal{I}$ for $\Delta t = 5 [E_{0}^{-1}]$. Notably, the amplitudes of $\mathcal{R}$ and $\mathcal{I}$  are significant when the system couples with the cold bath while negligible when the system couples with the hot bath. Eventually, as $\Delta t$ increases, the system thermalizes, and the dynamic phase ultimately vanishes.


A natural follow-up question is whether suppression of the coherences due to dephasing processes can restore the refrigerator regime. To investigate this regime, we now consider that the system couples with a unique dephasing noise source. We set then $\Gamma_{\phi,C}= \Gamma_{\phi,H}= \Gamma_{\phi}$, with $ \Gamma_{\phi} = 10 \Gamma_{\Sigma,H}$.

Surprisingly, as Fig.\ref{fig:4}(a) shows, despite there being no evidence of time-dependent oscillations (coherences) in $\langle Q_{C} \rangle$, both baths are still heated. To better understand 
the system dynamics, in Fig.\ref{fig:4}(b), we plot the time evolution of the coherence terms $\mathcal{R}$ and $\mathcal{I}$ varying the strenghts of dephasing $\Gamma_\phi$ for the time-interval 
$\Delta t=5[E_{0}^{-1}]$. From \ref{fig:4}(b), it is clear that the amplitude of $\mathcal{R}$ systematically diminishes when $\Gamma_{\phi}$ increases. However, $\mathcal{R}$ does not completely fade away 
but instead presents sharp peaks at integer values of $t/\Delta t$. Such discontinuities in $\mathcal{R}$ occur at times when the control parameter abruptly compresses or expands. The reason for 
such behavior is that when $\Delta \neq 0$, this protocol will always lead to jumping operators defined on a distinct basis depending on the isothermal stroke, $L_{k,C} \leftrightarrow L_{k,H}$. 
This result suggests that coherences can never vanish. In other words the cycle gives heating as a result of the isentropic phases, at least
when they are treated as a mere basis change.


\section{Fluctuations and stability}
\label{sec:4}

As discussed extensively in the literature, one of the key properties of small heat in engines is the importance of fluctuations.  To this aim, 
the open system dynamics is now addressed by using a stochastic representation of the Lindblad equation, also known as unraveling \cite{Carmichael_1993,Wiseman_2009,Manzano_2022}. By exploiting 
this approach, one is able to determine the probability distribution of  thermodynamics quantities, such as heat $Q_{\alpha}$. In this work, we employ the MCWF method to compute the energetic 
exchanges \cite{Dalibard_1992,Molmer_1996,Brun_2002,Gardiner_1992}. In Appendix \ref{ap:B}, for completeness, we give a brief description of this method.

Following this stochastic approach, for a specific $j$th-trajectory and a single $k$th-jump, it is possible to identify the energy exchange between the system and the bath as
\begin{equation}
	\begin{aligned}
	\Delta\mathcal{E}^{j}_{k,\alpha}&=\Big( \mathcal{E}^{j}_{\alpha}(t_{k} +\delta t)- \mathcal{E}^{j}_{\alpha}(t_{k}-\delta t) \Big),
	\label{eq:D_E_jk_1}
	\end{aligned}
\end{equation} with $\mathcal{E}^{j}_{k,\alpha}(t)=\langle \phi_{j}(t) | H_{\alpha} | \phi_{j}(t)\rangle$ and $t_{k}$ the time at which a single quantum jump occurs. As mentioned before, because 
of the absence of an input work, the change of internal energy equals the heat exchange, i.e., $\Delta\mathcal{E}^{j}_{k,\alpha} = Q^{j}_{k,\alpha}$. It is thus straightforward to show that the 
heat  $Q^{j}_{k,\alpha}$ exchanged by the system to the reservoir can be computed as 
\begin{equation}
	\begin{aligned}
	Q^{j}_{k,\alpha}&= -\Delta E_{\alpha} (\delta_{t_{k},t^{\uparrow}_k}-\delta_{t_{k},t^{\downarrow}_k}),
	\label{eq:D_E_jk_2}
	\end{aligned}
\end{equation} with $\delta_{t_{k},t^{\uparrow,\downarrow}_k}$ the delta of Kronecker and $t^{\uparrow}_k,t^{\downarrow}_k$ the times at which an ``up" or ``down" jump occurs. 
The system absorbs $+\delta_{t,t^{\uparrow}_k}$ (or emits $-\delta_{t,t^{\downarrow}_k}$) a single photon of energy $\Delta E_{\alpha}$ from (or to) the $\alpha$th-bath. Consequently, if we consider 
all $k$th-jumps occurring in a single $j$th-trajectory, it is possible to define the total heat exchange as
\begin{equation}
	\begin{aligned}
    Q^{j}_{\alpha}=  Q^{\text{abs},j}_{\alpha} + Q^{\text{loss},j}_{\alpha},
	\label{eq:D_E_jk_3}
	\end{aligned}
\end{equation} with $Q^{\text{abs},j}_{\alpha} =\sum_{k}^{N^{\uparrow,j}_{\alpha}}Q^{j}_{k,\alpha}$ and $Q^{\text{loss},j}_{\alpha}= \sum_{k}^{N^{\downarrow,j}_{ \alpha}}Q^{j}_{k,\alpha}$. Here, $ N^{\uparrow,j}_{\alpha}$ and $N^{\downarrow,j}_{ \alpha}$ are the total number of jumps up and down, respectively. Interestingly, from these previous definitions, the ME results can be easily recovered after averaging over $N_{\text{traj}}$ random trajectories, $\langle Q_{\alpha} \rangle = \sum_{j}^{N_{ \text{traj}}}Q^{j}_{\alpha} / N_{ \text{traj}}$.

Following Ref.\cite{Karimi_2020}, in Fig.\ref{fig:1}(c), we display a simple sketch of the possible measurement process of the energetic exchanges. The measurement device consists of a nanocalorimeter realized by a finite-size absorber at temperature $T^{\alpha}$ weakly coupled to an infinite bath. The design ensures that changes in the absorber temperature unveil whether a photon is absorbed or emitted. In this manner, we can measure the stochastic heat exchanges of the system by monitoring the absorber temperature.

We now proceed to analyze the stochastic properties of our Otto cycle, focusing on studying the FRs~\cite{Jarzynski_1997,Campisi_2011} and the TURs~\cite{Pietzonka_2016,Horowitz_2020}.
These last are important in assessing how the stability of the engine  is affected if one tries to increase the efficiency towards its optimal value.
The temperature variations are assumed to be small enough, which allows us to consider constant rates, not depending on the history of previous jumps.

\subsection{Fluctuation relations}
\label{sec:4.1}

Irreversible entropy production is one of the intrinsic characteristics of non-equilibrium systems. To study the fluctuations in the entropy production, namely $\Sigma$, one needs to treat $\Sigma$ as a random variable distributed according to a certain probability distribution $P(\Sigma)$. These distributions satisfy a set of fundamental symmetry relations known as FRs, which can generally be expressed as  \cite{Campisi_2011,Campisi_2011,Leggio_2013,Eposito_2009,Gupta_2020,Campisi_2014}
\begin{equation}
\begin{aligned}
	 \text{ln}\Big(\frac{P[\Sigma;\mathcal{V}]}{\tilde{P}[-\Sigma;\tilde{\mathcal{V}}]}\Big) = \Sigma.
\label{eq:QFT_1}
\end{aligned}
\end{equation} $\mathcal{V}$ describes the forward driving, while $\tilde{\mathcal{V}}$ corresponds to its time-reversed path. For our cycle, the protocols are simply defined as 
$\mathcal{V} : q_{\alpha}(t)$ and $\tilde{\mathcal{V}} : q_{\alpha}(-t)$. $P[\cdot]$ and $\tilde{P}[\cdot]$ denote the probability distributions for the forward and backward evolution, respectively. 
In particular, given our cycle, the entropy production can be simply computed as $\Sigma = \beta_{C} Q_{C} +  \beta_{H} Q_{H}$ and obey the FR
\begin{equation}
\begin{aligned}
	 \langle \text{e}^{-\Sigma} \rangle = \int_{\Sigma} d\Sigma P[\Sigma;\mathcal{V}]  \text{e}^{-\Sigma}  = 1.
\label{eq:JE}
\end{aligned}
\end{equation} Note that the integral is defined using the forward probability distribution.
It is clear that their convergence depends on stochastic sampling. For our particular cycle, and given the different protocols, one question that naturally arises is: what would be the optimal 
sampling size for each configuration? In what follows, we shall focus on answering this question by analyzing the role of system coherences in stochastic dynamics. 

In order to study the forward and backward evolution, we shall employ the MCWF method following the next steps. For each $j$th-trajectory, the system is initialized in the excited state $|e\rangle_{C}$. 
At the time $t_{0}$, we turn on the driving protocol $\mathcal{V}$ and generate the stochastic dynamics using the MCWF method until $t_{1}$. After this time, we swap the protocols 
$\mathcal{V}\to\tilde{\mathcal{V}}$ and let the system stochastically evolves until $t_{2}$. For each evolution, we collect all stochastic heat exchanges and store them in sequence 
$\mathcal{S} = \{ (Q^{1}_{C},Q^{1}_{H}), (Q^{2}_{C},Q^{2}_{H}), ..., (Q^{j}_{C},Q^{j}_{H}),..., (Q^{N_{\text{traj}}}_{C},Q^{N_{\text{traj}}}_{H}) \}$, with $N_{\text{traj}}$ the total number of random trajectories. 
\begin{figure}[hbt!]
  \centering
    \includegraphics[width=7cm]{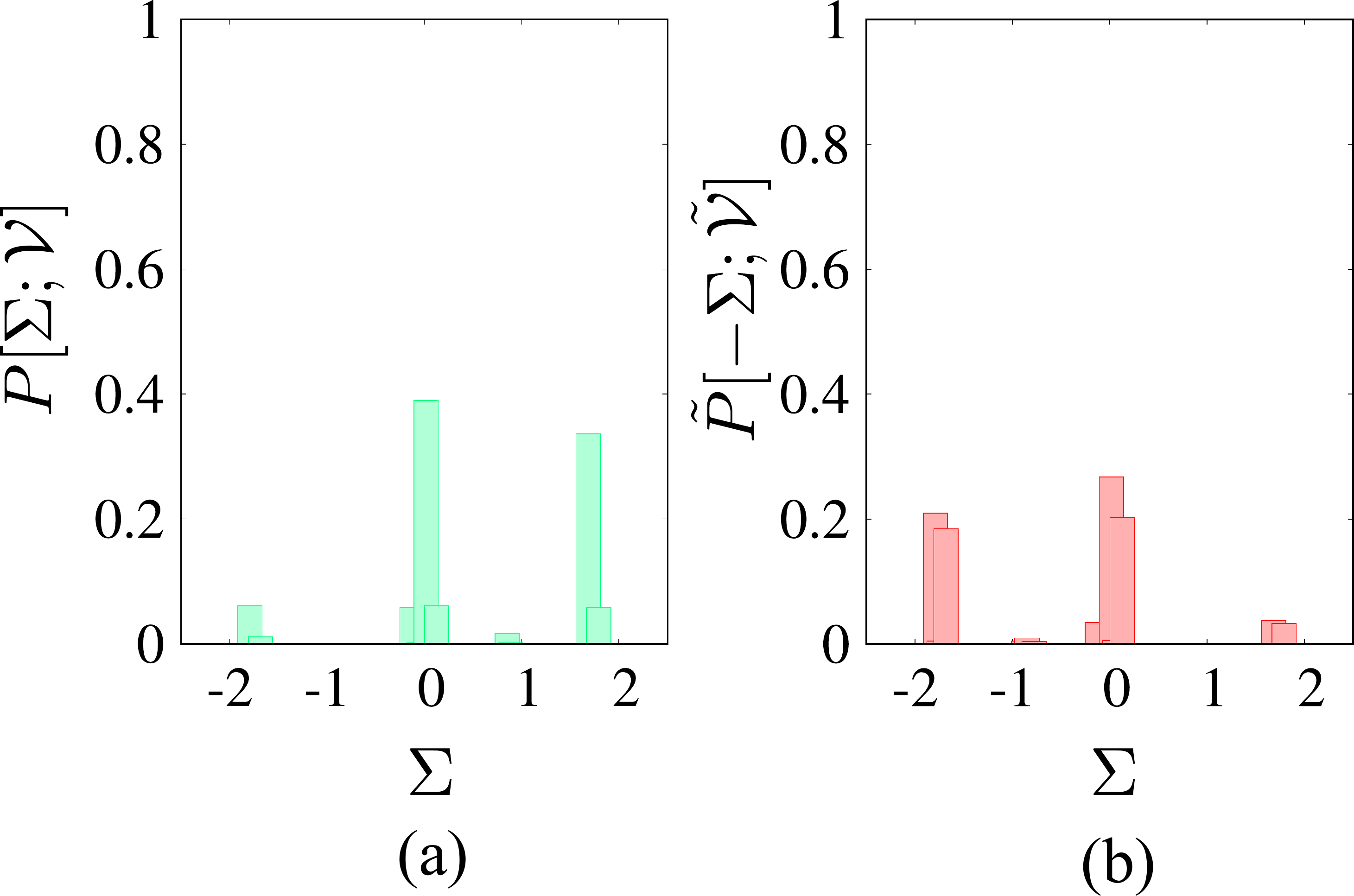}
   \caption{\textbf{Incoherent cycle}. Probability distributions of $\Sigma=\beta_{C}Q_{C} + \beta_{H}Q_{H}$ for the forward (a) and backward (b) evolution. We fixed $\Delta t= 20 [E_0]^{-1}$ and $\Gamma_{\phi}=0$.  All results are computed by means of  MCWF method and considering $10^5$ trajectories. The other numerical parameters are the same in Fig.\ref{fig:2}.}
   \label{fig:5}
\end{figure}

We start by considering the incoherent refrigerator, that is, $q_{C}=0.019[E_0]$ and $q_{H}=0.38[E_0]$, with $\Delta=0$. For simplicity, we fixed $\Gamma_\phi=0$ and $\Delta t =20[E_0^{-1}]$. 
\begin{figure}[hbt!]
  \centering
    \includegraphics[width=8cm]{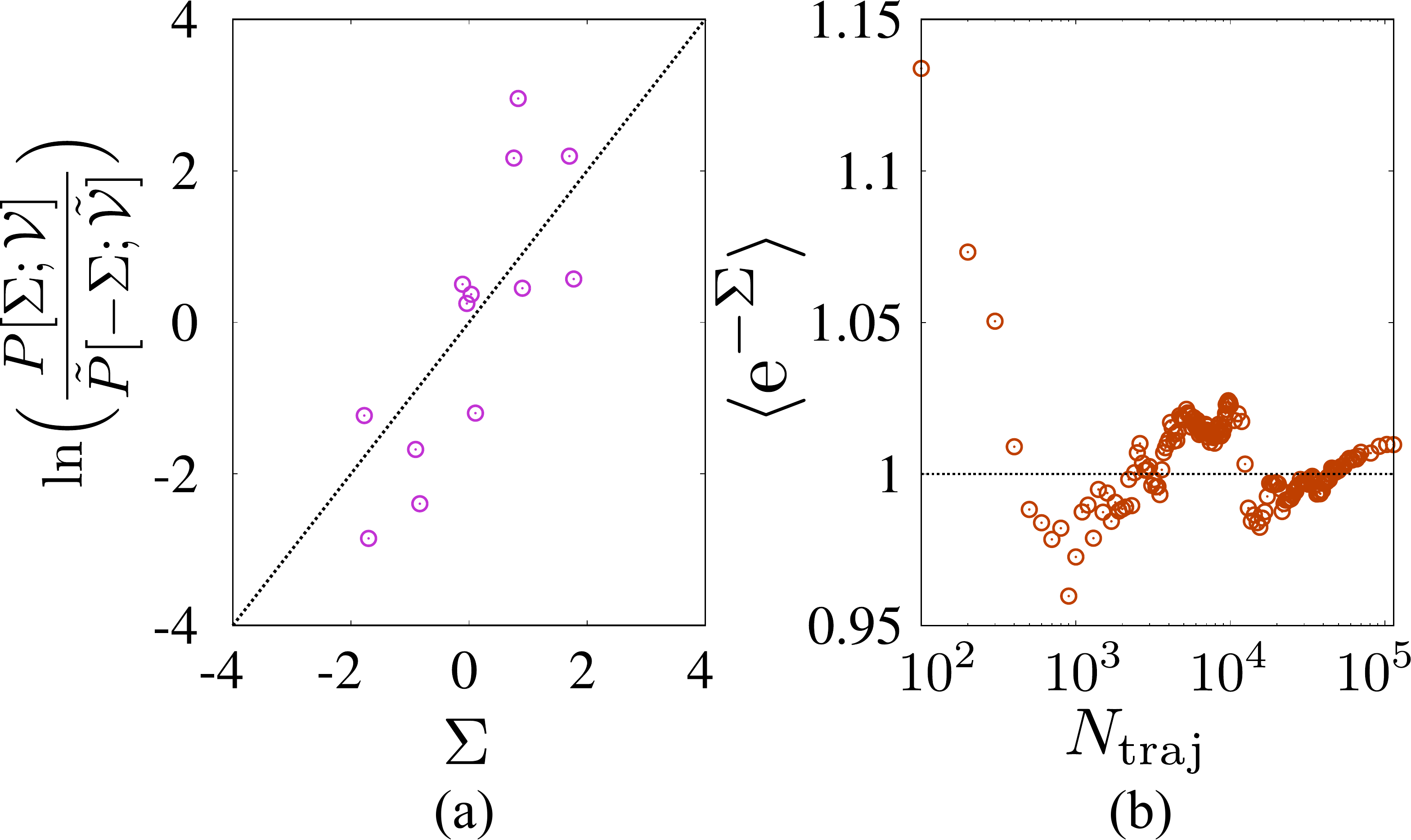}
   \caption{\textbf{Incoherent cycle}. (a) Plot of $\text{ln} (P[\Sigma;\mathcal{V}]/ \tilde{P}[-\Sigma;\tilde{\mathcal{V}}]$ as a function of $\Sigma=\beta_{C}Q_{C} + \beta_{H}Q_{H}$. Theoretical curve plot in dashed line. (b)  Plot of $\langle \text{e}^{\Sigma} \rangle$ as a function of number of trajectories $N_\text{traj}$. All results are computed by considering $10^5$ trajectories. We fixed $\Delta t= 20 [E_0]^{-1}$ and $\Gamma_{\phi}=0$. The other numerical parameters are the same in Fig.\ref{fig:2}.}
   \label{fig:6}
\end{figure} 

For illustrative purposes, in Fig.\ref{fig:5}, we plot the probability distribution of $\Sigma$ for the forward (a) and backward (b) evolution, both obtained after considering $10^5$ trajectories. 
As shown, both probabilities present a very intriguing distribution due to the small number of jumps occurring in each process, i.e., the accessible values of $\Sigma$ are few. 

The forward and backward distribution probabilities are centered at $\langle \Sigma \rangle_{\text{f}} = 0.53$ and $\langle \Sigma \rangle_{\text{b}}=-0.63$, respectively. Here, $\langle \Sigma \rangle_{\text{f}}$ and $\langle \Sigma \rangle_{\text{b}}$ are calculated by simply averaging $\Sigma$ over all random realizations of each process in the stationary regime. Likewise, it is possible to compute the variance by following the relation $\text{Var}(\Sigma)_{F,B} = \langle \Sigma^2 \rangle_{F,B} - \langle \Sigma \rangle_{F,B}^2 $. We get  $\text{Var}(\Sigma)_{\text{f}} = 1.10 $ and $\text{Var}(\Sigma)_{\text{b}}=0.98$, respectively. These results in turn suggest that the distribution $P[\Sigma;\mathcal{V}]$ is wider than  $\tilde{P}[-\Sigma;\tilde{\mathcal{V}}]$. Another measure of the difference between the two normalized distributions is relative entropy \cite{Jarzynski_2006}
\begin{equation}
\begin{aligned}
	 D \Big[ P[\Sigma;\mathcal{V}] \Big| \tilde{P}[-\Sigma;\tilde{\mathcal{V}}]  \Big] &=  \int_{\Sigma} d\Sigma P[\Sigma;\mathcal{V}]  \text{ln}\Big(\frac{P[\Sigma;\mathcal{V}]}{\tilde{P}[-\Sigma;\tilde{\mathcal{V}}]}\Big) \\
	 &= \langle \Sigma \rangle_{\text{f}}.  
\label{eq:Rel_entropy}
\end{aligned}
\end{equation} Since entropy production quantifies how dissipative a process is, Eq.\eqref{eq:Rel_entropy} reveals that the distinguishability between the two distributions would be more significant for more dissipative processes. Remarkably, Eq.\eqref{eq:Rel_entropy} only depends on the averaged value of $\langle \Sigma \rangle_{\text{f}}$, meaning the degree of distinguishability is independent of the unraveling.

The logarithmic ratio $\text{ln}(P[\Sigma;\mathcal{V}]/\tilde{P}[-\Sigma;\tilde{\mathcal{V}}])$ as a function of $\Sigma$ is shown  Fig.\ref{fig:6}(a). The spread of the numerical data around the theoretical expected value
decreases, as expected,  when the number of trajectories $N_{\text{traj}}$ increases. The results presented in Figs.\ref{fig:6}(a) and (b) suggest that while the logarithmic ratio 
$\text{ln}(P[\Sigma;\mathcal{V}]/\tilde{P}[-\Sigma;\tilde{\mathcal{V}}])$ requires a significant stochastic sampling, the integral $\langle \text{e}^{-\Sigma} \rangle$ may converge for smaller collection of realizations. According to Ref.\cite{Jarzynski_2006}, 
the convergence of the FR, $\langle \text{e}^{-\Sigma} \rangle=1$, relies on the averaged entropy production $\langle \Sigma \rangle_{\text{b}}$ of the backward processes. The optimal number of trajectories will, indeed, 
depend on $\langle \Sigma \rangle_{\text{b}}$ of form $N_{\text{op}} \sim \text{exp}[\langle \Sigma \rangle_{\text{b}}]$, meaning that the more dissipative the process, the smaller sampling is needed to ensure 
the convergence of the integral $\langle \text{e}^{-\Sigma} \rangle$.  
\begin{figure}[htb!]
  \centering
    \includegraphics[width=7cm]{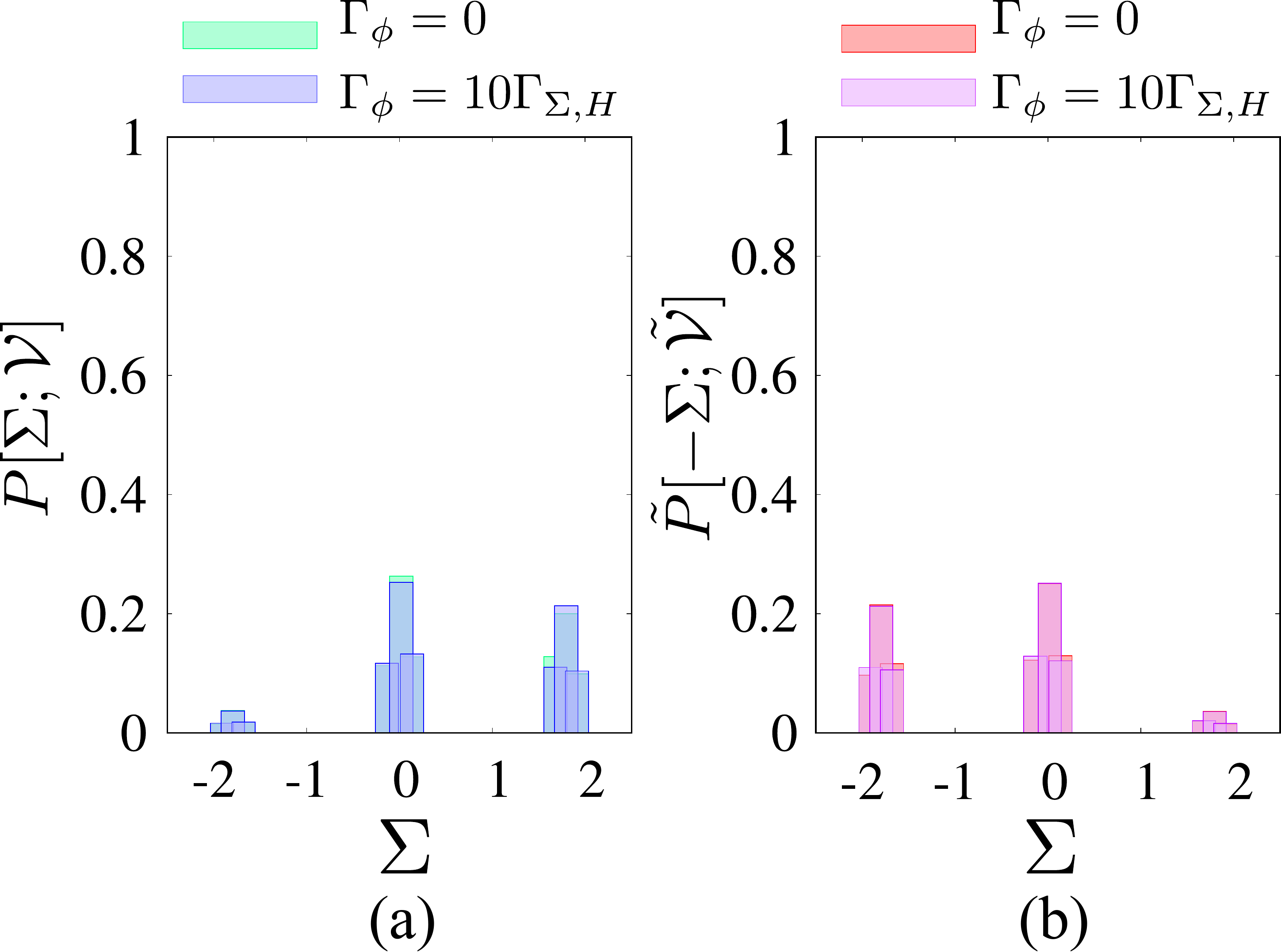}
   \caption{\textbf{Coherent cycle}. Probability distributions of $\Sigma=\beta_{C}Q_{C} + \beta_{H}Q_{H}$ for the forward (a) and backward (b) evolution. In both panels, we report the cases with ($\Gamma_{\phi}=10\Gamma_{\Sigma,H}$) and without ($\Gamma_{\phi}=0$) dephasing. We fixed $\Delta t= 20 [E_0]^{-1}$. All results are computed by means of  MCWF method and considering $10^5$ trajectories. The other numerical parameters are the same in Fig.\ref{fig:2}.}
   \label{fig:7}
\end{figure}
\begin{figure}[htb!]
  \centering
    \includegraphics[width=8cm]{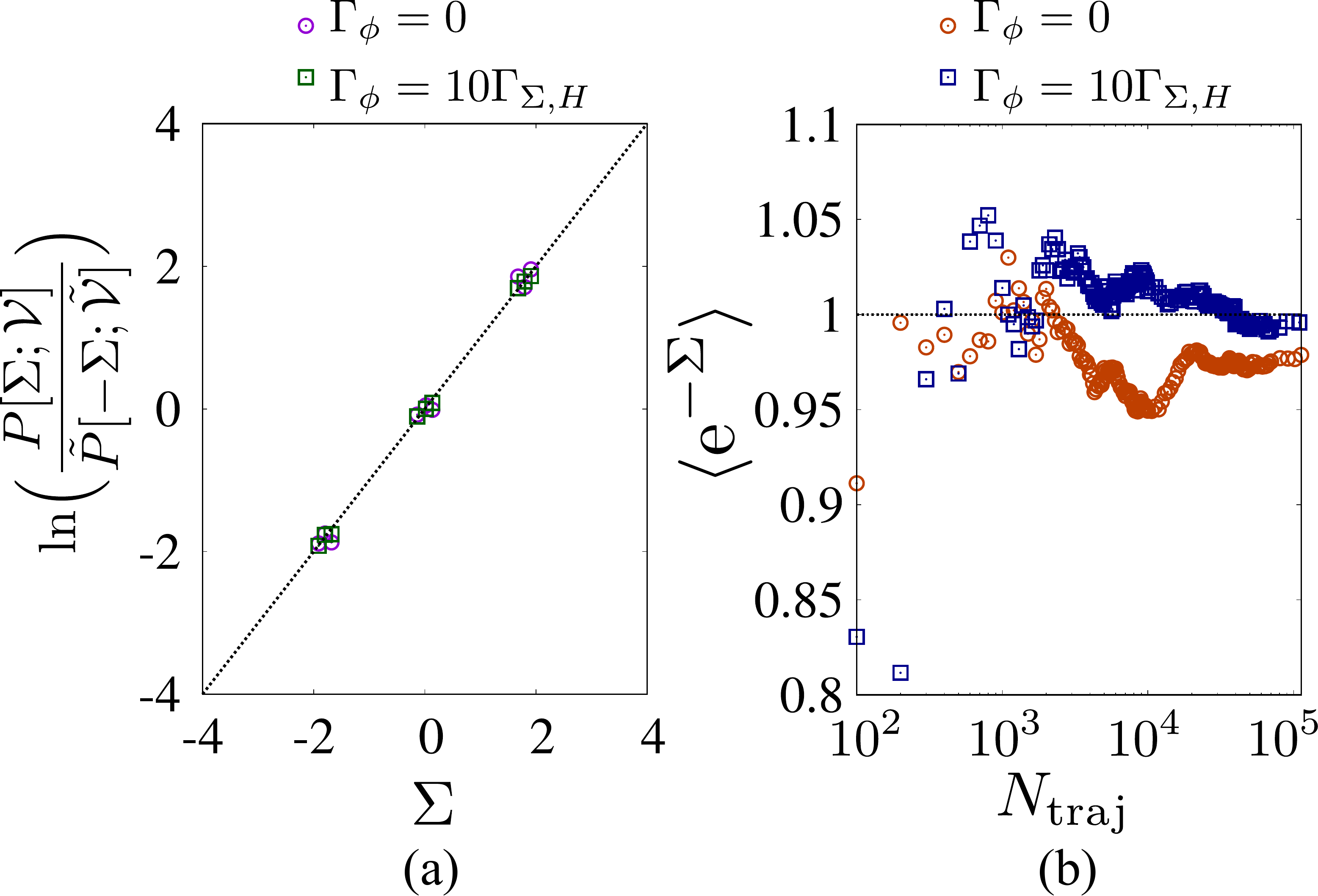}
   \caption{\textbf{Coherent cycle}. (a) Plot of $\text{ln} (P[\Sigma;\mathcal{V}]/ \tilde{P}[-\Sigma;\tilde{\mathcal{V}}]$ as a function of $\Sigma=\beta_{C}Q_{C} + \beta_{H}Q_{H}$. Theoretical curve plot in dashed line. (b)  Plot of $\langle \text{e}^{\Sigma} \rangle$ as a function of number of trajectories $N_\text{traj}$. All results are computed by considering $10^5$ trajectories. We fixed $\Delta t= 20 [E_0]^{-1}$ and $\Gamma_{\phi}=0$. The other numerical parameters are the same in Fig.\ref{fig:2}.}
   \label{fig:8}
\end{figure} 

We move outside the refrigerator regime with protocol $q_{C}=0 $ and $q_{H}= 20 \Delta$, for $\Delta = 0.019 [E_{0}]$ and $\Delta t= 20 [E_{0}]^{-1}$. We analyze the cases of 
no-dephasing ($\Gamma_{\phi}=0$) and strong-dephasing ($\Gamma_{\phi}=10\Gamma^{H}_{\Sigma}$). All numerical results are computed employing $10^5$ realizations.

The distribution probabilities $P[\Sigma;\mathcal{V}]$ and  $\tilde{P}[\Sigma;\mathcal{V}]$ plotted in Fig.\ref{fig:7} exhibit a similar structure to the ones obtained for the previous case.  
A striking outcome for the coherent case exposes in Fig.\ref{fig:8}. While the logarithmic ratio $\text{ln}(P[\Sigma;\mathcal{V}]/\tilde{P}[-\Sigma;\tilde{\mathcal{V}}])$ converges very well for the 
sampling of $N_{\text{traj}}$ realizations (Fig.\ref{fig:8}(a)), the integral $\langle  \text{e}^{-\Sigma} \rangle$ does not (Fig.\ref{fig:8}(b)). Yet, we find that $\langle  \text{e}^{-\Sigma} \rangle$ convergence becomes faster when the system couples with a dephasing noise source. 
The explanation for this behavior is rather simple. For such a purpose, it is necessary to compute the reversed entropy production for each case. After averaging over $N_{\text{traj}}$ trajectories, 
we get that $\langle \Sigma \rangle_{B,\Gamma_{\phi}=10\Gamma^{H}_{\Sigma}}=-0.65$ is slightly larger than $\langle \Sigma \rangle_{ B,\Gamma_{\phi}=0}=-0.60$. As expected, the  convergence 
is  faster on increasing the  dissipative process.

\subsection{Thermodynamic uncertainty relations}
\label{sec:4.2}

Thermodynamic fluctuations strongly affect  the dynamics and stability of nanoscale thermal machines. As well-known, thermodynamic uncertainty relations (TURs) impose strict restrictions on the 
fluctuations of thermodynamic currents, say for example $Q_{\alpha}$ \cite{Barato_2015,Horowitz_2020,Pietzonka_2016,Pietzonka_2017,Pietzonka_2018,Timpanaro_2019},
\begin{equation}
\begin{aligned}
	 \frac{\text{Var}(Q_{\alpha})}{\langle Q_{\alpha} \rangle^2 } \geq \frac{2}{\langle \Sigma \rangle},
\label{eq:TUR_1}
\end{aligned}
\end{equation} with $\text{Var}(Q_{\alpha})= \langle Q_{\alpha}^2 \rangle - \langle Q_{\alpha} \rangle ^2$ and $\langle \Sigma \rangle$ the averaged entropy production. Eq. \eqref{eq:TUR_1} expresses a trade-off between process precision, quantified by the signal-to-noise ratio (SNR), and dissipation, quantified through the entropy production. As a matter of fact, in order to reduce fluctuations in the heat exchange $Q_{\alpha}$ and stabilize the cycle, Eq.\eqref{eq:TUR_1} states that it is necessary to increase dissipation. 
\begin{figure}[!htb]
  \centering
   \includegraphics[width=7.5 cm]{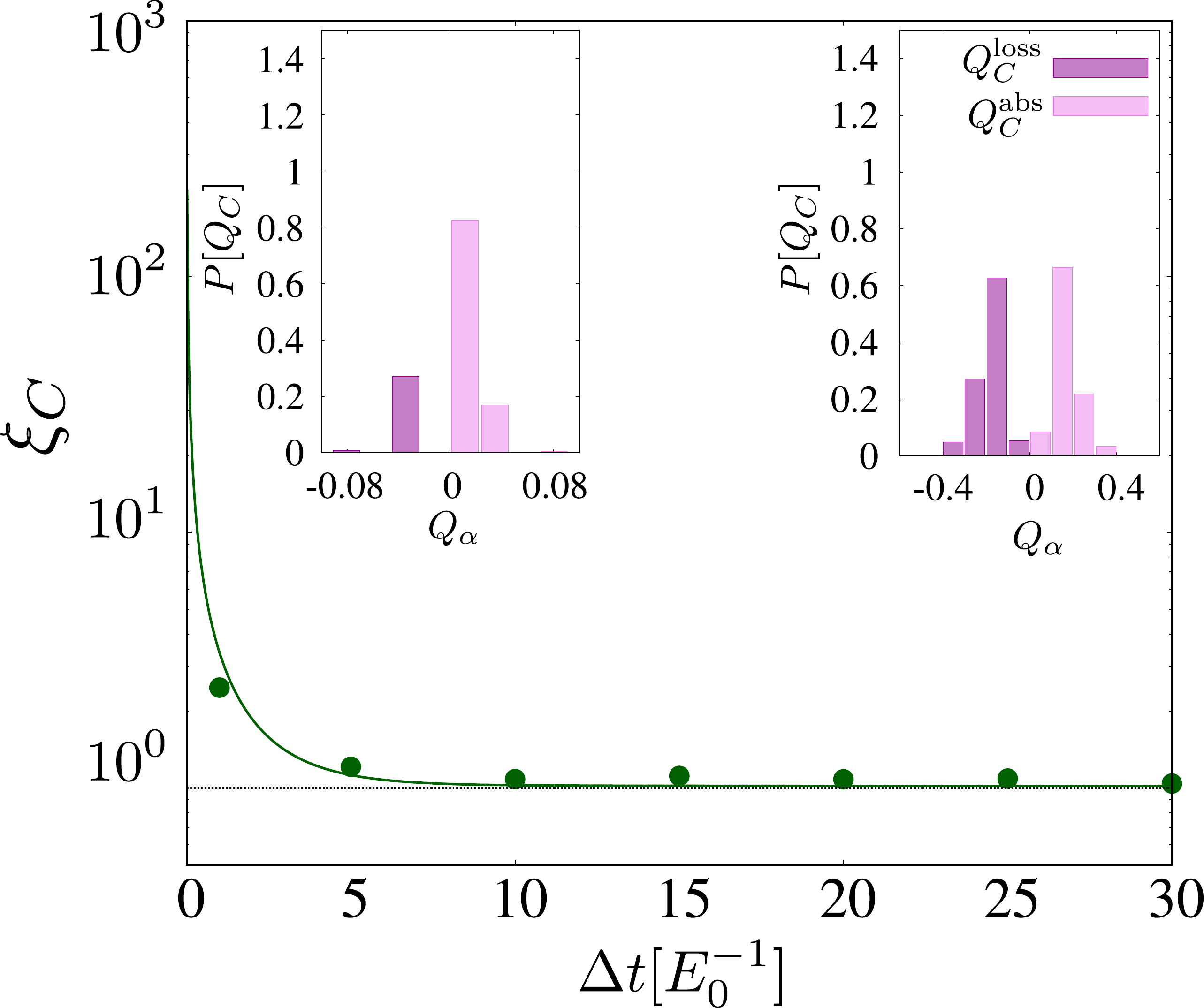}
   \caption{\textbf{Incoherent cycle}. (a) Numerical and analytical plots of the ratio $\xi_{C}$ as a function $\Delta t$ for the case of $\Gamma_{\phi}=0$. Numerical results are plotted using points, while analytical results are plotted using bold lines. Inset plots depict the distribution probability of $Q_{C}$ for $\Delta t [E_{0}^{-1}] = 1$ and $\Delta t [E_{0}^{-1}] = 20$, respectively.  All the numerical results are obtained considering $3.10^4$ trajectories. We fixed $\Gamma_{\phi}=0$. The other numerical parameters are the same in Fig.\ref{fig:2}.}
   \label{fig:9}
\end{figure} 

In what follows, we study analytically and numerically the behavior of the previous TUR bound for the coherent and incoherent cycles. Recall that, independently of the operating regime, it is possible to range the duration of the isothermal strokes by changing the time interval $\Delta t$. In terms of system precision and cycle stabilization, it is thus interesting to explore how the isothermal stroke duration may impact fluctuations of certain heat currents. From now on, and for the sake of simplicity, we will analyze ratio 
$$
\xi_{\alpha}=\Big(\frac{\text{Var}(Q_{\alpha})}{\langle Q_{\alpha} \rangle^2 }\Big) / \Big( \frac{2}{\langle \Sigma \rangle} \Big) \;.
$$ 
Eq.\eqref{eq:TUR_1} now reads $\xi_{\alpha}\geq1$.

We employ, also in this case, the MCWF method to compute the numerical results. The averaged entropy production computes as $\langle \Sigma\rangle = \beta_{C} \langle Q_{C}\rangle + 
\beta_{H} \langle Q_{H}\rangle$, with $\langle Q_{\alpha}\rangle$ defined in Eq.\eqref{eq:E_alpha_1}, while the variance is given by
\begin{equation}
	\begin{aligned}
	\text{Var} (Q_{\alpha}) &= \Big( \Delta E_{\alpha}\Big)^2 \Big[ \frac{1}{2} -  \Big( \mathcal{D}_{\text{f}} + \mathcal{D}_{\text{i}}  \Big)^2 \Big],
	\label{eq:Var_Q_f}
	\end{aligned}
\end{equation} with $\mathcal{D}_{\text{i},\text{f}}$ the solutions of Eq.\eqref{eq:SOE_5}. See Appendix \ref{ap:C} for further details.
\begin{figure}[!htb]
  \centering
   \includegraphics[width=7.5cm]{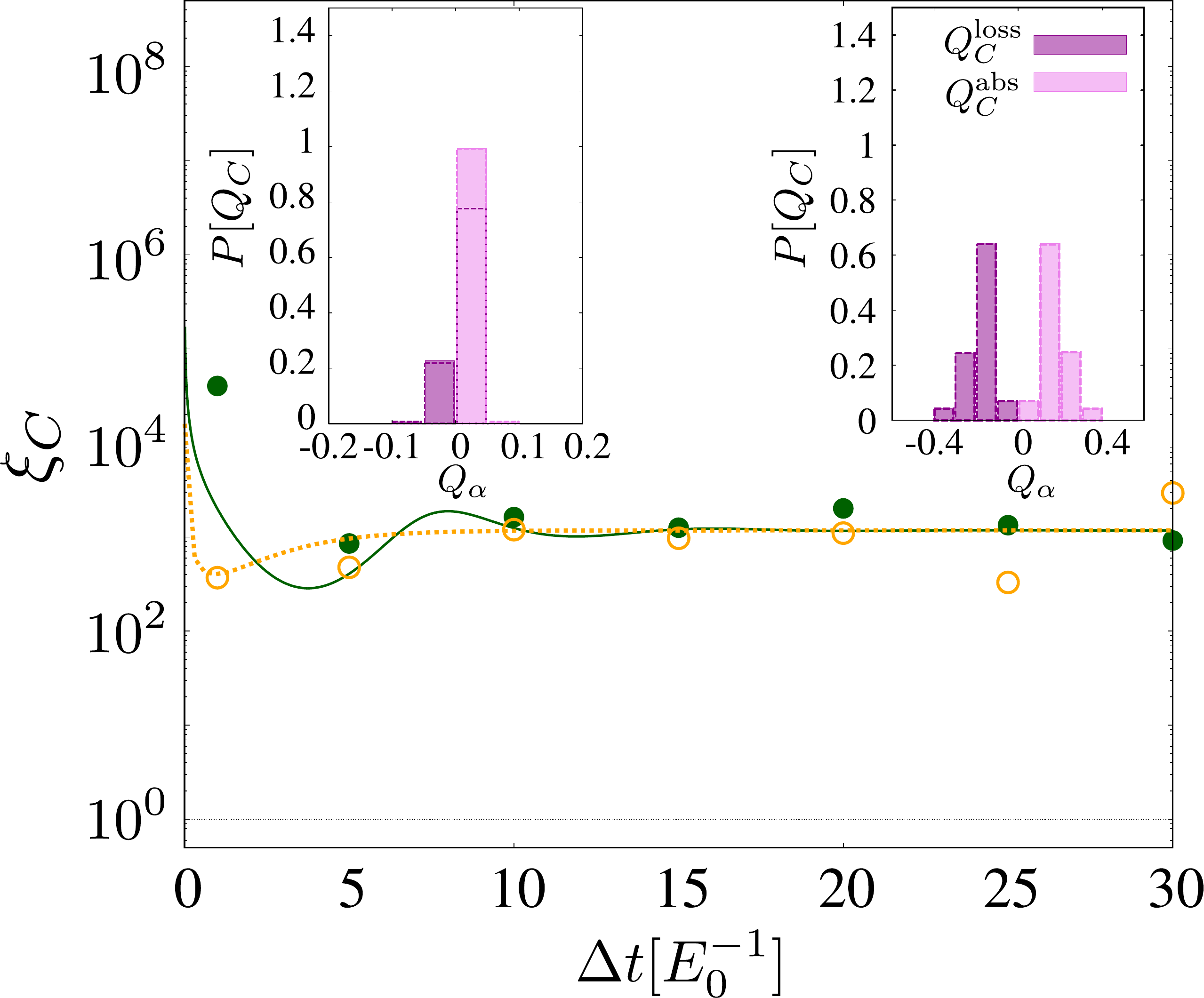}
   \caption{\textbf{Coherent cycle}. Numerical and analytical plots of the ratio $\xi_{C}$ as a function $\Delta t$  for the cases of $\Gamma_{\phi}=0$ (green) and $\Gamma_{\phi}=10\Gamma_{\Sigma,H}$ (orange). Numerical results are plotted using points, while analytical results are plotted using bold lines. Inset plots depict the distribution probability of $Q_{C}$ for $\Delta t [E_{0}^{-1}] = 1$ and $\Delta t [E_{0}^{-1}] = 20$, respectively. Histograms plotted in bold lines correspond to $\Gamma_{\phi}=0$, while histograms plotted in dashed lines correspond to $\Gamma_{\phi}=10\Gamma_{\Sigma,H}$. All the numerical results are obtained by considering $3.10^4$ trajectories. We fixed $\Gamma_{\phi}=0$. The other numerical parameters are the same in Fig.\ref{fig:2}.}
   \label{fig:10}
\end{figure} 

In Fig.\ref{fig:9}, we plot the ratio $\xi_{C}$ as a function of $\Delta t$ without including dephasing, $\Gamma_{\phi}=0$. Here, we consider the incoherent refrigerator, i.e., $q_{C}= 0.019[E_{0}]$,  $q_{H}= 0.38[E_{0}]$, and $\Delta = 0$. Due to the numerical cost, we only plot a few numerical points, each obtained using $3.10^6$ trajectories. Any deviation from the analytical curves is due to poor statistics. Note that
$\xi_{C}= \xi_{H}$ as shown in Appendix \ref{ap:D}.

It is interesting to note that the ratio $\xi_{C}$ is significant for small $\Delta t$, while it saturates for longer time intervals. As mentioned in Sec.\ref{sec:3}, this specific dynamic stems from the thermalization process itself and, as we shall show, reflects in the stochastic results. From the inset plots, it is clear that for short time intervals $\Delta t$, the number of jumps allowed in the system is insignificant. Nevertheless, when $\Delta t$ increases, the system experiences additional energetic exchanges, which enhances the statistics and reduces the standard deviation of the probability distribution $P[Q_{C}]$. In contrast, in the case of prolonged time intervals, $P[Q_{C}]$ remains unchanged since the system thermalizes, thus interrupting the energetic exchanges. 
\begin{figure}[!htb]
  \centering
   \includegraphics[width=7.5cm]{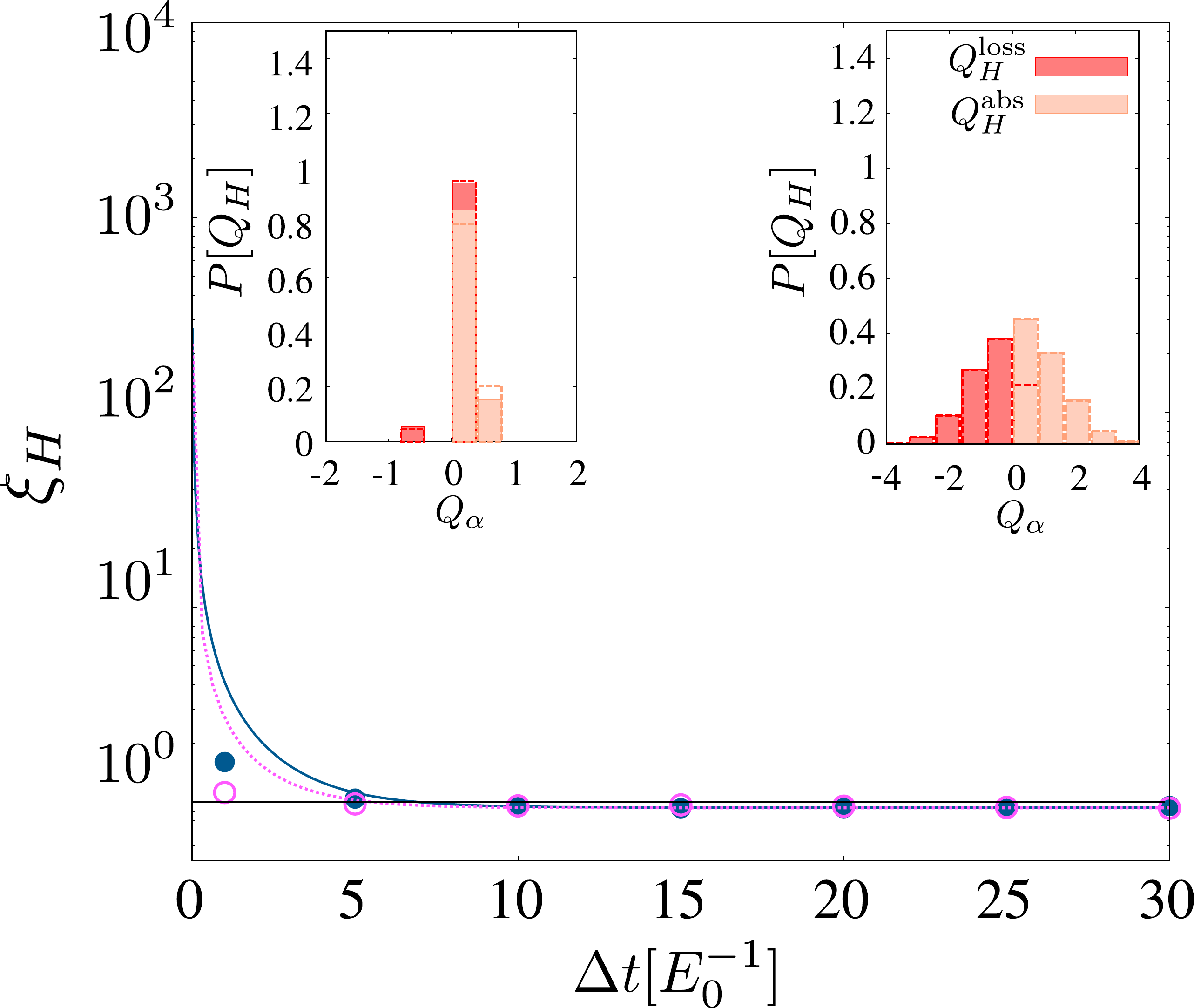}
   \caption{\textbf{Coherent cycle}. Numerical and analytical plots of the ratio $\xi_{H}$  as a function $\Delta t$ for the cases of $\Gamma_{\phi}=0$ (blue) and $\Gamma_{\phi}=10\Gamma_{\Sigma,H}$ (magenta). Numerical results are plotted using points, while analytical results are plotted using bold lines. Inset plots depict the distribution probability of $Q_{H}$ for $\Delta t [E_{0}^{-1}] = 1$ and $\Delta t [E_{0}^{-1}] = 20$, respectively. Histograms plotted in bold lines correspond to $\Gamma_{\phi}=0$, while histograms plotted in dashed lines correspond to $\Gamma_{\phi}=10\Gamma_{\Sigma,H}$. All the numerical results are obtained by considering $3.10^4$ trajectories. The other numerical parameters are the same in Fig.\ref{fig:2}.}
   \label{fig:11}
\end{figure} 

When considering the coherent cycle, we find that the process precision for the thermodynamic current $Q_{C}$ worsens, i.e., the ratio $\xi_{C}$ increases (see Fig.\ref{fig:10}). Although not shown, an additional numerical analysis unveiled that such an increment in $\xi_{C}$ is mainly due to a reduction in $\langle Q_{C}\rangle$ since the variance $\text{Var}(Q_{C})$ remains almost unchanged for both regimes. As clearly shown in Fig.\ref{fig:3}, the average heat $\langle Q_{C}\rangle$ decreases in the presence of coherences. Indeed, it is easy to prove that average heat exchange explicitly depends on coherences as $\langle Q_{C} \rangle = \Delta E_{C} \mathcal{R}_{\text{c}}$, see Appendix \ref{ap:D}.

The remarkable result emerging from the coherent case is that system coherences reduce the thermodynamic fluctuations of $Q_{H}$ below the classical bound for large values of $\Delta t$, see Fig.\ref{fig:11}. Similar results have been observed in previous works \cite{Timpanaro_2019,Ptaszynnski_2018,Arash_2021}. Interestingly, when a dephasing source noise couples to the system, we find that thermodynamic fluctuations remain unaffected. In fact, whereas $\Delta t \gg 1/(\Gamma_{\Sigma,\alpha}/2 + 2\Gamma_{\phi,\alpha}) $, we successfully prove that the ratios $\xi_{C}$ and  $\xi_{H}$ both compute as 
\begin{equation}
	\begin{aligned}
	\xi_{C,H} =  \frac{\Big[ \frac{1}{2} - \mathcal{R}_{\text{c,a}}^2 \Big]}{ \mathcal{R}_{\text{c,a}}^2 \Big( \Delta E_{C}\mathcal{R}_{\text{c}} - \Delta E_{H}\mathcal{R}_{\text{a}} \Big) }
	\label{eq:ratio_CH_coh_4},
	\end{aligned}
\end{equation} regardless of the value of $\Gamma_{\phi}$. See Appendix \ref{ap:D} for further details.

\section{Conclusions}
\label{sec:5}

We studied the dynamics of a fast-driven Otto cycle operating under different regimes. Specifically, by employing a Lindbladian approach, we successfully identified the refrigerator and non-refrigerator cycles. The results of this first study support previous ideas that optimal refrigeration can be realized by mimicking classical dynamics via a simple incoherent sudden cycle \cite{Pekola_2019}. In fact, we proposed an alternative fast-driving protocol for which the system behaves as an incoherent refrigerator, where all system coherences in the steady state regime are suppressed. The protocol consists of a sudden variation of the control parameter $q_{\alpha}$, followed by a complete cancelation of the detuning energy $\Delta$.

Further stochastic analysis revealed that the suppression of coherence not only restores the cooling in the cycle but also increases or decreases the quantum fluctuations of certain thermodynamic currents. Specifically, when analyzing the stochastic characteristics of the heat exchange $Q_{C}$, we found that the system precision worsens when system coherences are strong. Although, for the same cycle protocol, we observed the opposite behavior for the heat exchange $Q_{H}$. Here, the current instabilities reduce, even below the classical TUR bound \eqref{eq:TUR_1} \cite{Timpanaro_2019,Ptaszynnski_2018}. 

Interestingly, we found that, for all regimes, the cycle stabilizes when we fix a long time interval $\Delta t$. In other words, system precision improves when the isothermal strokes of the thermal machine operate for long periods. Here, the system thermalizes, and the entropy production reaches its maximum value. Since the energetic exchanges cease for long values of $\Delta t$, the probability distributions of $Q_{\alpha}$ remain unchanged, setting fixed values of the SNR $\text{Var}(Q_{\alpha})/\langle Q_{\alpha} \rangle^2$. 
 
We provided evidence that coherence may reduce the entropy production for irreversible processes. On the one hand, we found this mechanism minimizes typical random deviations in FRs due to poor statistics. On the contrary, when we analyzed the behavior of the integral $\langle \text{e}^{-\Sigma} \rangle$ as a function of the stochastic sampling, our results revealed the convergence of the equality $\langle \text{e}^{-\Sigma} \rangle=1$ presents a slower rate than for the incoherent cycle. These results correlate favorably with Ref.\cite{Jarzynski_2006}. 

Our research has highlighted the role of system coherences in small thermal machines. As shown in our work, the stochastic approach provides a powerful tool for investigating possible instabilities in the cycle. In fact, the evidence from this study points towards the idea that coherence plays a relevant role in thermodynamic fluctuations and fluctuation relations.

\section*{Acknowledgments}

We acknowledge Michele Campisi for insightful discussions. The work of R.F. has been supported by a Google Quantum Research Award. R.F. acknowledges that his research has been conducted within the framework of the Trieste Institute for Theoretical Quantum Technologies (TQT). E. P. acknowledges the QUANTERA project SiUCs and COST Action CA 21144 superqumap. E.P. also acknowledges ICTP for the hospitality.

\appendix
\renewcommand\thefigure{\thesection.\arabic{figure}} 
\section{\label{ap:A} Lindblad equation}
\setcounter{figure}{0} 

By replacing the jumps operators $L_{\beta,\alpha}$, $\beta=\uparrow,\downarrow,\phi$, into Eq.\eqref{eq:lindblad}, we get the following set of uncoupled equations
\begin{equation}
	\begin{aligned}
	\dot{\rho}_{gg} &= - \Gamma_{\Sigma,\alpha} \rho_{gg} + \Gamma_{\downarrow}^{\alpha},\\
	\dot{\rho}_{ge} &= - \Big(\frac{\Gamma_{\Sigma,\alpha}}{2} + 2\Gamma_{\phi,\alpha} \Big) \rho_{ge},
	\label{eq:rho_deph_1_noapp}
	\end{aligned}
\end{equation} with $\Gamma_{\Sigma,\alpha} = \Gamma_{\uparrow,\alpha} + \Gamma_{\downarrow,\alpha}$, $\alpha=H,C$ The solution of Eq.\eqref{eq:rho_deph_1_noapp} is thus
\begin{equation}
	\begin{aligned}
	{\rho}^{\text{f}}_{gg} &= \frac{\Gamma_{\downarrow,\alpha}}{\Gamma_{\Sigma,\alpha}} + \frac{ (-\Gamma_{\downarrow,\alpha}  + \Gamma_{\Sigma,\alpha} \rho_{gg}^{\text{i}}) }{\Gamma_{\Sigma,\alpha}} \text{e}^{-\Gamma_{\Sigma,\alpha} \Delta t},\\
	{\rho}^{\text{f}}_{ge} &=\text{e}^{-(\frac{\Gamma_{\Sigma,\alpha}}{2} + 2\Gamma^{\alpha}_\phi) \Delta t} \text{e}^{i\varphi_{\alpha}} {\rho}^{\text{i}}_{ge}.
	\label{eq:rho_deph_3_noapp}
	\end{aligned}
\end{equation} Here, $\varphi_{\alpha} = \int_{0}^{\Delta t} \Delta E_{\alpha} dt = \Delta E_{\alpha} \Delta t$ is the dynamic phase acquired in each leg of the cycle.

For the following calculations, it will be useful to parametrize the density matrix as: 
\begin{equation}
	\begin{aligned}
	\begin{pmatrix}
    \rho_{gg} & \rho_{ge} \\
    \rho_{ge}^{*} & 1-\rho_{gg} 
    \end{pmatrix} \rightarrow
    \begin{pmatrix}
    \mathcal{D}+\frac{1}{2} & \mathcal{R}+i\mathcal{I} \\
    \mathcal{R}-i\mathcal{I} & \frac{1}{2}-\mathcal{D}, 
    \end{pmatrix}
	\label{eq:rep}
	\end{aligned}
\end{equation} where $\mathcal{D} = \rho_{gg} - 1/2$, $\mathcal{R} = \text{Re} (\rho_{ge})$, and $\mathcal{I} = \text{Im} (\rho_{ge})$. Eq.\eqref{eq:rho_deph_3_noapp} thus transforms to a coupled system of equation of form
\begin{equation}
	\begin{aligned}
	\mathcal{D}_{\text{f}} &= \frac{\Gamma_{\downarrow,\alpha}}{\Gamma_{\Sigma,\alpha}} + \frac{ (-\Gamma_{\downarrow,\alpha}  + \Gamma_{\Sigma,\alpha} (\mathcal{D}_{\text{i}} + \frac{1}{2})) }{\Gamma_{\Sigma,\alpha}} \text{e}^{-\Gamma_{\Sigma,\alpha} \Delta t} - \frac{1}{2},\\
	\mathcal{R}_{\text{f}} &=\text{e}^{-(\frac{\Gamma_{\Sigma,\alpha}}{2} + 2\Gamma_{\phi,\alpha}) \Delta t} \Big( \mathcal{R}_{\text{i}} \cos(\varphi_{\alpha}) -  \mathcal{I}_{\text{i}} \sin(\varphi_{\alpha}) \Big),\\
	\mathcal{I}_{\text{f}} &=\text{e}^{-(\frac{\Gamma_{\Sigma,\alpha}}{2} + 2\Gamma_{\phi,\alpha}) \Delta t} \Big( \mathcal{R}_{\text{i}} \sin(\varphi_{\alpha}) +  \mathcal{I}_{\text{i}} \cos(\varphi_{\alpha}) \Big).
	\label{eq:SOE_3_new_noapp}
	\end{aligned}
\end{equation}

Taking into account each thermalization stroke, we obtain the following set of equations:
\begin{equation}
	\begin{aligned}
	\mathcal{D}_{\text{b}} &= \frac{\Gamma_{\downarrow,C}}{\Gamma_{\Sigma,C}} + \frac{ (-\Gamma_{\downarrow,C}  + \Gamma_{\Sigma,C} (\mathcal{D}_{\text{a}} + \frac{1}{2})) }{\Gamma_{\Sigma,C}} \text{e}^{-\Gamma_{\Sigma,C} \Delta t} - \frac{1}{2},\\
	\mathcal{R}_{\text{b}} &=\text{e}^{-(\frac{\Gamma_{\Sigma,C}}{2} + 2\Gamma_{\phi,C}) \Delta t} \Big( \mathcal{R}_{\text{a}} \cos(\varphi_{C}) -  \mathcal{I}_{\text{a}} \sin(\varphi_{C}) \Big),\\
	\mathcal{I}_{\text{b}} &=\text{e}^{-(\frac{\Gamma_{\Sigma,C}}{2} + 2\Gamma_{\phi,C}) \Delta t} \Big( \mathcal{R}_{\text{a}} \sin(\varphi_{C}) +  \mathcal{I}_{\text{a}} \cos(\varphi_{C}) \Big),\\
	\mathcal{D}_{\text{d}} &= \frac{\Gamma_{\downarrow,H}}{\Gamma_{\Sigma,H}} + \frac{ (-\Gamma_{\downarrow,H}  + \Gamma_{\Sigma,H} (\mathcal{D}_{\text{c}} + \frac{1}{2})) }{\Gamma_{\Sigma,H}} \text{e}^{-\Gamma_{\Sigma,H} \Delta t} - \frac{1}{2},\\
	\mathcal{R}_{\text{d}} &=\text{e}^{-(\frac{\Gamma_{\Sigma,H}}{2} + 2\Gamma_{\phi,H}) \Delta t} \Big( \mathcal{R}_{\text{c}} \cos(\varphi_{H}) -  \mathcal{I}_{\text{c}} \sin(\varphi_{H}) \Big),\\
	\mathcal{I}_{\text{d}} &=\text{e}^{-(\frac{\Gamma_{\Sigma,H}}{2} + 2\Gamma_{\phi,H}) \Delta t} \Big( \mathcal{R}_{\text{c}} \sin(\varphi_{H}) +  \mathcal{I}_{\text{c}} \cos(\varphi_{H}) \Big)
	\label{eq:SOE_4_new_noapp}
	\end{aligned}
\end{equation}

The isentropic processes are modeled by simply imposing the continuity condition of the density matrix, i.e., $\rho^\text{f} = \rho^\text{i}$. The relation between the elements of the final and initial density matrix is given by $\rho_{kl}^\text{f} = \sum_{k'l'} \rho_{k'l'}^\text{i} \langle k | k' \rangle \langle l' | l \rangle$ with $kl$ and $k'l'$  the indexes of the new and old bases, respectively. In our case, the transformation $\rho_{kl}^\text{f}  \to \rho_{k'l'}^\text{i} $ must done for the strokes $\text{b} \to \text{c}$ and $\text{d} \to \text{a}$. For our cycle, we then get
\begin{equation}
	\begin{aligned}
	\mathcal{D}_{\text{c}} &= \Big(\eta_{C} \eta_{H} + \sqrt{1-\eta_{C}^2}\sqrt{1-\eta_{H}^2} \Big) \mathcal{D}_{\text{b}} \\
	&+ \Big(\eta_C \sqrt{1-\eta_{H}^2} - \eta_H \sqrt{1-\eta_{C}^2} \Big)\mathcal{R}_{\text{b}},\\
	\mathcal{R}_{\text{c}} &= \Big(- \eta_{C}\sqrt{1-\eta_{H}}^2 + \eta_{H}\sqrt{1-\eta_{C}}^2 \Big) \mathcal{D}_{\text{b}} \\
	&+ \Big(\eta_{C} \eta_{H} + \sqrt{1-\eta_{C}^2}\sqrt{1-\eta_{H}^2} \Big) \mathcal{R}_{\text{b}},\\
	\mathcal{I}_{\text{c}} &= \mathcal{I}_{\text{b}}\\
	\mathcal{D}_{\text{a}} &= \Big(\eta_{C} \eta_{H} + \sqrt{1-\eta_{C}^2}\sqrt{1-\eta_{H}^2} \Big) \mathcal{D}_{\text{d}} \\
	&+ \Big(-\eta_C \sqrt{1-\eta_{H}^2} + \eta_H \sqrt{1-\eta_{C}^2} \Big)\mathcal{R}_{\text{d}},\\
	\mathcal{R}_{\text{a}} &= \Big( \eta_{C}\sqrt{1-\eta_{H}}^2 - \eta_{H}\sqrt{1-\eta_{C}}^2 \Big) \mathcal{D}_{\text{d}} \\
	&+ \Big(\eta_{C} \eta_{H} + \sqrt{1-\eta_{C}^2}\sqrt{1-\eta_{H}^2} \Big) \mathcal{R}_{\text{d}},\\
	\mathcal{I}_{\text{a}} &= \mathcal{I}_{\text{d}},
	\label{eq:isen}
	\end{aligned}
\end{equation}

\renewcommand\thefigure{\thesection.\arabic{figure}} 
\section{\label{ap:B} Monte Carlo Wave Function method}
\setcounter{figure}{0}

Without going into further details, the  MCWF method consists of the following two elements. For a specific $j$th-realization of the stochastic process, the system is described as a pure state $|\phi_{j}(t) \rangle$, which at time $t+\delta t$ evolves as
\begin{equation}
|\phi_{j}(t+\delta t)\rangle =  \left\{
\begin{tabular}{ ccc } 
 $ \frac{C_{0,\alpha}|\phi_{j}(t) \rangle }{ \sqrt{\delta p_{\alpha}(t)}} $ & with probability & $\delta p_{\alpha}(t)$ \\
          &         &         \\ 
 $ \frac{C_{k,\alpha}|\phi_{j}(t) \rangle }{ \sqrt{\delta p_{k}(t)}} $ & with probability & $\delta p_{k,\alpha}(t)$ \\
 \end{tabular}
				\right.
 \label{eq:MCWF}
\end{equation} The evolution operators are defined as $C_{0,\alpha}= 1- i \delta t  H/\hbar$ and  $C_{k,\alpha} = \sqrt{\delta t} L_{k,\alpha}$, with $H=H_{\alpha} - i \hbar \sum_{m} C^{\dag}_{k,\alpha} C_{k,\alpha}$ and $L_{k,\alpha}$ the jump operators given in Eq.\eqref{eq:lindblad}. Since  $C_{0,\alpha}$ and $C_{k,\alpha}$ are non Hermitians, for each case, the wavefunction evolution is followed by a normalization given by the norms $\sqrt{\delta p_{\alpha}(t)}$ and $\sqrt{\delta p_{k}(t)}$, with $\delta p_{\alpha}(t) = \sum_{k} \delta p_{k,\alpha}(t)$ and $\delta p_{m,\alpha}(t) = \delta t \langle \phi_{j}(t) | C^{\dag}_{k,\alpha}C_{k,\alpha} | \phi_{j}(t)\rangle$. The time step $\delta t $ is chosen to be sufficiently small in order to satisfy $\delta p_{\alpha} \ll 1$. 

Eq.\eqref{eq:MCWF} basically states that, given a specific $j$th-trajectory, one part of the system evolves as a smooth function governed by a non-hermitian Hamiltonian $H$, while, between $[t,t+\delta t]$, the system $|\phi_{j}(t)\rangle$ can experience random jumps occurring with probability $\delta p_{k,\alpha}(t)$ and defined by the jump operators $C_{k,\alpha}$.

\renewcommand\thefigure{\thesection.\arabic{figure}} 
\section{\label{ap:C} Variance}
\setcounter{figure}{0} 

To obtain an analytical expression of variance $\text{Var}(Q_{\alpha})$, we need to go a step further. We start by defining the cumulant generating function of $E_{\alpha}$, 
\begin{equation}
	\begin{aligned}
	C_{E_{\alpha}}(\lambda) &= \text{ln} \langle e^{\lambda H_{\alpha}}  \rangle.
	\label{eq:K_lambda}
	\end{aligned}
\end{equation} In this way, we have
\begin{equation}
	\begin{aligned}
	\langle E_{\alpha} \rangle &= \frac{\partial C_{E_{\alpha}}(\lambda) }{\partial \lambda} ,\\
	\text{Var} (E_{\alpha}) &= \frac{\partial^2 C_{E_{\alpha}}(\lambda) }{\partial^2 \lambda}.
	\label{eq:K_n}
	\end{aligned}
\end{equation} After evaluating the Hamiltonian $H_{\alpha}$ into $C_{E_{\alpha}}(\lambda)$, Eq.\eqref{eq:K_n} changes to
\begin{equation}
	\begin{aligned}
	\langle E_{\alpha} \rangle &=  -\frac{\Delta E_{\alpha}}{2} \Big(2\rho_{gg}-1\Big) ,\\
	\text{Var} (E_{\alpha}) &= \Big( \frac{\Delta E_{\alpha}}{2} \Big)^2 - \langle E_{\alpha} \rangle^2 .
	\label{eq:K_n_1}
	\end{aligned}
\end{equation} 

The mean value of $ \langle Q_{\alpha} \rangle = \langle E^{\text{f}}_{\alpha} \rangle- \langle E^{\text{i}}_{\alpha} \rangle$ is thus $ \langle Q_{\alpha} \rangle =  \Delta E_{\alpha} \Big(\rho^{\text{f}}_{gg}-\rho^{\text{i}}_{gg}\Big)= \Delta E_{\alpha}\Big(\mathcal{D}_{\text{f}}-\mathcal{D}_{\text{i}}\Big)$with $\mathcal{D}^{\text{f,i}} = \rho^{\text{f,i}}_{gg} -1/2$. This last expression perfectly matches with Eq.\eqref{eq:Q_CIncoh}.

In the case of $\text{Var} (Q_{\alpha})$, we employ the identity $\text{Var} (a X + b Y )= a^2 \text{Var} (X) + b^2 \text{Var} (Y) + 2 ab \text{Cov}(X,Y)$, obtaining
\begin{equation}
	\begin{aligned}
	\text{Var} (Q_{\alpha}) &= \text{Var} (E^{\text{f}}_{\alpha}) + \text{Var} (E^{\text{i}}_{\alpha}) - 2 \text{Cov}(E^{\text{f}}_{\alpha},E^{\text{i}}_{\alpha}),\\
	&= 2 \Big( \frac{\Delta E_{\alpha}}{2} \Big)^2 - \langle E^{\text{f}}_{\alpha} \rangle^2 - \langle E^{\text{i}}_{\alpha} \rangle^2 - 2 \text{Cov}(E^{\text{f}}_{\alpha},E^{\text{i}}_{\alpha}),\\
	&=  2 \Big( \frac{\Delta E_{\alpha}}{2} \Big)^2 - \Big(\frac{\Delta E_{\alpha}}{2} \Big)^2 \Big(2\rho^{\text{f}}_{gg}-1\Big)^2 \\
	&- \Big(\frac{\Delta E_{\alpha}}{2} \Big)^2 \Big(2\rho^{\text{i}}_{gg}-1\Big)^2 - 2  \text{Cov}(E^{\text{f}}_{\alpha},E^{\text{i}}_{\alpha}),\\
	&= 2 \Big( \frac{\Delta E_{\alpha}}{2} \Big)^2 - \Big( \Delta E_{\alpha} \Big)^2 \Big( \mathcal{D}_{\text{f}}^2 + \mathcal{D}_{\text{i}}^2 \Big) \\
	&-  2\text{Cov}(E^{\text{f}}_{\alpha},E^{\text{i}}_{\alpha}) .
	\label{eq:Var_Q_1}
	\end{aligned}
\end{equation} 
The $\text{Cov}(E^{\text{f}}_{\alpha},E^{\text{i}}_{\alpha})$ can be simply computed as follows
\begin{equation}
	\begin{aligned}
	\text{Cov}(E^{\text{f}}_{\alpha},E^{\text{i}}_{\alpha}) &= \frac{\partial^{2} C_{E^{\text{i}}_{\alpha},E^{\text{f}}_{\alpha}}(\lambda_{\text{i}},\lambda_{\text{f}}) }{\partial \lambda_{\text{i}}\partial \lambda_{\text{f}}},
	\label{eq:Cov_1}
	\end{aligned}
\end{equation} with 
\begin{equation}
	\begin{aligned}
	C_{E^{\text{i}}_{\alpha},E^{\text{f}}_{\alpha}}(\lambda_{\text{i}},\lambda_{\text{f}}) &= \text{ln} \langle e^{\lambda_{\text{i}} H_{\alpha} \delta_{t,t_{\text{i}}} + \lambda_{\text{f}} H_{\alpha} \delta_{t,t_{\text{f}}} }  \rangle.
	\label{eq:K_lambda_2}
	\end{aligned}
\end{equation}  We get then
\begin{equation}
	\begin{aligned}
	\text{Cov}(E^{\text{f}}_{\alpha},E^{\text{i}}_{\alpha}) &= (\Delta E_{\alpha})^2 \mathcal{D}_{\text{i}}\mathcal{D}_{\text{f}},\\
	\label{eq:Cov_2}
	\end{aligned}
\end{equation} 

Finally, the variance can be written as
\begin{equation}
	\begin{aligned}
	\text{Var} (Q_{\alpha}) &= 2 \Big( \frac{\Delta E_{\alpha}}{2} \Big)^2 - \Big(\Delta E_{\alpha} \Big)^2 \Big( \mathcal{D}_{\text{f}}^2 + \mathcal{D}_{\text{i}}^2 \Big) \\
	&-  2\Big( \Delta E_{\alpha} \Big)^2 \mathcal{D}_{\text{i}}\mathcal{D}_{\text{f}} ,\\
	&= 2 \Big( \frac{\Delta E_{\alpha}}{2} \Big)^2 - \Big( \Delta E_{\alpha} \Big)^2 \Big( \mathcal{D}_{\text{f}} + \mathcal{D}_{\text{i}}  \Big)^2,\\
	\text{Var} (Q_{\alpha}) &= \Big( \Delta E_{\alpha}\Big)^2 \Big[ \frac{1}{2} -  \Big( \mathcal{D}_{\text{f}} + \mathcal{D}_{\text{i}}  \Big)^2 \Big].
	\label{eq:Var_Q_2}
	\end{aligned}
\end{equation} 

\renewcommand\thefigure{\thesection.\arabic{figure}} 
\section{\label{ap:D} TUR bound}
\setcounter{figure}{0} 

We start by considering the incoherent case. Since $\eta_{C},\eta_{H} = 1$, the continuity conditions \eqref{eq:isen} reduce to $\mathcal{D}_{\text{b}}=\mathcal{D}_{\text{c}}$ and $\mathcal{D}_{\text{d}}=\mathcal{D}_{\text{a}}$ . The variance for both currents $Q_{C}$ and $Q_{H}$ can be simply computed as
\begin{equation}
	\begin{aligned}
	\text{Var} (Q_{C,H}) &= \Big( \Delta E_{C,H}\Big)^2 \Big[ \frac{1}{2} -  \Big( \mathcal{D}_{\text{b}} + \mathcal{D}_{\text{a}}  \Big)^2 \Big]
	\label{eq:Var_CQ_incoh}
	\end{aligned}
\end{equation} 

From Eqs.\eqref{eq:Var_CQ_incoh}  and \eqref{eq:Q_CIncoh}, it is easy to proof that the ratios $\xi_{C}$ and $\xi_{H}$ are equals and compute as
\begin{equation}
	\begin{aligned}
	\xi_{C,H} &= \frac{1}{\Delta E_{H} - \Delta E_{C}} \frac{\Big[ \frac{1}{2} -  \Big( \mathcal{D}_{\text{b}} + \mathcal{D}_{\text{a}}  \Big)^2 \Big]}{\Big(\mathcal{D}_{\text{b}}-\mathcal{D}_{\text{a}}\Big)^2},
	\label{eq:eta_CH_incoh}
	\end{aligned}
\end{equation} with $\mathcal{D}_{\text{b}}-\mathcal{D}_{\text{a}}=\mathcal{F}[\Gamma^{C}_{\downarrow},\Gamma^{H}_{\downarrow},\Delta t]$ defined in Eq.\eqref{eq:f_incoh}.

We now move forward and consider the coherent case. For simplicity, we approximate $\eta_{H} \approx 1$ as $q_{H} \ll \Delta$. Under this assumption, and following the results in Eq.\eqref{eq:isen}, the density matrix elements $\mathcal{D}$ and $\mathcal{R}$ for stroke $\text{a} \to \text{b}$ transform to $\mathcal{D}_{\text{a}} \approx \mathcal{R}_{\text{d}}$ and $\mathcal{D}_{\text{b}} \approx \mathcal{R}_{\text{c}}$, while for the stroke  $\text{c} \to \text{d}$, we have $\mathcal{D}_{\text{c}} \approx -\mathcal{R}_{\text{b}}$ and $\mathcal{D}_{\text{d}} \approx -\mathcal{R}_{\text{a}}$. The variances $\text{Var} (Q_{C})$  and $\text{Var} (Q_{H})$ can be thus expressed as
\begin{equation}
	\begin{aligned}
    \text{Var} (Q_{C}) &= \Big( \Delta E_{C}\Big)^2 \Big[ \frac{1}{2} -  \Big( \mathcal{R}_{\text{c}} + \mathcal{R}_{\text{d}}  \Big)^2 \Big],\\
	\text{Var} (Q_{H}) &= \Big( \Delta E_{H}\Big)^2 \Big[ \frac{1}{2} -  \Big( \mathcal{R}_{\text{a}} + \mathcal{R}_{\text{b}}  \Big)^2 \Big].
	\label{eq:Var_Q_CH_coh_1}
	\end{aligned}
\end{equation} If we consider the case of large time interval $\Delta t \gg 1/ (\Gamma_{\Sigma,\alpha}/2 + 2\Gamma_{\phi,\alpha})$, it is thus possible to neglect the terms $\text{e}^{-(\frac{\Gamma_{\Sigma,\alpha}}{2} + 2\Gamma_{\phi,\alpha}) \Delta t} \cos(\varphi_{\alpha} ) \approx 0$ and $\text{e}^{-(\frac{\Gamma_{\Sigma,\alpha}}{2} + 2\Gamma_{\phi,\alpha}) \Delta t} \sin(\varphi_{\alpha} ) \approx 0$ in Eq.\eqref{eq:SOE_4_new_noapp}. In this manner, the continuity conditions reduce to $\mathcal{R}_{\text{b}},\mathcal{R}_{\text{d}}\approx 0$. Eq.\eqref{eq:Var_Q_CH_coh_1} now reads
\begin{equation}
	\begin{aligned}
    \text{Var} (Q_{C,H}) &= \Big( \Delta E_{C,H}\Big)^2 \Big[ \frac{1}{2} -  \mathcal{R}_{\text{c,a}}^2\Big].
	\label{eq:Var_Q_CH_coh_2}
	\end{aligned}
\end{equation} Likewise, the averaged heats $\langle Q_{H} \rangle$ and $\langle Q_{C} \rangle$ in Eq.\eqref{eq:E_alpha_1} transform as
$\langle Q_{C} \rangle= \Delta E_{C} (\mathcal{R}_{\text{c}} -\mathcal{R}_{\text{d}} )$ and $\langle Q_{H} \rangle=\Delta E_{H} (\mathcal{R}_{\text{b}} -\mathcal{R}_{\text{a}} )$. In particular, we get
\begin{equation}
	\begin{aligned}
	\langle Q_{C} \rangle = \Delta E_{C} \mathcal{R}_{\text{c}}, \,
    \langle Q_{H} \rangle = -\Delta E_{H} \mathcal{R}_{\text{a}}.
	\label{eq:Q_all_2}
	\end{aligned}
\end{equation} Given the last expressions, averaged entropy production expresses as $\langle \Sigma \rangle = \Delta E_{C}\mathcal{R}_{\text{c}} - \Delta E_{H}\mathcal{R}_{\text{a}}$.

The ratios $\xi_{C,H}$ can finally be written as
\begin{equation}
	\begin{aligned}
	\xi_{C,H} =  \frac{\Big[ \frac{1}{2} - \mathcal{R}_{\text{c,a}}^2 \Big]}{ \mathcal{R}_{\text{c,a}}^2 \Big( \Delta E_{C}\mathcal{R}_{\text{c}} - \Delta E_{H}\mathcal{R}_{\text{a}} \Big) }.
	\label{eq:ratio_H0_2}
	\end{aligned}
\end{equation} 
\hspace{0.5cm}

\bibliography{references}

\begin{thebibliography}{95}%
\makeatletter
\providecommand \@ifxundefined [1]{%
 \@ifx{#1\undefined}
}%
\providecommand \@ifnum [1]{%
 \ifnum #1\expandafter \@firstoftwo
 \else \expandafter \@secondoftwo
 \fi
}%
\providecommand \@ifx [1]{%
 \ifx #1\expandafter \@firstoftwo
 \else \expandafter \@secondoftwo
 \fi
}%
\providecommand \natexlab [1]{#1}%
\providecommand \enquote  [1]{``#1''}%
\providecommand \bibnamefont  [1]{#1}%
\providecommand \bibfnamefont [1]{#1}%
\providecommand \citenamefont [1]{#1}%
\providecommand \href@noop [0]{\@secondoftwo}%
\providecommand \href [0]{\begingroup \@sanitize@url \@href}%
\providecommand \@href[1]{\@@startlink{#1}\@@href}%
\providecommand \@@href[1]{\endgroup#1\@@endlink}%
\providecommand \@sanitize@url [0]{\catcode `\\12\catcode `\$12\catcode
  `\&12\catcode `\#12\catcode `\^12\catcode `\_12\catcode `\%12\relax}%
\providecommand \@@startlink[1]{}%
\providecommand \@@endlink[0]{}%
\providecommand \url  [0]{\begingroup\@sanitize@url \@url }%
\providecommand \@url [1]{\endgroup\@href {#1}{\urlprefix }}%
\providecommand \urlprefix  [0]{URL }%
\providecommand \Eprint [0]{\href }%
\providecommand \doibase [0]{http://dx.doi.org/}%
\providecommand \selectlanguage [0]{\@gobble}%
\providecommand \bibinfo  [0]{\@secondoftwo}%
\providecommand \bibfield  [0]{\@secondoftwo}%
\providecommand \translation [1]{[#1]}%
\providecommand \BibitemOpen [0]{}%
\providecommand \bibitemStop [0]{}%
\providecommand \bibitemNoStop [0]{.\EOS\space}%
\providecommand \EOS [0]{\spacefactor3000\relax}%
\providecommand \BibitemShut  [1]{\csname bibitem#1\endcsname}%
\let\auto@bib@innerbib\@empty
\bibitem [{\citenamefont {Vinjanampathy}\ and\ \citenamefont
  {Anders}(2016)}]{Vinjanampathy_2016}%
  \BibitemOpen
  \bibfield  {author} {\bibinfo {author} {\bibfnamefont {S.}~\bibnamefont
  {Vinjanampathy}}\ and\ \bibinfo {author} {\bibfnamefont {J.}~\bibnamefont
  {Anders}},\ }\href {\doibase 10.1080/00107514.2016.1201896} {\bibfield
  {journal} {\bibinfo  {journal} {Contemporary Physics}\ }\textbf {\bibinfo
  {volume} {57}},\ \bibinfo {pages} {545} (\bibinfo {year} {2016})},\ \Eprint
  {http://arxiv.org/abs/https://doi.org/10.1080/00107514.2016.1201896}
  {https://doi.org/10.1080/00107514.2016.1201896} \BibitemShut {NoStop}%
\bibitem [{\citenamefont {Ali}\ \emph {et~al.}(2020)\citenamefont {Ali},
  \citenamefont {Huang},\ and\ \citenamefont {Zhang}}]{Ali_2020}%
  \BibitemOpen
  \bibfield  {author} {\bibinfo {author} {\bibfnamefont {M.~M.}\ \bibnamefont
  {Ali}}, \bibinfo {author} {\bibfnamefont {W.-M.}\ \bibnamefont {Huang}}, \
  and\ \bibinfo {author} {\bibfnamefont {W.-M.}\ \bibnamefont {Zhang}},\ }\href
  {\doibase 10.1038/s41598-020-70450-y} {\bibfield  {journal} {\bibinfo
  {journal} {Scientific Reports}\ }\textbf {\bibinfo {volume} {10}},\ \bibinfo
  {pages} {13500} (\bibinfo {year} {2020})}\BibitemShut {NoStop}%
\bibitem [{\citenamefont {Kosloff}(2013)}]{Kosloff_2013}%
  \BibitemOpen
  \bibfield  {author} {\bibinfo {author} {\bibfnamefont {R.}~\bibnamefont
  {Kosloff}},\ }\href {\doibase 10.3390/e15062100} {\bibfield  {journal}
  {\bibinfo  {journal} {Entropy}\ }\textbf {\bibinfo {volume} {15}},\ \bibinfo
  {pages} {2100} (\bibinfo {year} {2013})}\BibitemShut {NoStop}%
\bibitem [{\citenamefont {Pekola}(2015)}]{Pekola_2015}%
  \BibitemOpen
  \bibfield  {author} {\bibinfo {author} {\bibfnamefont {J.~P.}\ \bibnamefont
  {Pekola}},\ }\href {\doibase 10.1038/nphys3169} {\bibfield  {journal}
  {\bibinfo  {journal} {Nature Physics}\ }\textbf {\bibinfo {volume} {11}},\
  \bibinfo {pages} {118} (\bibinfo {year} {2015})}\BibitemShut {NoStop}%
\bibitem [{\citenamefont {Uzdin}\ \emph {et~al.}(2015)\citenamefont {Uzdin},
  \citenamefont {Levy},\ and\ \citenamefont {Kosloff}}]{Uzdin_2015}%
  \BibitemOpen
  \bibfield  {author} {\bibinfo {author} {\bibfnamefont {R.}~\bibnamefont
  {Uzdin}}, \bibinfo {author} {\bibfnamefont {A.}~\bibnamefont {Levy}}, \ and\
  \bibinfo {author} {\bibfnamefont {R.}~\bibnamefont {Kosloff}},\ }\href
  {\doibase 10.1103/PhysRevX.5.031044} {\bibfield  {journal} {\bibinfo
  {journal} {Phys. Rev. X}\ }\textbf {\bibinfo {volume} {5}},\ \bibinfo {pages}
  {031044} (\bibinfo {year} {2015})}\BibitemShut {NoStop}%
\bibitem [{\citenamefont {H\"anggi}\ and\ \citenamefont
  {Marchesoni}(2009)}]{Hanggi_2009}%
  \BibitemOpen
  \bibfield  {author} {\bibinfo {author} {\bibfnamefont {P.}~\bibnamefont
  {H\"anggi}}\ and\ \bibinfo {author} {\bibfnamefont {F.}~\bibnamefont
  {Marchesoni}},\ }\href {\doibase 10.1103/RevModPhys.81.387} {\bibfield
  {journal} {\bibinfo  {journal} {Rev. Mod. Phys.}\ }\textbf {\bibinfo {volume}
  {81}},\ \bibinfo {pages} {387} (\bibinfo {year} {2009})}\BibitemShut
  {NoStop}%
\bibitem [{\citenamefont {Goold}\ \emph {et~al.}(2016)\citenamefont {Goold},
  \citenamefont {Huber}, \citenamefont {Riera}, \citenamefont {del Rio},\ and\
  \citenamefont {Skrzypczyk}}]{Goold_2016}%
  \BibitemOpen
  \bibfield  {author} {\bibinfo {author} {\bibfnamefont {J.}~\bibnamefont
  {Goold}}, \bibinfo {author} {\bibfnamefont {M.}~\bibnamefont {Huber}},
  \bibinfo {author} {\bibfnamefont {A.}~\bibnamefont {Riera}}, \bibinfo
  {author} {\bibfnamefont {L.}~\bibnamefont {del Rio}}, \ and\ \bibinfo
  {author} {\bibfnamefont {P.}~\bibnamefont {Skrzypczyk}},\ }\href {\doibase
  10.1088/1751-8113/49/14/143001} {\bibfield  {journal} {\bibinfo  {journal}
  {Journal of Physics A: Mathematical and Theoretical}\ }\textbf {\bibinfo
  {volume} {49}},\ \bibinfo {pages} {143001} (\bibinfo {year}
  {2016})}\BibitemShut {NoStop}%
\bibitem [{\citenamefont {Bouton}\ \emph {et~al.}(2021)\citenamefont {Bouton},
  \citenamefont {Nettersheim}, \citenamefont {Burgardt}, \citenamefont {Adam},
  \citenamefont {Lutz},\ and\ \citenamefont {Widera}}]{Bouton_2021}%
  \BibitemOpen
  \bibfield  {author} {\bibinfo {author} {\bibfnamefont {Q.}~\bibnamefont
  {Bouton}}, \bibinfo {author} {\bibfnamefont {J.}~\bibnamefont {Nettersheim}},
  \bibinfo {author} {\bibfnamefont {S.}~\bibnamefont {Burgardt}}, \bibinfo
  {author} {\bibfnamefont {D.}~\bibnamefont {Adam}}, \bibinfo {author}
  {\bibfnamefont {E.}~\bibnamefont {Lutz}}, \ and\ \bibinfo {author}
  {\bibfnamefont {A.}~\bibnamefont {Widera}},\ }\href {\doibase
  10.1038/s41467-021-22222-z} {\bibfield  {journal} {\bibinfo  {journal}
  {Nature Communications}\ }\textbf {\bibinfo {volume} {12}},\ \bibinfo {pages}
  {2063} (\bibinfo {year} {2021})}\BibitemShut {NoStop}%
\bibitem [{\citenamefont {Peterson}\ \emph {et~al.}(2019)\citenamefont
  {Peterson}, \citenamefont {Batalh\~ao}, \citenamefont {Herrera},
  \citenamefont {Souza}, \citenamefont {Sarthour}, \citenamefont {Oliveira},\
  and\ \citenamefont {Serra}}]{Peterson_2019}%
  \BibitemOpen
  \bibfield  {author} {\bibinfo {author} {\bibfnamefont {J.~P.~S.}\
  \bibnamefont {Peterson}}, \bibinfo {author} {\bibfnamefont {T.~B.}\
  \bibnamefont {Batalh\~ao}}, \bibinfo {author} {\bibfnamefont
  {M.}~\bibnamefont {Herrera}}, \bibinfo {author} {\bibfnamefont {A.~M.}\
  \bibnamefont {Souza}}, \bibinfo {author} {\bibfnamefont {R.~S.}\ \bibnamefont
  {Sarthour}}, \bibinfo {author} {\bibfnamefont {I.~S.}\ \bibnamefont
  {Oliveira}}, \ and\ \bibinfo {author} {\bibfnamefont {R.~M.}\ \bibnamefont
  {Serra}},\ }\href {\doibase 10.1103/PhysRevLett.123.240601} {\bibfield
  {journal} {\bibinfo  {journal} {Phys. Rev. Lett.}\ }\textbf {\bibinfo
  {volume} {123}},\ \bibinfo {pages} {240601} (\bibinfo {year}
  {2019})}\BibitemShut {NoStop}%
\bibitem [{\citenamefont {Ono}\ \emph {et~al.}(2020)\citenamefont {Ono},
  \citenamefont {Shevchenko}, \citenamefont {Mori}, \citenamefont {Moriyama},\
  and\ \citenamefont {Nori}}]{Ono_2020}%
  \BibitemOpen
  \bibfield  {author} {\bibinfo {author} {\bibfnamefont {K.}~\bibnamefont
  {Ono}}, \bibinfo {author} {\bibfnamefont {S.~N.}\ \bibnamefont {Shevchenko}},
  \bibinfo {author} {\bibfnamefont {T.}~\bibnamefont {Mori}}, \bibinfo {author}
  {\bibfnamefont {S.}~\bibnamefont {Moriyama}}, \ and\ \bibinfo {author}
  {\bibfnamefont {F.}~\bibnamefont {Nori}},\ }\href {\doibase
  10.1103/PhysRevLett.125.166802} {\bibfield  {journal} {\bibinfo  {journal}
  {Phys. Rev. Lett.}\ }\textbf {\bibinfo {volume} {125}},\ \bibinfo {pages}
  {166802} (\bibinfo {year} {2020})}\BibitemShut {NoStop}%
\bibitem [{\citenamefont {Ji}\ \emph {et~al.}(2022)\citenamefont {Ji},
  \citenamefont {Chai}, \citenamefont {Wang}, \citenamefont {Guo},
  \citenamefont {Rong}, \citenamefont {Shi}, \citenamefont {Ren}, \citenamefont
  {Wang},\ and\ \citenamefont {Du}}]{Ji_2022}%
  \BibitemOpen
  \bibfield  {author} {\bibinfo {author} {\bibfnamefont {W.}~\bibnamefont
  {Ji}}, \bibinfo {author} {\bibfnamefont {Z.}~\bibnamefont {Chai}}, \bibinfo
  {author} {\bibfnamefont {M.}~\bibnamefont {Wang}}, \bibinfo {author}
  {\bibfnamefont {Y.}~\bibnamefont {Guo}}, \bibinfo {author} {\bibfnamefont
  {X.}~\bibnamefont {Rong}}, \bibinfo {author} {\bibfnamefont {F.}~\bibnamefont
  {Shi}}, \bibinfo {author} {\bibfnamefont {C.}~\bibnamefont {Ren}}, \bibinfo
  {author} {\bibfnamefont {Y.}~\bibnamefont {Wang}}, \ and\ \bibinfo {author}
  {\bibfnamefont {J.}~\bibnamefont {Du}},\ }\href {\doibase
  10.1103/PhysRevLett.128.090602} {\bibfield  {journal} {\bibinfo  {journal}
  {Phys. Rev. Lett.}\ }\textbf {\bibinfo {volume} {128}},\ \bibinfo {pages}
  {090602} (\bibinfo {year} {2022})}\BibitemShut {NoStop}%
\bibitem [{\citenamefont {von Lindenfels}\ \emph {et~al.}(2019)\citenamefont
  {von Lindenfels}, \citenamefont {Gr\"ab}, \citenamefont {Schmiegelow},
  \citenamefont {Kaushal}, \citenamefont {Schulz}, \citenamefont {Mitchison},
  \citenamefont {Goold}, \citenamefont {Schmidt-Kaler},\ and\ \citenamefont
  {Poschinger}}]{Lindenfels_2019}%
  \BibitemOpen
  \bibfield  {author} {\bibinfo {author} {\bibfnamefont {D.}~\bibnamefont {von
  Lindenfels}}, \bibinfo {author} {\bibfnamefont {O.}~\bibnamefont {Gr\"ab}},
  \bibinfo {author} {\bibfnamefont {C.~T.}\ \bibnamefont {Schmiegelow}},
  \bibinfo {author} {\bibfnamefont {V.}~\bibnamefont {Kaushal}}, \bibinfo
  {author} {\bibfnamefont {J.}~\bibnamefont {Schulz}}, \bibinfo {author}
  {\bibfnamefont {M.~T.}\ \bibnamefont {Mitchison}}, \bibinfo {author}
  {\bibfnamefont {J.}~\bibnamefont {Goold}}, \bibinfo {author} {\bibfnamefont
  {F.}~\bibnamefont {Schmidt-Kaler}}, \ and\ \bibinfo {author} {\bibfnamefont
  {U.~G.}\ \bibnamefont {Poschinger}},\ }\href {\doibase
  10.1103/PhysRevLett.123.080602} {\bibfield  {journal} {\bibinfo  {journal}
  {Phys. Rev. Lett.}\ }\textbf {\bibinfo {volume} {123}},\ \bibinfo {pages}
  {080602} (\bibinfo {year} {2019})}\BibitemShut {NoStop}%
\bibitem [{\citenamefont {Zou}\ \emph {et~al.}(2017)\citenamefont {Zou},
  \citenamefont {Jiang}, \citenamefont {Mei}, \citenamefont {Guo},\ and\
  \citenamefont {Du}}]{Zou_2017}%
  \BibitemOpen
  \bibfield  {author} {\bibinfo {author} {\bibfnamefont {Y.}~\bibnamefont
  {Zou}}, \bibinfo {author} {\bibfnamefont {Y.}~\bibnamefont {Jiang}}, \bibinfo
  {author} {\bibfnamefont {Y.}~\bibnamefont {Mei}}, \bibinfo {author}
  {\bibfnamefont {X.}~\bibnamefont {Guo}}, \ and\ \bibinfo {author}
  {\bibfnamefont {S.}~\bibnamefont {Du}},\ }\href {\doibase
  10.1103/PhysRevLett.119.050602} {\bibfield  {journal} {\bibinfo  {journal}
  {Phys. Rev. Lett.}\ }\textbf {\bibinfo {volume} {119}},\ \bibinfo {pages}
  {050602} (\bibinfo {year} {2017})}\BibitemShut {NoStop}%
\bibitem [{\citenamefont {Brantut}\ \emph {et~al.}(2013)\citenamefont
  {Brantut}, \citenamefont {Grenier}, \citenamefont {Meineke}, \citenamefont
  {Stadler}, \citenamefont {Krinner}, \citenamefont {Kollath}, \citenamefont
  {Esslinger},\ and\ \citenamefont {Georges}}]{Brantut_2013}%
  \BibitemOpen
  \bibfield  {author} {\bibinfo {author} {\bibfnamefont {J.-P.}\ \bibnamefont
  {Brantut}}, \bibinfo {author} {\bibfnamefont {C.}~\bibnamefont {Grenier}},
  \bibinfo {author} {\bibfnamefont {J.}~\bibnamefont {Meineke}}, \bibinfo
  {author} {\bibfnamefont {D.}~\bibnamefont {Stadler}}, \bibinfo {author}
  {\bibfnamefont {S.}~\bibnamefont {Krinner}}, \bibinfo {author} {\bibfnamefont
  {C.}~\bibnamefont {Kollath}}, \bibinfo {author} {\bibfnamefont
  {T.}~\bibnamefont {Esslinger}}, \ and\ \bibinfo {author} {\bibfnamefont
  {A.}~\bibnamefont {Georges}},\ }\href {\doibase 10.1126/science.1242308}
  {\bibfield  {journal} {\bibinfo  {journal} {Science}\ }\textbf {\bibinfo
  {volume} {342}},\ \bibinfo {pages} {713} (\bibinfo {year}
  {2013})}\BibitemShut {NoStop}%
\bibitem [{\citenamefont {Klatzow}\ \emph {et~al.}(2019)\citenamefont
  {Klatzow}, \citenamefont {Becker}, \citenamefont {Ledingham}, \citenamefont
  {Weinzetl}, \citenamefont {Kaczmarek}, \citenamefont {Saunders},
  \citenamefont {Nunn}, \citenamefont {Walmsley}, \citenamefont {Uzdin},\ and\
  \citenamefont {Poem}}]{Klatzow_2019}%
  \BibitemOpen
  \bibfield  {author} {\bibinfo {author} {\bibfnamefont {J.}~\bibnamefont
  {Klatzow}}, \bibinfo {author} {\bibfnamefont {J.~N.}\ \bibnamefont {Becker}},
  \bibinfo {author} {\bibfnamefont {P.~M.}\ \bibnamefont {Ledingham}}, \bibinfo
  {author} {\bibfnamefont {C.}~\bibnamefont {Weinzetl}}, \bibinfo {author}
  {\bibfnamefont {K.~T.}\ \bibnamefont {Kaczmarek}}, \bibinfo {author}
  {\bibfnamefont {D.~J.}\ \bibnamefont {Saunders}}, \bibinfo {author}
  {\bibfnamefont {J.}~\bibnamefont {Nunn}}, \bibinfo {author} {\bibfnamefont
  {I.~A.}\ \bibnamefont {Walmsley}}, \bibinfo {author} {\bibfnamefont
  {R.}~\bibnamefont {Uzdin}}, \ and\ \bibinfo {author} {\bibfnamefont
  {E.}~\bibnamefont {Poem}},\ }\href {\doibase 10.1103/PhysRevLett.122.110601}
  {\bibfield  {journal} {\bibinfo  {journal} {Phys. Rev. Lett.}\ }\textbf
  {\bibinfo {volume} {122}},\ \bibinfo {pages} {110601} (\bibinfo {year}
  {2019})}\BibitemShut {NoStop}%
\bibitem [{\citenamefont {Guthrie}\ \emph {et~al.}(2022)\citenamefont
  {Guthrie}, \citenamefont {Satrya}, \citenamefont {Chang}, \citenamefont
  {Menczel}, \citenamefont {Nori},\ and\ \citenamefont
  {Pekola}}]{Guthrie_2022}%
  \BibitemOpen
  \bibfield  {author} {\bibinfo {author} {\bibfnamefont {A.}~\bibnamefont
  {Guthrie}}, \bibinfo {author} {\bibfnamefont {C.~D.}\ \bibnamefont {Satrya}},
  \bibinfo {author} {\bibfnamefont {Y.-C.}\ \bibnamefont {Chang}}, \bibinfo
  {author} {\bibfnamefont {P.}~\bibnamefont {Menczel}}, \bibinfo {author}
  {\bibfnamefont {F.}~\bibnamefont {Nori}}, \ and\ \bibinfo {author}
  {\bibfnamefont {J.~P.}\ \bibnamefont {Pekola}},\ }\href {\doibase
  10.1103/PhysRevApplied.17.064022} {\bibfield  {journal} {\bibinfo  {journal}
  {Phys. Rev. Applied}\ }\textbf {\bibinfo {volume} {17}},\ \bibinfo {pages}
  {064022} (\bibinfo {year} {2022})}\BibitemShut {NoStop}%
\bibitem [{\citenamefont {Ronzani}\ \emph {et~al.}(2018)\citenamefont
  {Ronzani}, \citenamefont {Karimi}, \citenamefont {Senior}, \citenamefont
  {Chang}, \citenamefont {Peltonen}, \citenamefont {Chen},\ and\ \citenamefont
  {Pekola}}]{Ronzani_2018}%
  \BibitemOpen
  \bibfield  {author} {\bibinfo {author} {\bibfnamefont {A.}~\bibnamefont
  {Ronzani}}, \bibinfo {author} {\bibfnamefont {B.}~\bibnamefont {Karimi}},
  \bibinfo {author} {\bibfnamefont {J.}~\bibnamefont {Senior}}, \bibinfo
  {author} {\bibfnamefont {Y.-C.}\ \bibnamefont {Chang}}, \bibinfo {author}
  {\bibfnamefont {J.~T.}\ \bibnamefont {Peltonen}}, \bibinfo {author}
  {\bibfnamefont {C.}~\bibnamefont {Chen}}, \ and\ \bibinfo {author}
  {\bibfnamefont {J.~P.}\ \bibnamefont {Pekola}},\ }\href {\doibase
  10.1038/s41567-018-0199-4} {\bibfield  {journal} {\bibinfo  {journal} {Nature
  Physics}\ }\textbf {\bibinfo {volume} {14}},\ \bibinfo {pages} {991}
  (\bibinfo {year} {2018})}\BibitemShut {NoStop}%
\bibitem [{\citenamefont {Pekola}\ and\ \citenamefont
  {Khaymovich}(2019)}]{Pekola_2019}%
  \BibitemOpen
  \bibfield  {author} {\bibinfo {author} {\bibfnamefont {J.}~\bibnamefont
  {Pekola}}\ and\ \bibinfo {author} {\bibfnamefont {I.}~\bibnamefont
  {Khaymovich}},\ }\href {\doibase 10.1146/annurev-conmatphys-033117-054120}
  {\bibfield  {journal} {\bibinfo  {journal} {Annual Review of Condensed Matter
  Physics}\ }\textbf {\bibinfo {volume} {10}},\ \bibinfo {pages} {193}
  (\bibinfo {year} {2019})},\ \Eprint
  {http://arxiv.org/abs/https://doi.org/10.1146/annurev-conmatphys-033117-054120}
  {https://doi.org/10.1146/annurev-conmatphys-033117-054120} \BibitemShut
  {NoStop}%
\bibitem [{\citenamefont {Karimi}\ and\ \citenamefont
  {Pekola}(2016)}]{Karimi_2016}%
  \BibitemOpen
  \bibfield  {author} {\bibinfo {author} {\bibfnamefont {B.}~\bibnamefont
  {Karimi}}\ and\ \bibinfo {author} {\bibfnamefont {J.~P.}\ \bibnamefont
  {Pekola}},\ }\href {\doibase 10.1103/PhysRevB.94.184503} {\bibfield
  {journal} {\bibinfo  {journal} {Phys. Rev. B}\ }\textbf {\bibinfo {volume}
  {94}},\ \bibinfo {pages} {184503} (\bibinfo {year} {2016})}\BibitemShut
  {NoStop}%
\bibitem [{\citenamefont {Maslennikov}\ \emph {et~al.}(2019)\citenamefont
  {Maslennikov}, \citenamefont {Ding}, \citenamefont {Habl{\"u}tzel},
  \citenamefont {Gan}, \citenamefont {Roulet}, \citenamefont {Nimmrichter},
  \citenamefont {Dai}, \citenamefont {Scarani},\ and\ \citenamefont
  {Matsukevich}}]{Maslennikov_2019}%
  \BibitemOpen
  \bibfield  {author} {\bibinfo {author} {\bibfnamefont {G.}~\bibnamefont
  {Maslennikov}}, \bibinfo {author} {\bibfnamefont {S.}~\bibnamefont {Ding}},
  \bibinfo {author} {\bibfnamefont {R.}~\bibnamefont {Habl{\"u}tzel}}, \bibinfo
  {author} {\bibfnamefont {J.}~\bibnamefont {Gan}}, \bibinfo {author}
  {\bibfnamefont {A.}~\bibnamefont {Roulet}}, \bibinfo {author} {\bibfnamefont
  {S.}~\bibnamefont {Nimmrichter}}, \bibinfo {author} {\bibfnamefont
  {J.}~\bibnamefont {Dai}}, \bibinfo {author} {\bibfnamefont {V.}~\bibnamefont
  {Scarani}}, \ and\ \bibinfo {author} {\bibfnamefont {D.}~\bibnamefont
  {Matsukevich}},\ }\href {\doibase 10.1038/s41467-018-08090-0} {\bibfield
  {journal} {\bibinfo  {journal} {Nature Communications}\ }\textbf {\bibinfo
  {volume} {10}},\ \bibinfo {pages} {202} (\bibinfo {year} {2019})}\BibitemShut
  {NoStop}%
\bibitem [{\citenamefont {Niskanen}\ \emph {et~al.}(2007)\citenamefont
  {Niskanen}, \citenamefont {Nakamura},\ and\ \citenamefont
  {Pekola}}]{Niskanen_2007}%
  \BibitemOpen
  \bibfield  {author} {\bibinfo {author} {\bibfnamefont {A.~O.}\ \bibnamefont
  {Niskanen}}, \bibinfo {author} {\bibfnamefont {Y.}~\bibnamefont {Nakamura}},
  \ and\ \bibinfo {author} {\bibfnamefont {J.~P.}\ \bibnamefont {Pekola}},\
  }\href {\doibase 10.1103/PhysRevB.76.174523} {\bibfield  {journal} {\bibinfo
  {journal} {Phys. Rev. B}\ }\textbf {\bibinfo {volume} {76}},\ \bibinfo
  {pages} {174523} (\bibinfo {year} {2007})}\BibitemShut {NoStop}%
\bibitem [{\citenamefont {Roßnagel}\ \emph {et~al.}(2016)\citenamefont
  {Roßnagel}, \citenamefont {Dawkins}, \citenamefont {Tolazzi}, \citenamefont
  {Abah}, \citenamefont {Lutz}, \citenamefont {Schmidt-Kaler},\ and\
  \citenamefont {Singer}}]{Robnagel_2016}%
  \BibitemOpen
  \bibfield  {author} {\bibinfo {author} {\bibfnamefont {J.}~\bibnamefont
  {Roßnagel}}, \bibinfo {author} {\bibfnamefont {S.~T.}\ \bibnamefont
  {Dawkins}}, \bibinfo {author} {\bibfnamefont {K.~N.}\ \bibnamefont
  {Tolazzi}}, \bibinfo {author} {\bibfnamefont {O.}~\bibnamefont {Abah}},
  \bibinfo {author} {\bibfnamefont {E.}~\bibnamefont {Lutz}}, \bibinfo {author}
  {\bibfnamefont {F.}~\bibnamefont {Schmidt-Kaler}}, \ and\ \bibinfo {author}
  {\bibfnamefont {K.}~\bibnamefont {Singer}},\ }\href {\doibase
  10.1126/science.aad6320} {\bibfield  {journal} {\bibinfo  {journal}
  {Science}\ }\textbf {\bibinfo {volume} {352}},\ \bibinfo {pages} {325}
  (\bibinfo {year} {2016})}\BibitemShut {NoStop}%
\bibitem [{\citenamefont {Van~Horne}\ \emph {et~al.}(2020)\citenamefont
  {Van~Horne}, \citenamefont {Yum}, \citenamefont {Dutta}, \citenamefont
  {H{\"a}nggi}, \citenamefont {Gong}, \citenamefont {Poletti},\ and\
  \citenamefont {Mukherjee}}]{Horne_2020}%
  \BibitemOpen
  \bibfield  {author} {\bibinfo {author} {\bibfnamefont {N.}~\bibnamefont
  {Van~Horne}}, \bibinfo {author} {\bibfnamefont {D.}~\bibnamefont {Yum}},
  \bibinfo {author} {\bibfnamefont {T.}~\bibnamefont {Dutta}}, \bibinfo
  {author} {\bibfnamefont {P.}~\bibnamefont {H{\"a}nggi}}, \bibinfo {author}
  {\bibfnamefont {J.}~\bibnamefont {Gong}}, \bibinfo {author} {\bibfnamefont
  {D.}~\bibnamefont {Poletti}}, \ and\ \bibinfo {author} {\bibfnamefont
  {M.}~\bibnamefont {Mukherjee}},\ }\href {\doibase 10.1038/s41534-020-0264-6}
  {\bibfield  {journal} {\bibinfo  {journal} {npj Quantum Information}\
  }\textbf {\bibinfo {volume} {6}},\ \bibinfo {pages} {37} (\bibinfo {year}
  {2020})}\BibitemShut {NoStop}%
\bibitem [{\citenamefont {Zhang}\ \emph {et~al.}(2014)\citenamefont {Zhang},
  \citenamefont {Bariani},\ and\ \citenamefont {Meystre}}]{Zhang_2014}%
  \BibitemOpen
  \bibfield  {author} {\bibinfo {author} {\bibfnamefont {K.}~\bibnamefont
  {Zhang}}, \bibinfo {author} {\bibfnamefont {F.}~\bibnamefont {Bariani}}, \
  and\ \bibinfo {author} {\bibfnamefont {P.}~\bibnamefont {Meystre}},\ }\href
  {\doibase 10.1103/PhysRevLett.112.150602} {\bibfield  {journal} {\bibinfo
  {journal} {Phys. Rev. Lett.}\ }\textbf {\bibinfo {volume} {112}},\ \bibinfo
  {pages} {150602} (\bibinfo {year} {2014})}\BibitemShut {NoStop}%
\bibitem [{\citenamefont {Solfanelli}\ \emph {et~al.}(2022)\citenamefont
  {Solfanelli}, \citenamefont {Santini},\ and\ \citenamefont
  {Campisi}}]{Solfanelli_2022}%
  \BibitemOpen
  \bibfield  {author} {\bibinfo {author} {\bibfnamefont {A.}~\bibnamefont
  {Solfanelli}}, \bibinfo {author} {\bibfnamefont {A.}~\bibnamefont {Santini}},
  \ and\ \bibinfo {author} {\bibfnamefont {M.}~\bibnamefont {Campisi}},\ }\href
  {\doibase 10.1116/5.0091121} {\bibfield  {journal} {\bibinfo  {journal} {AVS
  Quantum Science}\ }\textbf {\bibinfo {volume} {4}},\ \bibinfo {pages}
  {026802} (\bibinfo {year} {2022})},\ \Eprint
  {http://arxiv.org/abs/https://doi.org/10.1116/5.0091121}
  {https://doi.org/10.1116/5.0091121} \BibitemShut {NoStop}%
\bibitem [{\citenamefont {Solfanelli}\ \emph {et~al.}(2021)\citenamefont
  {Solfanelli}, \citenamefont {Santini},\ and\ \citenamefont
  {Campisi}}]{Solfanelli_2021}%
  \BibitemOpen
  \bibfield  {author} {\bibinfo {author} {\bibfnamefont {A.}~\bibnamefont
  {Solfanelli}}, \bibinfo {author} {\bibfnamefont {A.}~\bibnamefont {Santini}},
  \ and\ \bibinfo {author} {\bibfnamefont {M.}~\bibnamefont {Campisi}},\ }\href
  {\doibase 10.1103/PRXQuantum.2.030353} {\bibfield  {journal} {\bibinfo
  {journal} {PRX Quantum}\ }\textbf {\bibinfo {volume} {2}},\ \bibinfo {pages}
  {030353} (\bibinfo {year} {2021})}\BibitemShut {NoStop}%
\bibitem [{\citenamefont {Esposito}\ \emph
  {et~al.}(2009{\natexlab{a}})\citenamefont {Esposito}, \citenamefont
  {Harbola},\ and\ \citenamefont {Mukamel}}]{Esposito_2009}%
  \BibitemOpen
  \bibfield  {author} {\bibinfo {author} {\bibfnamefont {M.}~\bibnamefont
  {Esposito}}, \bibinfo {author} {\bibfnamefont {U.}~\bibnamefont {Harbola}}, \
  and\ \bibinfo {author} {\bibfnamefont {S.}~\bibnamefont {Mukamel}},\ }\href
  {\doibase 10.1103/RevModPhys.81.1665} {\bibfield  {journal} {\bibinfo
  {journal} {Rev. Mod. Phys.}\ }\textbf {\bibinfo {volume} {81}},\ \bibinfo
  {pages} {1665} (\bibinfo {year} {2009}{\natexlab{a}})}\BibitemShut {NoStop}%
\bibitem [{\citenamefont {Campisi}\ \emph {et~al.}(2011)\citenamefont
  {Campisi}, \citenamefont {H\"anggi},\ and\ \citenamefont
  {Talkner}}]{Campisi_2011}%
  \BibitemOpen
  \bibfield  {author} {\bibinfo {author} {\bibfnamefont {M.}~\bibnamefont
  {Campisi}}, \bibinfo {author} {\bibfnamefont {P.}~\bibnamefont {H\"anggi}}, \
  and\ \bibinfo {author} {\bibfnamefont {P.}~\bibnamefont {Talkner}},\ }\href
  {\doibase 10.1103/RevModPhys.83.771} {\bibfield  {journal} {\bibinfo
  {journal} {Rev. Mod. Phys.}\ }\textbf {\bibinfo {volume} {83}},\ \bibinfo
  {pages} {771} (\bibinfo {year} {2011})}\BibitemShut {NoStop}%
\bibitem [{\citenamefont {Verley}\ \emph {et~al.}(2014)\citenamefont {Verley},
  \citenamefont {Esposito}, \citenamefont {Willaert},\ and\ \citenamefont
  {Van~den Broeck}}]{Verley_2014}%
  \BibitemOpen
  \bibfield  {author} {\bibinfo {author} {\bibfnamefont {G.}~\bibnamefont
  {Verley}}, \bibinfo {author} {\bibfnamefont {M.}~\bibnamefont {Esposito}},
  \bibinfo {author} {\bibfnamefont {T.}~\bibnamefont {Willaert}}, \ and\
  \bibinfo {author} {\bibfnamefont {C.}~\bibnamefont {Van~den Broeck}},\ }\href
  {\doibase 10.1038/ncomms5721} {\bibfield  {journal} {\bibinfo  {journal}
  {Nature Communications}\ }\textbf {\bibinfo {volume} {5}},\ \bibinfo {pages}
  {4721} (\bibinfo {year} {2014})}\BibitemShut {NoStop}%
\bibitem [{\citenamefont {Horowitz}(2012)}]{Horowitz_2012}%
  \BibitemOpen
  \bibfield  {author} {\bibinfo {author} {\bibfnamefont {J.~M.}\ \bibnamefont
  {Horowitz}},\ }\href {\doibase 10.1103/PhysRevE.85.031110} {\bibfield
  {journal} {\bibinfo  {journal} {Phys. Rev. E}\ }\textbf {\bibinfo {volume}
  {85}},\ \bibinfo {pages} {031110} (\bibinfo {year} {2012})}\BibitemShut
  {NoStop}%
\bibitem [{\citenamefont {Seifert}(2005)}]{Seifert_2005}%
  \BibitemOpen
  \bibfield  {author} {\bibinfo {author} {\bibfnamefont {U.}~\bibnamefont
  {Seifert}},\ }\href {\doibase 10.1103/PhysRevLett.95.040602} {\bibfield
  {journal} {\bibinfo  {journal} {Phys. Rev. Lett.}\ }\textbf {\bibinfo
  {volume} {95}},\ \bibinfo {pages} {040602} (\bibinfo {year}
  {2005})}\BibitemShut {NoStop}%
\bibitem [{\citenamefont {Seifert}(2008)}]{Seifert_2008}%
  \BibitemOpen
  \bibfield  {author} {\bibinfo {author} {\bibfnamefont {U.}~\bibnamefont
  {Seifert}},\ }\href {\doibase 10.1140/epjb/e2008-00001-9} {\bibfield
  {journal} {\bibinfo  {journal} {The European Physical Journal B}\ }\textbf
  {\bibinfo {volume} {64}},\ \bibinfo {pages} {423} (\bibinfo {year}
  {2008})}\BibitemShut {NoStop}%
\bibitem [{\citenamefont {Dalibard}\ \emph {et~al.}(1992)\citenamefont
  {Dalibard}, \citenamefont {Castin},\ and\ \citenamefont
  {M\o{}lmer}}]{Dalibard_1992}%
  \BibitemOpen
  \bibfield  {author} {\bibinfo {author} {\bibfnamefont {J.}~\bibnamefont
  {Dalibard}}, \bibinfo {author} {\bibfnamefont {Y.}~\bibnamefont {Castin}}, \
  and\ \bibinfo {author} {\bibfnamefont {K.}~\bibnamefont {M\o{}lmer}},\ }\href
  {\doibase 10.1103/PhysRevLett.68.580} {\bibfield  {journal} {\bibinfo
  {journal} {Phys. Rev. Lett.}\ }\textbf {\bibinfo {volume} {68}},\ \bibinfo
  {pages} {580} (\bibinfo {year} {1992})}\BibitemShut {NoStop}%
\bibitem [{\citenamefont {M{\o}lmer}\ and\ \citenamefont
  {Castin}(1996)}]{Molmer_1996}%
  \BibitemOpen
  \bibfield  {author} {\bibinfo {author} {\bibfnamefont {K.}~\bibnamefont
  {M{\o}lmer}}\ and\ \bibinfo {author} {\bibfnamefont {Y.}~\bibnamefont
  {Castin}},\ }\href {\doibase 10.1088/1355-5111/8/1/007} {\bibfield  {journal}
  {\bibinfo  {journal} {Quantum and Semiclassical Optics: Journal of the
  European Optical Society Part B}\ }\textbf {\bibinfo {volume} {8}},\ \bibinfo
  {pages} {49} (\bibinfo {year} {1996})}\BibitemShut {NoStop}%
\bibitem [{\citenamefont {Brun}(2002)}]{Brun_2002}%
  \BibitemOpen
  \bibfield  {author} {\bibinfo {author} {\bibfnamefont {T.~A.}\ \bibnamefont
  {Brun}},\ }\href {\doibase 10.1119/1.1475328} {\bibfield  {journal} {\bibinfo
   {journal} {American Journal of Physics}\ }\textbf {\bibinfo {volume} {70}},\
  \bibinfo {pages} {719} (\bibinfo {year} {2002})},\ \Eprint
  {http://arxiv.org/abs/https://doi.org/10.1119/1.1475328}
  {https://doi.org/10.1119/1.1475328} \BibitemShut {NoStop}%
\bibitem [{\citenamefont {Gardiner}\ \emph {et~al.}(1992)\citenamefont
  {Gardiner}, \citenamefont {Parkins},\ and\ \citenamefont
  {Zoller}}]{Gardiner_1992}%
  \BibitemOpen
  \bibfield  {author} {\bibinfo {author} {\bibfnamefont {C.~W.}\ \bibnamefont
  {Gardiner}}, \bibinfo {author} {\bibfnamefont {A.~S.}\ \bibnamefont
  {Parkins}}, \ and\ \bibinfo {author} {\bibfnamefont {P.}~\bibnamefont
  {Zoller}},\ }\href {\doibase 10.1103/PhysRevA.46.4363} {\bibfield  {journal}
  {\bibinfo  {journal} {Phys. Rev. A}\ }\textbf {\bibinfo {volume} {46}},\
  \bibinfo {pages} {4363} (\bibinfo {year} {1992})}\BibitemShut {NoStop}%
\bibitem [{\citenamefont {Manzano}\ and\ \citenamefont
  {Zambrini}(2022)}]{Manzano_2022}%
  \BibitemOpen
  \bibfield  {author} {\bibinfo {author} {\bibfnamefont {G.}~\bibnamefont
  {Manzano}}\ and\ \bibinfo {author} {\bibfnamefont {R.}~\bibnamefont
  {Zambrini}},\ }\href {\doibase 10.1116/5.0079886} {\bibfield  {journal}
  {\bibinfo  {journal} {AVS Quantum Science}\ }\textbf {\bibinfo {volume}
  {4}},\ \bibinfo {pages} {025302} (\bibinfo {year} {2022})},\ \Eprint
  {http://arxiv.org/abs/https://doi.org/10.1116/5.0079886}
  {https://doi.org/10.1116/5.0079886} \BibitemShut {NoStop}%
\bibitem [{\citenamefont {Minev}\ \emph {et~al.}(2019)\citenamefont {Minev},
  \citenamefont {Mundhada}, \citenamefont {Shankar}, \citenamefont {Reinhold},
  \citenamefont {Guti{\'e}rrez-J{\'a}uregui}, \citenamefont {Schoelkopf},
  \citenamefont {Mirrahimi}, \citenamefont {Carmichael},\ and\ \citenamefont
  {Devoret}}]{Minev_2019}%
  \BibitemOpen
  \bibfield  {author} {\bibinfo {author} {\bibfnamefont {Z.~K.}\ \bibnamefont
  {Minev}}, \bibinfo {author} {\bibfnamefont {S.~O.}\ \bibnamefont {Mundhada}},
  \bibinfo {author} {\bibfnamefont {S.}~\bibnamefont {Shankar}}, \bibinfo
  {author} {\bibfnamefont {P.}~\bibnamefont {Reinhold}}, \bibinfo {author}
  {\bibfnamefont {R.}~\bibnamefont {Guti{\'e}rrez-J{\'a}uregui}}, \bibinfo
  {author} {\bibfnamefont {R.~J.}\ \bibnamefont {Schoelkopf}}, \bibinfo
  {author} {\bibfnamefont {M.}~\bibnamefont {Mirrahimi}}, \bibinfo {author}
  {\bibfnamefont {H.~J.}\ \bibnamefont {Carmichael}}, \ and\ \bibinfo {author}
  {\bibfnamefont {M.~H.}\ \bibnamefont {Devoret}},\ }\href {\doibase
  10.1038/s41586-019-1287-z} {\bibfield  {journal} {\bibinfo  {journal}
  {Nature}\ }\textbf {\bibinfo {volume} {570}},\ \bibinfo {pages} {200}
  (\bibinfo {year} {2019})}\BibitemShut {NoStop}%
\bibitem [{\citenamefont {Murch}\ \emph {et~al.}(2013)\citenamefont {Murch},
  \citenamefont {Weber}, \citenamefont {Macklin},\ and\ \citenamefont
  {Siddiqi}}]{Murch_2013}%
  \BibitemOpen
  \bibfield  {author} {\bibinfo {author} {\bibfnamefont {K.~W.}\ \bibnamefont
  {Murch}}, \bibinfo {author} {\bibfnamefont {S.~J.}\ \bibnamefont {Weber}},
  \bibinfo {author} {\bibfnamefont {C.}~\bibnamefont {Macklin}}, \ and\
  \bibinfo {author} {\bibfnamefont {I.}~\bibnamefont {Siddiqi}},\ }\href
  {\doibase 10.1038/nature12539} {\bibfield  {journal} {\bibinfo  {journal}
  {Nature}\ }\textbf {\bibinfo {volume} {502}},\ \bibinfo {pages} {211}
  (\bibinfo {year} {2013})}\BibitemShut {NoStop}%
\bibitem [{\citenamefont {Vijay}\ \emph {et~al.}(2011)\citenamefont {Vijay},
  \citenamefont {Slichter},\ and\ \citenamefont {Siddiqi}}]{Vijay_2011}%
  \BibitemOpen
  \bibfield  {author} {\bibinfo {author} {\bibfnamefont {R.}~\bibnamefont
  {Vijay}}, \bibinfo {author} {\bibfnamefont {D.~H.}\ \bibnamefont {Slichter}},
  \ and\ \bibinfo {author} {\bibfnamefont {I.}~\bibnamefont {Siddiqi}},\ }\href
  {\doibase 10.1103/PhysRevLett.106.110502} {\bibfield  {journal} {\bibinfo
  {journal} {Phys. Rev. Lett.}\ }\textbf {\bibinfo {volume} {106}},\ \bibinfo
  {pages} {110502} (\bibinfo {year} {2011})}\BibitemShut {NoStop}%
\bibitem [{\citenamefont {Karimi}\ and\ \citenamefont
  {Pekola}(2020)}]{Karimi_2020}%
  \BibitemOpen
  \bibfield  {author} {\bibinfo {author} {\bibfnamefont {B.}~\bibnamefont
  {Karimi}}\ and\ \bibinfo {author} {\bibfnamefont {J.~P.}\ \bibnamefont
  {Pekola}},\ }\href {\doibase 10.1103/PhysRevLett.124.170601} {\bibfield
  {journal} {\bibinfo  {journal} {Phys. Rev. Lett.}\ }\textbf {\bibinfo
  {volume} {124}},\ \bibinfo {pages} {170601} (\bibinfo {year}
  {2020})}\BibitemShut {NoStop}%
\bibitem [{\citenamefont {Friedman}\ \emph {et~al.}(2018)\citenamefont
  {Friedman}, \citenamefont {Agarwalla},\ and\ \citenamefont
  {Segal}}]{Friedman_2018}%
  \BibitemOpen
  \bibfield  {author} {\bibinfo {author} {\bibfnamefont {H.~M.}\ \bibnamefont
  {Friedman}}, \bibinfo {author} {\bibfnamefont {B.~K.}\ \bibnamefont
  {Agarwalla}}, \ and\ \bibinfo {author} {\bibfnamefont {D.}~\bibnamefont
  {Segal}},\ }\href {\doibase 10.1088/1367-2630/aad5fc} {\bibfield  {journal}
  {\bibinfo  {journal} {New Journal of Physics}\ }\textbf {\bibinfo {volume}
  {20}},\ \bibinfo {pages} {083026} (\bibinfo {year} {2018})}\BibitemShut
  {NoStop}%
\bibitem [{\citenamefont {Campisi}\ \emph {et~al.}(2015)\citenamefont
  {Campisi}, \citenamefont {Pekola},\ and\ \citenamefont
  {Fazio}}]{Campisi_2015}%
  \BibitemOpen
  \bibfield  {author} {\bibinfo {author} {\bibfnamefont {M.}~\bibnamefont
  {Campisi}}, \bibinfo {author} {\bibfnamefont {J.}~\bibnamefont {Pekola}}, \
  and\ \bibinfo {author} {\bibfnamefont {R.}~\bibnamefont {Fazio}},\ }\href
  {\doibase 10.1088/1367-2630/17/3/035012} {\bibfield  {journal} {\bibinfo
  {journal} {New Journal of Physics}\ }\textbf {\bibinfo {volume} {17}},\
  \bibinfo {pages} {035012} (\bibinfo {year} {2015})}\BibitemShut {NoStop}%
\bibitem [{\citenamefont {Mart{\'\i}nez}\ \emph {et~al.}(2016)\citenamefont
  {Mart{\'\i}nez}, \citenamefont {Rold{\'a}n}, \citenamefont {Dinis},
  \citenamefont {Petrov}, \citenamefont {Parrondo},\ and\ \citenamefont
  {Rica}}]{Martinez_2016}%
  \BibitemOpen
  \bibfield  {author} {\bibinfo {author} {\bibfnamefont {I.~A.}\ \bibnamefont
  {Mart{\'\i}nez}}, \bibinfo {author} {\bibfnamefont {{\'E}.}~\bibnamefont
  {Rold{\'a}n}}, \bibinfo {author} {\bibfnamefont {L.}~\bibnamefont {Dinis}},
  \bibinfo {author} {\bibfnamefont {D.}~\bibnamefont {Petrov}}, \bibinfo
  {author} {\bibfnamefont {J.~M.~R.}\ \bibnamefont {Parrondo}}, \ and\ \bibinfo
  {author} {\bibfnamefont {R.~A.}\ \bibnamefont {Rica}},\ }\href {\doibase
  10.1038/nphys3518} {\bibfield  {journal} {\bibinfo  {journal} {Nature
  Physics}\ }\textbf {\bibinfo {volume} {12}},\ \bibinfo {pages} {67} (\bibinfo
  {year} {2016})}\BibitemShut {NoStop}%
\bibitem [{\citenamefont {Leggio}\ \emph {et~al.}(2013)\citenamefont {Leggio},
  \citenamefont {Napoli}, \citenamefont {Messina},\ and\ \citenamefont
  {Breuer}}]{Leggio_2013}%
  \BibitemOpen
  \bibfield  {author} {\bibinfo {author} {\bibfnamefont {B.}~\bibnamefont
  {Leggio}}, \bibinfo {author} {\bibfnamefont {A.}~\bibnamefont {Napoli}},
  \bibinfo {author} {\bibfnamefont {A.}~\bibnamefont {Messina}}, \ and\
  \bibinfo {author} {\bibfnamefont {H.-P.}\ \bibnamefont {Breuer}},\ }\href
  {\doibase 10.1103/PhysRevA.88.042111} {\bibfield  {journal} {\bibinfo
  {journal} {Phys. Rev. A}\ }\textbf {\bibinfo {volume} {88}},\ \bibinfo
  {pages} {042111} (\bibinfo {year} {2013})}\BibitemShut {NoStop}%
\bibitem [{\citenamefont {Esposito}\ \emph
  {et~al.}(2009{\natexlab{b}})\citenamefont {Esposito}, \citenamefont
  {Harbola},\ and\ \citenamefont {Mukamel}}]{Eposito_2009}%
  \BibitemOpen
  \bibfield  {author} {\bibinfo {author} {\bibfnamefont {M.}~\bibnamefont
  {Esposito}}, \bibinfo {author} {\bibfnamefont {U.}~\bibnamefont {Harbola}}, \
  and\ \bibinfo {author} {\bibfnamefont {S.}~\bibnamefont {Mukamel}},\ }\href
  {\doibase 10.1103/RevModPhys.81.1665} {\bibfield  {journal} {\bibinfo
  {journal} {Rev. Mod. Phys.}\ }\textbf {\bibinfo {volume} {81}},\ \bibinfo
  {pages} {1665} (\bibinfo {year} {2009}{\natexlab{b}})}\BibitemShut {NoStop}%
\bibitem [{\citenamefont {Gupta}\ \emph {et~al.}(2020)\citenamefont {Gupta},
  \citenamefont {Plata},\ and\ \citenamefont {Pal}}]{Gupta_2020}%
  \BibitemOpen
  \bibfield  {author} {\bibinfo {author} {\bibfnamefont {D.}~\bibnamefont
  {Gupta}}, \bibinfo {author} {\bibfnamefont {C.~A.}\ \bibnamefont {Plata}}, \
  and\ \bibinfo {author} {\bibfnamefont {A.}~\bibnamefont {Pal}},\ }\href
  {\doibase 10.1103/PhysRevLett.124.110608} {\bibfield  {journal} {\bibinfo
  {journal} {Phys. Rev. Lett.}\ }\textbf {\bibinfo {volume} {124}},\ \bibinfo
  {pages} {110608} (\bibinfo {year} {2020})}\BibitemShut {NoStop}%
\bibitem [{\citenamefont {Manzano}\ \emph {et~al.}(2018)\citenamefont
  {Manzano}, \citenamefont {Horowitz},\ and\ \citenamefont
  {Parrondo}}]{Manzano_2018}%
  \BibitemOpen
  \bibfield  {author} {\bibinfo {author} {\bibfnamefont {G.}~\bibnamefont
  {Manzano}}, \bibinfo {author} {\bibfnamefont {J.~M.}\ \bibnamefont
  {Horowitz}}, \ and\ \bibinfo {author} {\bibfnamefont {J.~M.~R.}\ \bibnamefont
  {Parrondo}},\ }\href {\doibase 10.1103/PhysRevX.8.031037} {\bibfield
  {journal} {\bibinfo  {journal} {Phys. Rev. X}\ }\textbf {\bibinfo {volume}
  {8}},\ \bibinfo {pages} {031037} (\bibinfo {year} {2018})}\BibitemShut
  {NoStop}%
\bibitem [{\citenamefont {Jarzynski}(1997)}]{Jarzynski_1997}%
  \BibitemOpen
  \bibfield  {author} {\bibinfo {author} {\bibfnamefont {C.}~\bibnamefont
  {Jarzynski}},\ }\href {\doibase 10.1103/PhysRevLett.78.2690} {\bibfield
  {journal} {\bibinfo  {journal} {Phys. Rev. Lett.}\ }\textbf {\bibinfo
  {volume} {78}},\ \bibinfo {pages} {2690} (\bibinfo {year}
  {1997})}\BibitemShut {NoStop}%
\bibitem [{\citenamefont {Yukawa}(2000)}]{Yukawa_2000}%
  \BibitemOpen
  \bibfield  {author} {\bibinfo {author} {\bibfnamefont {S.}~\bibnamefont
  {Yukawa}},\ }\href {\doibase 10.1143/JPSJ.69.2367} {\bibfield  {journal}
  {\bibinfo  {journal} {Journal of the Physical Society of Japan}\ }\textbf
  {\bibinfo {volume} {69}},\ \bibinfo {pages} {2367} (\bibinfo {year}
  {2000})},\ \Eprint
  {http://arxiv.org/abs/https://doi.org/10.1143/JPSJ.69.2367}
  {https://doi.org/10.1143/JPSJ.69.2367} \BibitemShut {NoStop}%
\bibitem [{\citenamefont {Buffoni}\ and\ \citenamefont
  {Campisi}(2022)}]{Buffoni_2022}%
  \BibitemOpen
  \bibfield  {author} {\bibinfo {author} {\bibfnamefont {L.}~\bibnamefont
  {Buffoni}}\ and\ \bibinfo {author} {\bibfnamefont {M.}~\bibnamefont
  {Campisi}},\ }\href {\doibase 10.1007/s10955-022-02877-8} {\bibfield
  {journal} {\bibinfo  {journal} {Journal of Statistical Physics}\ }\textbf
  {\bibinfo {volume} {186}},\ \bibinfo {pages} {31} (\bibinfo {year}
  {2022})}\BibitemShut {NoStop}%
\bibitem [{\citenamefont {Barato}\ and\ \citenamefont
  {Seifert}(2015)}]{Barato_2015}%
  \BibitemOpen
  \bibfield  {author} {\bibinfo {author} {\bibfnamefont {A.~C.}\ \bibnamefont
  {Barato}}\ and\ \bibinfo {author} {\bibfnamefont {U.}~\bibnamefont
  {Seifert}},\ }\href {\doibase 10.1103/PhysRevLett.114.158101} {\bibfield
  {journal} {\bibinfo  {journal} {Phys. Rev. Lett.}\ }\textbf {\bibinfo
  {volume} {114}},\ \bibinfo {pages} {158101} (\bibinfo {year}
  {2015})}\BibitemShut {NoStop}%
\bibitem [{\citenamefont {Horowitz}\ and\ \citenamefont
  {Gingrich}(2020)}]{Horowitz_2020}%
  \BibitemOpen
  \bibfield  {author} {\bibinfo {author} {\bibfnamefont {J.~M.}\ \bibnamefont
  {Horowitz}}\ and\ \bibinfo {author} {\bibfnamefont {T.~R.}\ \bibnamefont
  {Gingrich}},\ }\href {\doibase 10.1038/s41567-019-0702-6} {\bibfield
  {journal} {\bibinfo  {journal} {Nature Physics}\ }\textbf {\bibinfo {volume}
  {16}},\ \bibinfo {pages} {15} (\bibinfo {year} {2020})}\BibitemShut {NoStop}%
\bibitem [{\citenamefont {Pietzonka}\ \emph {et~al.}(2016)\citenamefont
  {Pietzonka}, \citenamefont {Barato},\ and\ \citenamefont
  {Seifert}}]{Pietzonka_2016}%
  \BibitemOpen
  \bibfield  {author} {\bibinfo {author} {\bibfnamefont {P.}~\bibnamefont
  {Pietzonka}}, \bibinfo {author} {\bibfnamefont {A.~C.}\ \bibnamefont
  {Barato}}, \ and\ \bibinfo {author} {\bibfnamefont {U.}~\bibnamefont
  {Seifert}},\ }\href {\doibase 10.1103/PhysRevE.93.052145} {\bibfield
  {journal} {\bibinfo  {journal} {Phys. Rev. E}\ }\textbf {\bibinfo {volume}
  {93}},\ \bibinfo {pages} {052145} (\bibinfo {year} {2016})}\BibitemShut
  {NoStop}%
\bibitem [{\citenamefont {Pietzonka}\ \emph {et~al.}(2017)\citenamefont
  {Pietzonka}, \citenamefont {Ritort},\ and\ \citenamefont
  {Seifert}}]{Pietzonka_2017}%
  \BibitemOpen
  \bibfield  {author} {\bibinfo {author} {\bibfnamefont {P.}~\bibnamefont
  {Pietzonka}}, \bibinfo {author} {\bibfnamefont {F.}~\bibnamefont {Ritort}}, \
  and\ \bibinfo {author} {\bibfnamefont {U.}~\bibnamefont {Seifert}},\ }\href
  {\doibase 10.1103/PhysRevE.96.012101} {\bibfield  {journal} {\bibinfo
  {journal} {Phys. Rev. E}\ }\textbf {\bibinfo {volume} {96}},\ \bibinfo
  {pages} {012101} (\bibinfo {year} {2017})}\BibitemShut {NoStop}%
\bibitem [{\citenamefont {Pietzonka}\ and\ \citenamefont
  {Seifert}(2018)}]{Pietzonka_2018}%
  \BibitemOpen
  \bibfield  {author} {\bibinfo {author} {\bibfnamefont {P.}~\bibnamefont
  {Pietzonka}}\ and\ \bibinfo {author} {\bibfnamefont {U.}~\bibnamefont
  {Seifert}},\ }\href {\doibase 10.1103/PhysRevLett.120.190602} {\bibfield
  {journal} {\bibinfo  {journal} {Phys. Rev. Lett.}\ }\textbf {\bibinfo
  {volume} {120}},\ \bibinfo {pages} {190602} (\bibinfo {year}
  {2018})}\BibitemShut {NoStop}%
\bibitem [{\citenamefont {Gingrich}\ \emph {et~al.}(2016)\citenamefont
  {Gingrich}, \citenamefont {Horowitz}, \citenamefont {Perunov},\ and\
  \citenamefont {England}}]{Gingrich_2016}%
  \BibitemOpen
  \bibfield  {author} {\bibinfo {author} {\bibfnamefont {T.~R.}\ \bibnamefont
  {Gingrich}}, \bibinfo {author} {\bibfnamefont {J.~M.}\ \bibnamefont
  {Horowitz}}, \bibinfo {author} {\bibfnamefont {N.}~\bibnamefont {Perunov}}, \
  and\ \bibinfo {author} {\bibfnamefont {J.~L.}\ \bibnamefont {England}},\
  }\href {\doibase 10.1103/PhysRevLett.116.120601} {\bibfield  {journal}
  {\bibinfo  {journal} {Phys. Rev. Lett.}\ }\textbf {\bibinfo {volume} {116}},\
  \bibinfo {pages} {120601} (\bibinfo {year} {2016})}\BibitemShut {NoStop}%
\bibitem [{\citenamefont {Liu}\ and\ \citenamefont {Segal}(2019)}]{Liu_2019}%
  \BibitemOpen
  \bibfield  {author} {\bibinfo {author} {\bibfnamefont {J.}~\bibnamefont
  {Liu}}\ and\ \bibinfo {author} {\bibfnamefont {D.}~\bibnamefont {Segal}},\
  }\href {\doibase 10.1103/PhysRevE.99.062141} {\bibfield  {journal} {\bibinfo
  {journal} {Phys. Rev. E}\ }\textbf {\bibinfo {volume} {99}},\ \bibinfo
  {pages} {062141} (\bibinfo {year} {2019})}\BibitemShut {NoStop}%
\bibitem [{\citenamefont {Rignon-Bret}\ \emph {et~al.}(2021)\citenamefont
  {Rignon-Bret}, \citenamefont {Guarnieri}, \citenamefont {Goold},\ and\
  \citenamefont {Mitchison}}]{Bret_2021}%
  \BibitemOpen
  \bibfield  {author} {\bibinfo {author} {\bibfnamefont {A.}~\bibnamefont
  {Rignon-Bret}}, \bibinfo {author} {\bibfnamefont {G.}~\bibnamefont
  {Guarnieri}}, \bibinfo {author} {\bibfnamefont {J.}~\bibnamefont {Goold}}, \
  and\ \bibinfo {author} {\bibfnamefont {M.~T.}\ \bibnamefont {Mitchison}},\
  }\href {\doibase 10.1103/PhysRevE.103.012133} {\bibfield  {journal} {\bibinfo
   {journal} {Phys. Rev. E}\ }\textbf {\bibinfo {volume} {103}},\ \bibinfo
  {pages} {012133} (\bibinfo {year} {2021})}\BibitemShut {NoStop}%
\bibitem [{\citenamefont {Holubec}\ and\ \citenamefont
  {Ryabov}(2017)}]{Holubec_2017}%
  \BibitemOpen
  \bibfield  {author} {\bibinfo {author} {\bibfnamefont {V.}~\bibnamefont
  {Holubec}}\ and\ \bibinfo {author} {\bibfnamefont {A.}~\bibnamefont
  {Ryabov}},\ }\href {\doibase 10.1103/PhysRevE.96.030102} {\bibfield
  {journal} {\bibinfo  {journal} {Phys. Rev. E}\ }\textbf {\bibinfo {volume}
  {96}},\ \bibinfo {pages} {030102} (\bibinfo {year} {2017})}\BibitemShut
  {NoStop}%
\bibitem [{\citenamefont {Holubec}(2014)}]{Holubec_2014}%
  \BibitemOpen
  \bibfield  {author} {\bibinfo {author} {\bibfnamefont {V.}~\bibnamefont
  {Holubec}},\ }\href {\doibase 10.1088/1742-5468/2014/05/P05022} {\bibfield
  {journal} {\bibinfo  {journal} {Journal of Statistical Mechanics: Theory and
  Experiment}\ }\textbf {\bibinfo {volume} {2014}},\ \bibinfo {pages} {P05022}
  (\bibinfo {year} {2014})}\BibitemShut {NoStop}%
\bibitem [{\citenamefont {Souza}\ \emph {et~al.}(2022)\citenamefont {Souza},
  \citenamefont {Manzano}, \citenamefont {Fazio},\ and\ \citenamefont
  {Iemini}}]{Souza_2022}%
  \BibitemOpen
  \bibfield  {author} {\bibinfo {author} {\bibfnamefont {L.~d.~S.}\
  \bibnamefont {Souza}}, \bibinfo {author} {\bibfnamefont {G.}~\bibnamefont
  {Manzano}}, \bibinfo {author} {\bibfnamefont {R.}~\bibnamefont {Fazio}}, \
  and\ \bibinfo {author} {\bibfnamefont {F.}~\bibnamefont {Iemini}},\ }\href
  {\doibase 10.1103/PhysRevE.106.014143} {\bibfield  {journal} {\bibinfo
  {journal} {Phys. Rev. E}\ }\textbf {\bibinfo {volume} {106}},\ \bibinfo
  {pages} {014143} (\bibinfo {year} {2022})}\BibitemShut {NoStop}%
\bibitem [{\citenamefont {Latune}\ \emph {et~al.}(2021)\citenamefont {Latune},
  \citenamefont {Sinayskiy},\ and\ \citenamefont {Petruccione}}]{Latune_2021}%
  \BibitemOpen
  \bibfield  {author} {\bibinfo {author} {\bibfnamefont {C.~L.}\ \bibnamefont
  {Latune}}, \bibinfo {author} {\bibfnamefont {I.}~\bibnamefont {Sinayskiy}}, \
  and\ \bibinfo {author} {\bibfnamefont {F.}~\bibnamefont {Petruccione}},\
  }\href {\doibase 10.1140/epjs/s11734-021-00085-1} {\bibfield  {journal}
  {\bibinfo  {journal} {The European Physical Journal Special Topics}\ }\textbf
  {\bibinfo {volume} {230}},\ \bibinfo {pages} {841} (\bibinfo {year}
  {2021})}\BibitemShut {NoStop}%
\bibitem [{\citenamefont {Streltsov}\ \emph {et~al.}(2017)\citenamefont
  {Streltsov}, \citenamefont {Adesso},\ and\ \citenamefont
  {Plenio}}]{Streltsov_2017}%
  \BibitemOpen
  \bibfield  {author} {\bibinfo {author} {\bibfnamefont {A.}~\bibnamefont
  {Streltsov}}, \bibinfo {author} {\bibfnamefont {G.}~\bibnamefont {Adesso}}, \
  and\ \bibinfo {author} {\bibfnamefont {M.~B.}\ \bibnamefont {Plenio}},\
  }\href {\doibase 10.1103/RevModPhys.89.041003} {\bibfield  {journal}
  {\bibinfo  {journal} {Rev. Mod. Phys.}\ }\textbf {\bibinfo {volume} {89}},\
  \bibinfo {pages} {041003} (\bibinfo {year} {2017})}\BibitemShut {NoStop}%
\bibitem [{\citenamefont {Park}\ \emph {et~al.}(2013)\citenamefont {Park},
  \citenamefont {Kim}, \citenamefont {Sagawa},\ and\ \citenamefont
  {Kim}}]{Park_2013}%
  \BibitemOpen
  \bibfield  {author} {\bibinfo {author} {\bibfnamefont {J.~J.}\ \bibnamefont
  {Park}}, \bibinfo {author} {\bibfnamefont {K.-H.}\ \bibnamefont {Kim}},
  \bibinfo {author} {\bibfnamefont {T.}~\bibnamefont {Sagawa}}, \ and\ \bibinfo
  {author} {\bibfnamefont {S.~W.}\ \bibnamefont {Kim}},\ }\href {\doibase
  10.1103/PhysRevLett.111.230402} {\bibfield  {journal} {\bibinfo  {journal}
  {Phys. Rev. Lett.}\ }\textbf {\bibinfo {volume} {111}},\ \bibinfo {pages}
  {230402} (\bibinfo {year} {2013})}\BibitemShut {NoStop}%
\bibitem [{\citenamefont {Brandner}\ \emph {et~al.}(2017)\citenamefont
  {Brandner}, \citenamefont {Bauer},\ and\ \citenamefont
  {Seifert}}]{Brandner_2017}%
  \BibitemOpen
  \bibfield  {author} {\bibinfo {author} {\bibfnamefont {K.}~\bibnamefont
  {Brandner}}, \bibinfo {author} {\bibfnamefont {M.}~\bibnamefont {Bauer}}, \
  and\ \bibinfo {author} {\bibfnamefont {U.}~\bibnamefont {Seifert}},\ }\href
  {\doibase 10.1103/PhysRevLett.119.170602} {\bibfield  {journal} {\bibinfo
  {journal} {Phys. Rev. Lett.}\ }\textbf {\bibinfo {volume} {119}},\ \bibinfo
  {pages} {170602} (\bibinfo {year} {2017})}\BibitemShut {NoStop}%
\bibitem [{\citenamefont {Hammam}\ \emph {et~al.}(2021)\citenamefont {Hammam},
  \citenamefont {Hassouni}, \citenamefont {Fazio},\ and\ \citenamefont
  {Manzano}}]{Hammam_2021}%
  \BibitemOpen
  \bibfield  {author} {\bibinfo {author} {\bibfnamefont {K.}~\bibnamefont
  {Hammam}}, \bibinfo {author} {\bibfnamefont {Y.}~\bibnamefont {Hassouni}},
  \bibinfo {author} {\bibfnamefont {R.}~\bibnamefont {Fazio}}, \ and\ \bibinfo
  {author} {\bibfnamefont {G.}~\bibnamefont {Manzano}},\ }\href {\doibase
  10.1088/1367-2630/abeb47} {\bibfield  {journal} {\bibinfo  {journal} {New
  Journal of Physics}\ }\textbf {\bibinfo {volume} {23}},\ \bibinfo {pages}
  {043024} (\bibinfo {year} {2021})}\BibitemShut {NoStop}%
\bibitem [{\citenamefont {Manzano}\ \emph {et~al.}(2019)\citenamefont
  {Manzano}, \citenamefont {Silva},\ and\ \citenamefont
  {Parrondo}}]{Manzano_2019}%
  \BibitemOpen
  \bibfield  {author} {\bibinfo {author} {\bibfnamefont {G.}~\bibnamefont
  {Manzano}}, \bibinfo {author} {\bibfnamefont {R.}~\bibnamefont {Silva}}, \
  and\ \bibinfo {author} {\bibfnamefont {J.~M.~R.}\ \bibnamefont {Parrondo}},\
  }\href {\doibase 10.1103/PhysRevE.99.042135} {\bibfield  {journal} {\bibinfo
  {journal} {Phys. Rev. E}\ }\textbf {\bibinfo {volume} {99}},\ \bibinfo
  {pages} {042135} (\bibinfo {year} {2019})}\BibitemShut {NoStop}%
\bibitem [{\citenamefont {Camati}\ \emph {et~al.}(2019)\citenamefont {Camati},
  \citenamefont {Santos},\ and\ \citenamefont {Serra}}]{Camati_2019}%
  \BibitemOpen
  \bibfield  {author} {\bibinfo {author} {\bibfnamefont {P.~A.}\ \bibnamefont
  {Camati}}, \bibinfo {author} {\bibfnamefont {J.~F.~G.}\ \bibnamefont
  {Santos}}, \ and\ \bibinfo {author} {\bibfnamefont {R.~M.}\ \bibnamefont
  {Serra}},\ }\href {\doibase 10.1103/PhysRevA.99.062103} {\bibfield  {journal}
  {\bibinfo  {journal} {Phys. Rev. A}\ }\textbf {\bibinfo {volume} {99}},\
  \bibinfo {pages} {062103} (\bibinfo {year} {2019})}\BibitemShut {NoStop}%
\bibitem [{\citenamefont {Killoran}\ \emph {et~al.}(2015)\citenamefont
  {Killoran}, \citenamefont {Huelga},\ and\ \citenamefont
  {Plenio}}]{Killoran_2015}%
  \BibitemOpen
  \bibfield  {author} {\bibinfo {author} {\bibfnamefont {N.}~\bibnamefont
  {Killoran}}, \bibinfo {author} {\bibfnamefont {S.~F.}\ \bibnamefont
  {Huelga}}, \ and\ \bibinfo {author} {\bibfnamefont {M.~B.}\ \bibnamefont
  {Plenio}},\ }\href {\doibase 10.1063/1.4932307} {\bibfield  {journal}
  {\bibinfo  {journal} {The Journal of Chemical Physics}\ }\textbf {\bibinfo
  {volume} {143}},\ \bibinfo {pages} {155102} (\bibinfo {year} {2015})},\
  \Eprint {http://arxiv.org/abs/https://doi.org/10.1063/1.4932307}
  {https://doi.org/10.1063/1.4932307} \BibitemShut {NoStop}%
\bibitem [{\citenamefont {Scully}\ \emph {et~al.}(2011)\citenamefont {Scully},
  \citenamefont {Chapin}, \citenamefont {Dorfman}, \citenamefont {Kim},\ and\
  \citenamefont {Svidzinsky}}]{Scully_2011}%
  \BibitemOpen
  \bibfield  {author} {\bibinfo {author} {\bibfnamefont {M.~O.}\ \bibnamefont
  {Scully}}, \bibinfo {author} {\bibfnamefont {K.~R.}\ \bibnamefont {Chapin}},
  \bibinfo {author} {\bibfnamefont {K.~E.}\ \bibnamefont {Dorfman}}, \bibinfo
  {author} {\bibfnamefont {M.~B.}\ \bibnamefont {Kim}}, \ and\ \bibinfo
  {author} {\bibfnamefont {A.}~\bibnamefont {Svidzinsky}},\ }\href {\doibase
  10.1073/pnas.111023410} {\bibfield  {journal} {\bibinfo  {journal}
  {Proceedings of the National Academy of Sciences}\ }\textbf {\bibinfo
  {volume} {108}},\ \bibinfo {pages} {15097} (\bibinfo {year}
  {2011})}\BibitemShut {NoStop}%
\bibitem [{\citenamefont {Holubec}\ and\ \citenamefont
  {Novotn{\'y}}(2018)}]{Holubec_2018}%
  \BibitemOpen
  \bibfield  {author} {\bibinfo {author} {\bibfnamefont {V.}~\bibnamefont
  {Holubec}}\ and\ \bibinfo {author} {\bibfnamefont {T.}~\bibnamefont
  {Novotn{\'y}}},\ }\href {\doibase 10.1007/s10909-018-1960-x} {\bibfield
  {journal} {\bibinfo  {journal} {Journal of Low Temperature Physics}\ }\textbf
  {\bibinfo {volume} {192}},\ \bibinfo {pages} {147} (\bibinfo {year}
  {2018})}\BibitemShut {NoStop}%
\bibitem [{\citenamefont {Kilgour}\ and\ \citenamefont
  {Segal}(2018)}]{Kilgour_2018}%
  \BibitemOpen
  \bibfield  {author} {\bibinfo {author} {\bibfnamefont {M.}~\bibnamefont
  {Kilgour}}\ and\ \bibinfo {author} {\bibfnamefont {D.}~\bibnamefont
  {Segal}},\ }\href {\doibase 10.1103/PhysRevE.98.012117} {\bibfield  {journal}
  {\bibinfo  {journal} {Phys. Rev. E}\ }\textbf {\bibinfo {volume} {98}},\
  \bibinfo {pages} {012117} (\bibinfo {year} {2018})}\BibitemShut {NoStop}%
\bibitem [{\citenamefont {Du}\ and\ \citenamefont {Zhang}(2018)}]{Du_2018}%
  \BibitemOpen
  \bibfield  {author} {\bibinfo {author} {\bibfnamefont {J.-Y.}\ \bibnamefont
  {Du}}\ and\ \bibinfo {author} {\bibfnamefont {F.-L.}\ \bibnamefont {Zhang}},\
  }\href {\doibase 10.1088/1367-2630/aac688} {\bibfield  {journal} {\bibinfo
  {journal} {New Journal of Physics}\ }\textbf {\bibinfo {volume} {20}},\
  \bibinfo {pages} {063005} (\bibinfo {year} {2018})}\BibitemShut {NoStop}%
\bibitem [{\citenamefont {Correa}\ \emph {et~al.}(2014)\citenamefont {Correa},
  \citenamefont {Palao}, \citenamefont {Alonso},\ and\ \citenamefont
  {Adesso}}]{Correa_2014}%
  \BibitemOpen
  \bibfield  {author} {\bibinfo {author} {\bibfnamefont {L.~A.}\ \bibnamefont
  {Correa}}, \bibinfo {author} {\bibfnamefont {J.}~\bibnamefont {Palao}},
  \bibinfo {author} {\bibfnamefont {D.}~\bibnamefont {Alonso}}, \ and\ \bibinfo
  {author} {\bibfnamefont {G.}~\bibnamefont {Adesso}},\ }\href {\doibase
  10.1038/srep03949} {\bibfield  {journal} {\bibinfo  {journal} {Scientific
  Reports}\ }\textbf {\bibinfo {volume} {4}},\ \bibinfo {pages} {3949}
  (\bibinfo {year} {2014})}\BibitemShut {NoStop}%
\bibitem [{\citenamefont {Agarwalla}\ and\ \citenamefont
  {Segal}(2018)}]{Agarwalla_2018}%
  \BibitemOpen
  \bibfield  {author} {\bibinfo {author} {\bibfnamefont {B.~K.}\ \bibnamefont
  {Agarwalla}}\ and\ \bibinfo {author} {\bibfnamefont {D.}~\bibnamefont
  {Segal}},\ }\href {\doibase 10.1103/PhysRevB.98.155438} {\bibfield  {journal}
  {\bibinfo  {journal} {Phys. Rev. B}\ }\textbf {\bibinfo {volume} {98}},\
  \bibinfo {pages} {155438} (\bibinfo {year} {2018})}\BibitemShut {NoStop}%
\bibitem [{\citenamefont {Kalaee}\ \emph
  {et~al.}(2021{\natexlab{a}})\citenamefont {Kalaee}, \citenamefont {Wacker},\
  and\ \citenamefont {Potts}}]{Kalaee_2021}%
  \BibitemOpen
  \bibfield  {author} {\bibinfo {author} {\bibfnamefont {A.~A.~S.}\
  \bibnamefont {Kalaee}}, \bibinfo {author} {\bibfnamefont {A.}~\bibnamefont
  {Wacker}}, \ and\ \bibinfo {author} {\bibfnamefont {P.~P.}\ \bibnamefont
  {Potts}},\ }\href {\doibase 10.1103/PhysRevE.104.L012103} {\bibfield
  {journal} {\bibinfo  {journal} {Phys. Rev. E}\ }\textbf {\bibinfo {volume}
  {104}},\ \bibinfo {pages} {L012103} (\bibinfo {year}
  {2021}{\natexlab{a}})}\BibitemShut {NoStop}%
\bibitem [{\citenamefont {Menczel}\ \emph {et~al.}(2021)\citenamefont
  {Menczel}, \citenamefont {Loisa}, \citenamefont {Brandner},\ and\
  \citenamefont {Flindt}}]{Menczel_2021}%
  \BibitemOpen
  \bibfield  {author} {\bibinfo {author} {\bibfnamefont {P.}~\bibnamefont
  {Menczel}}, \bibinfo {author} {\bibfnamefont {E.}~\bibnamefont {Loisa}},
  \bibinfo {author} {\bibfnamefont {K.}~\bibnamefont {Brandner}}, \ and\
  \bibinfo {author} {\bibfnamefont {C.}~\bibnamefont {Flindt}},\ }\href
  {\doibase 10.1088/1751-8121/ac0c8f} {\bibfield  {journal} {\bibinfo
  {journal} {Journal of Physics A: Mathematical and Theoretical}\ }\textbf
  {\bibinfo {volume} {54}},\ \bibinfo {pages} {314002} (\bibinfo {year}
  {2021})}\BibitemShut {NoStop}%
\bibitem [{\citenamefont {Liu}\ and\ \citenamefont {Segal}(2021)}]{Liu_2021}%
  \BibitemOpen
  \bibfield  {author} {\bibinfo {author} {\bibfnamefont {J.}~\bibnamefont
  {Liu}}\ and\ \bibinfo {author} {\bibfnamefont {D.}~\bibnamefont {Segal}},\
  }\href {\doibase 10.1103/PhysRevE.103.032138} {\bibfield  {journal} {\bibinfo
   {journal} {Phys. Rev. E}\ }\textbf {\bibinfo {volume} {103}},\ \bibinfo
  {pages} {032138} (\bibinfo {year} {2021})}\BibitemShut {NoStop}%
\bibitem [{\citenamefont {Timpanaro}\ \emph {et~al.}(2019)\citenamefont
  {Timpanaro}, \citenamefont {Guarnieri}, \citenamefont {Goold},\ and\
  \citenamefont {Landi}}]{Timpanaro_2019}%
  \BibitemOpen
  \bibfield  {author} {\bibinfo {author} {\bibfnamefont {A.~M.}\ \bibnamefont
  {Timpanaro}}, \bibinfo {author} {\bibfnamefont {G.}~\bibnamefont
  {Guarnieri}}, \bibinfo {author} {\bibfnamefont {J.}~\bibnamefont {Goold}}, \
  and\ \bibinfo {author} {\bibfnamefont {G.~T.}\ \bibnamefont {Landi}},\ }\href
  {\doibase 10.1103/PhysRevLett.123.090604} {\bibfield  {journal} {\bibinfo
  {journal} {Phys. Rev. Lett.}\ }\textbf {\bibinfo {volume} {123}},\ \bibinfo
  {pages} {090604} (\bibinfo {year} {2019})}\BibitemShut {NoStop}%
\bibitem [{\citenamefont {Kalaee}\ \emph
  {et~al.}(2021{\natexlab{b}})\citenamefont {Kalaee}, \citenamefont {Wacker},\
  and\ \citenamefont {Potts}}]{Arash_2021}%
  \BibitemOpen
  \bibfield  {author} {\bibinfo {author} {\bibfnamefont {A.~A.~S.}\
  \bibnamefont {Kalaee}}, \bibinfo {author} {\bibfnamefont {A.}~\bibnamefont
  {Wacker}}, \ and\ \bibinfo {author} {\bibfnamefont {P.~P.}\ \bibnamefont
  {Potts}},\ }\href {\doibase 10.1103/PhysRevE.104.L012103} {\bibfield
  {journal} {\bibinfo  {journal} {Phys. Rev. E}\ }\textbf {\bibinfo {volume}
  {104}},\ \bibinfo {pages} {L012103} (\bibinfo {year}
  {2021}{\natexlab{b}})}\BibitemShut {NoStop}%
\bibitem [{\citenamefont {Ptaszy\ifmmode~\acute{n}\else
  \'{n}\fi{}ski}(2018)}]{Ptaszynnski_2018}%
  \BibitemOpen
  \bibfield  {author} {\bibinfo {author} {\bibfnamefont {K.}~\bibnamefont
  {Ptaszy\ifmmode~\acute{n}\else \'{n}\fi{}ski}},\ }\href {\doibase
  10.1103/PhysRevB.98.085425} {\bibfield  {journal} {\bibinfo  {journal} {Phys.
  Rev. B}\ }\textbf {\bibinfo {volume} {98}},\ \bibinfo {pages} {085425}
  (\bibinfo {year} {2018})}\BibitemShut {NoStop}%
\bibitem [{\citenamefont {Erdman}\ \emph {et~al.}(2019)\citenamefont {Erdman},
  \citenamefont {Cavina}, \citenamefont {Fazio}, \citenamefont {Taddei},\ and\
  \citenamefont {Giovannetti}}]{Erdman_2019}%
  \BibitemOpen
  \bibfield  {author} {\bibinfo {author} {\bibfnamefont {P.~A.}\ \bibnamefont
  {Erdman}}, \bibinfo {author} {\bibfnamefont {V.}~\bibnamefont {Cavina}},
  \bibinfo {author} {\bibfnamefont {R.}~\bibnamefont {Fazio}}, \bibinfo
  {author} {\bibfnamefont {F.}~\bibnamefont {Taddei}}, \ and\ \bibinfo {author}
  {\bibfnamefont {V.}~\bibnamefont {Giovannetti}},\ }\href {\doibase
  10.1088/1367-2630/ab4dca} {\bibfield  {journal} {\bibinfo  {journal} {New
  Journal of Physics}\ }\textbf {\bibinfo {volume} {21}},\ \bibinfo {pages}
  {103049} (\bibinfo {year} {2019})}\BibitemShut {NoStop}%
\bibitem [{\citenamefont {Cavina}\ \emph {et~al.}(2021)\citenamefont {Cavina},
  \citenamefont {Erdman}, \citenamefont {Abiuso}, \citenamefont {Tolomeo},\
  and\ \citenamefont {Giovannetti}}]{Cavina_2021}%
  \BibitemOpen
  \bibfield  {author} {\bibinfo {author} {\bibfnamefont {V.}~\bibnamefont
  {Cavina}}, \bibinfo {author} {\bibfnamefont {P.~A.}\ \bibnamefont {Erdman}},
  \bibinfo {author} {\bibfnamefont {P.}~\bibnamefont {Abiuso}}, \bibinfo
  {author} {\bibfnamefont {L.}~\bibnamefont {Tolomeo}}, \ and\ \bibinfo
  {author} {\bibfnamefont {V.}~\bibnamefont {Giovannetti}},\ }\href {\doibase
  10.1103/PhysRevA.104.032226} {\bibfield  {journal} {\bibinfo  {journal}
  {Phys. Rev. A}\ }\textbf {\bibinfo {volume} {104}},\ \bibinfo {pages}
  {032226} (\bibinfo {year} {2021})}\BibitemShut {NoStop}%
\bibitem [{\citenamefont {Pekola}\ \emph {et~al.}(2019)\citenamefont {Pekola},
  \citenamefont {Karimi}, \citenamefont {Thomas},\ and\ \citenamefont
  {Averin}}]{Pekola_2019_2}%
  \BibitemOpen
  \bibfield  {author} {\bibinfo {author} {\bibfnamefont {J.~P.}\ \bibnamefont
  {Pekola}}, \bibinfo {author} {\bibfnamefont {B.}~\bibnamefont {Karimi}},
  \bibinfo {author} {\bibfnamefont {G.}~\bibnamefont {Thomas}}, \ and\ \bibinfo
  {author} {\bibfnamefont {D.~V.}\ \bibnamefont {Averin}},\ }\href {\doibase
  10.1103/PhysRevB.100.085405} {\bibfield  {journal} {\bibinfo  {journal}
  {Phys. Rev. B}\ }\textbf {\bibinfo {volume} {100}},\ \bibinfo {pages}
  {085405} (\bibinfo {year} {2019})}\BibitemShut {NoStop}%
\bibitem [{\citenamefont {Oliver}\ \emph {et~al.}(2005)\citenamefont {Oliver},
  \citenamefont {Yu}, \citenamefont {Lee}, \citenamefont {Berggren},
  \citenamefont {Levitov},\ and\ \citenamefont {Orlando}}]{oliver_2005}%
  \BibitemOpen
  \bibfield  {author} {\bibinfo {author} {\bibfnamefont {W.~D.}\ \bibnamefont
  {Oliver}}, \bibinfo {author} {\bibfnamefont {Y.}~\bibnamefont {Yu}}, \bibinfo
  {author} {\bibfnamefont {J.~C.}\ \bibnamefont {Lee}}, \bibinfo {author}
  {\bibfnamefont {K.~K.}\ \bibnamefont {Berggren}}, \bibinfo {author}
  {\bibfnamefont {L.~S.}\ \bibnamefont {Levitov}}, \ and\ \bibinfo {author}
  {\bibfnamefont {T.~P.}\ \bibnamefont {Orlando}},\ }\href {\doibase
  10.1126/science.1119678} {\bibfield  {journal} {\bibinfo  {journal}
  {Science}\ }\textbf {\bibinfo {volume} {310}},\ \bibinfo {pages} {1653}
  (\bibinfo {year} {2005})}\BibitemShut {NoStop}%
\bibitem [{\citenamefont {Oliver}\ and\ \citenamefont
  {Valenzuela}(2009)}]{oliver_2009}%
  \BibitemOpen
  \bibfield  {author} {\bibinfo {author} {\bibfnamefont {W.~D.}\ \bibnamefont
  {Oliver}}\ and\ \bibinfo {author} {\bibfnamefont {S.~O.}\ \bibnamefont
  {Valenzuela}},\ }\href {\doibase 10.1007/s11128-009-0108-y} {\bibfield
  {journal} {\bibinfo  {journal} {Quantum Information Processing}\ }\textbf
  {\bibinfo {volume} {8}},\ \bibinfo {pages} {261} (\bibinfo {year}
  {2009})}\BibitemShut {NoStop}%
\bibitem [{\citenamefont {Koch}\ \emph {et~al.}(2007)\citenamefont {Koch},
  \citenamefont {Yu}, \citenamefont {Gambetta}, \citenamefont {Houck},
  \citenamefont {Schuster}, \citenamefont {Majer}, \citenamefont {Blais},
  \citenamefont {Devoret}, \citenamefont {Girvin},\ and\ \citenamefont
  {Schoelkopf}}]{Koch_2007}%
  \BibitemOpen
  \bibfield  {author} {\bibinfo {author} {\bibfnamefont {J.}~\bibnamefont
  {Koch}}, \bibinfo {author} {\bibfnamefont {T.~M.}\ \bibnamefont {Yu}},
  \bibinfo {author} {\bibfnamefont {J.}~\bibnamefont {Gambetta}}, \bibinfo
  {author} {\bibfnamefont {A.~A.}\ \bibnamefont {Houck}}, \bibinfo {author}
  {\bibfnamefont {D.~I.}\ \bibnamefont {Schuster}}, \bibinfo {author}
  {\bibfnamefont {J.}~\bibnamefont {Majer}}, \bibinfo {author} {\bibfnamefont
  {A.}~\bibnamefont {Blais}}, \bibinfo {author} {\bibfnamefont {M.~H.}\
  \bibnamefont {Devoret}}, \bibinfo {author} {\bibfnamefont {S.~M.}\
  \bibnamefont {Girvin}}, \ and\ \bibinfo {author} {\bibfnamefont {R.~J.}\
  \bibnamefont {Schoelkopf}},\ }\href {\doibase 10.1103/PhysRevA.76.042319}
  {\bibfield  {journal} {\bibinfo  {journal} {Phys. Rev. A}\ }\textbf {\bibinfo
  {volume} {76}},\ \bibinfo {pages} {042319} (\bibinfo {year}
  {2007})}\BibitemShut {NoStop}%
\bibitem [{\citenamefont {Berns}\ \emph {et~al.}(2008)\citenamefont {Berns},
  \citenamefont {Rudner}, \citenamefont {Valenzuela}, \citenamefont {Berggren},
  \citenamefont {Oliver}, \citenamefont {Levitov},\ and\ \citenamefont
  {Orlando}}]{berns_2008}%
  \BibitemOpen
  \bibfield  {author} {\bibinfo {author} {\bibfnamefont {D.~M.}\ \bibnamefont
  {Berns}}, \bibinfo {author} {\bibfnamefont {M.~S.}\ \bibnamefont {Rudner}},
  \bibinfo {author} {\bibfnamefont {S.~O.}\ \bibnamefont {Valenzuela}},
  \bibinfo {author} {\bibfnamefont {K.~K.}\ \bibnamefont {Berggren}}, \bibinfo
  {author} {\bibfnamefont {W.~D.}\ \bibnamefont {Oliver}}, \bibinfo {author}
  {\bibfnamefont {L.~S.}\ \bibnamefont {Levitov}}, \ and\ \bibinfo {author}
  {\bibfnamefont {T.~P.}\ \bibnamefont {Orlando}},\ }\href
  {https://doi.org/10.1038/nature07262} {\bibfield  {journal} {\bibinfo
  {journal} {Nature}\ }\textbf {\bibinfo {volume} {455}},\ \bibinfo {pages} {51
  EP } (\bibinfo {year} {2008})}\BibitemShut {NoStop}%
\bibitem [{\citenamefont {Clarke}\ and\ \citenamefont
  {Wilhelm}(2008)}]{Clarke_2008}%
  \BibitemOpen
  \bibfield  {author} {\bibinfo {author} {\bibfnamefont {J.}~\bibnamefont
  {Clarke}}\ and\ \bibinfo {author} {\bibfnamefont {F.~K.}\ \bibnamefont
  {Wilhelm}},\ }\href {\doibase 10.1038/nature07128} {\bibfield  {journal}
  {\bibinfo  {journal} {Nature}\ }\textbf {\bibinfo {volume} {453}},\ \bibinfo
  {pages} {1031} (\bibinfo {year} {2008})}\BibitemShut {NoStop}%
\bibitem [{\citenamefont {Solfanelli}\ \emph {et~al.}(2020)\citenamefont
  {Solfanelli}, \citenamefont {Falsetti},\ and\ \citenamefont
  {Campisi}}]{Solfanelli_2020}%
  \BibitemOpen
  \bibfield  {author} {\bibinfo {author} {\bibfnamefont {A.}~\bibnamefont
  {Solfanelli}}, \bibinfo {author} {\bibfnamefont {M.}~\bibnamefont
  {Falsetti}}, \ and\ \bibinfo {author} {\bibfnamefont {M.}~\bibnamefont
  {Campisi}},\ }\href {\doibase 10.1103/PhysRevB.101.054513} {\bibfield
  {journal} {\bibinfo  {journal} {Phys. Rev. B}\ }\textbf {\bibinfo {volume}
  {101}},\ \bibinfo {pages} {054513} (\bibinfo {year} {2020})}\BibitemShut
  {NoStop}%
\bibitem [{\citenamefont {Carmichael}(1993)}]{Carmichael_1993}%
  \BibitemOpen
  \bibfield  {author} {\bibinfo {author} {\bibfnamefont {H.}~\bibnamefont
  {Carmichael}},\ }\href@noop {} {\emph {\bibinfo {title} {An open systems
  approach to quantum optics, Lecture Notes in Physics Monographs}}}\ (\bibinfo
   {publisher} {Springer},\ \bibinfo {address} {Berlin, Heidelberg},\ \bibinfo
  {year} {1993})\BibitemShut {NoStop}%
\bibitem [{\citenamefont {Wiseman}\ and\ \citenamefont
  {Milburn}(2009)}]{Wiseman_2009}%
  \BibitemOpen
  \bibfield  {author} {\bibinfo {author} {\bibfnamefont {H.~M.}\ \bibnamefont
  {Wiseman}}\ and\ \bibinfo {author} {\bibfnamefont {G.~J.}\ \bibnamefont
  {Milburn}},\ }\href {\doibase 10.1017/CBO9780511813948} {\emph {\bibinfo
  {title} {Quantum Measurement and Control}}}\ (\bibinfo  {publisher}
  {Cambridge University Press},\ \bibinfo {year} {2009})\BibitemShut {NoStop}%
\bibitem [{\citenamefont {Campisi}(2014)}]{Campisi_2014}%
  \BibitemOpen
  \bibfield  {author} {\bibinfo {author} {\bibfnamefont {M.}~\bibnamefont
  {Campisi}},\ }\href {\doibase 10.1088/1751-8113/47/24/245001} {\bibfield
  {journal} {\bibinfo  {journal} {Journal of Physics A: Mathematical and
  Theoretical}\ }\textbf {\bibinfo {volume} {47}},\ \bibinfo {pages} {245001}
  (\bibinfo {year} {2014})}\BibitemShut {NoStop}%
\bibitem [{\citenamefont {Jarzynski}(2006)}]{Jarzynski_2006}%
  \BibitemOpen
  \bibfield  {author} {\bibinfo {author} {\bibfnamefont {C.}~\bibnamefont
  {Jarzynski}},\ }\href {\doibase 10.1103/PhysRevE.73.046105} {\bibfield
  {journal} {\bibinfo  {journal} {Phys. Rev. E}\ }\textbf {\bibinfo {volume}
  {73}},\ \bibinfo {pages} {046105} (\bibinfo {year} {2006})}\BibitemShut
  {NoStop}%
\end{thebibliography}%

\end{document}